\documentclass[11pt,a4paper]{article}

\usepackage{cite}
\usepackage{stackrel}
\usepackage{graphicx}
\usepackage{amsmath}
\usepackage{amssymb,MnSymbol}
\usepackage{amsfonts}
\usepackage{dsfont}
\usepackage{mathtools}
\usepackage{array}
\usepackage{rotating}
\usepackage{bbold,amsfonts}
\allowdisplaybreaks

\usepackage{comment}
\usepackage{soul}

\newcommand\expOO{ {\cal O}_H }
\newcommand\expO{ {O}_H }

\usepackage{stackengine,scalerel}
\newcommand\gl{\mathrel{\hstretch{1.5}{%
  \stackanchor[1pt]{\scriptscriptstyle>}{\scriptscriptstyle<}}}}
\stackMath

\usepackage[utf8]{inputenc}
\usepackage{bm}
\usepackage{xcolor}
\usepackage{float}
\usepackage{braket}
\usepackage{placeins}
\usepackage[height=8.8in,width=6.45in]{geometry}
\usepackage[font=small,labelfont=bf]{caption}
\usepackage[hidelinks]{hyperref}
\usepackage{booktabs,float,slashed}
\usepackage{standalone}
\usepackage{caption}
\usepackage{subcaption}
\usepackage{makecell}

\numberwithin{equation}{section}

\newcommand{\eisen}[1]{E^*\left(#1;\tau\right)}
\newcommand{\average}[1]{\langle #1\rangle}
\newcommand{\tFo}[4]{{}_2F_1\left(#1,#2;#3\vert #4\right)}
\newcommand{\tFt}[6]{{}_3F_2\left(#1,#2,#3;#4,#5\vert #6\right)}
\newcommand{\dd}{\mathrm{d}}
\newcommand{\Ymn}{Y_{mn}(\tau)}
\newcommand{\NPfn}[1]{\mathcal{E}(#1;\tau)}
\newcommand{\summn}{\sum_{(m,n)\neq (0,0)}}
\newcommand{\intRes}{\int_{{\rm Re}(s)=\frac{1}{2}}}
\newcommand{\intRepsilon}{\int_{{\rm Re}(s)=1+\epsilon}}
\newcommand{\Dfn}[2]{D_#1\left( #2;\tau\right)}

\newcommand{\Omax}[1]{\widehat{\mathcal{O}}_{#1}}

\def\half{{\scriptstyle \frac 12}}
\def\cN{{\mathcal{N}}}

\def\faH{{ H}}

\newcommand{\gym}{g_{_{\rm YM}}}

\newcommand{\HHLL}{{\rm HHLL}}

\usepackage{tikz}
\usetikzlibrary{calc}
\usepackage{color}
\usetikzlibrary{arrows.meta}
\usetikzlibrary{decorations.pathmorphing,arrows.meta,shadows,calc}

\def\sstar{{\star}}

\begin{document}

\begin{titlepage}

\vspace*{10mm}
\begin{center}
{\LARGE \bf 
Dynamics of Heavy Operators in $\mathcal{N}=4$ SYM: 
\vspace{0.2cm}

Integrated Correlators and AdS Bubbles}
\vspace*{15mm}

{\Large Francesco Aprile$^{(a)}$, Daniele Dorigoni$^{(b)}$, and Rudolfs Treilis$^{(c)}$ }

\vspace*{8mm}

$(a)$ Departamento de Fisica Teorica \& IPARCOS,\\ Facultad de Ciencias Fisicas, Universidad Complutense, 28040 Madrid\\
			\vskip 0.3cm
$(b)$ Centre for Particle Theory \& Department of Mathematical Sciences Durham University, \\
Lower Mountjoy, Stockton Road, Durham DH1 3LE, UK
\vskip 0.3cm

$(c)$ Dipartimento SMFI, Università di Parma,
Viale G.P. Usberti 7/A, 43100, Parma, Italy
    
\vspace*{0.8cm}

\end{center}

\begin{abstract}

We study integrated correlation functions of half-BPS operators in $SU(N)$ $\mathcal{N} = 4$ 
supersymmetric Yang-Mills theory (SYM) involving two superconformal primary operators in the stress-tensor multiplet 
and two identical maximal-trace operators of arbitrary $R$-charge $p$.   
Thanks to $\mathcal{N}=4$ SYM electro-magnetic duality these integrated correlators have recently 
been computed as exact functions of $N$, $p$, and of the Yang-Mills complexified coupling $\tau$.
Using a combination of tools from ${\rm SL}(2,\mathbb{Z})$ spectral theory and resurgence analysis, we study the 
landscape of large-$N$ and/or large-charge expansions for these correlators. 
In particular, we find novel non-perturbative effects in the limit where $N\rightarrow \infty$ with $p/N^2$ fixed. From a holographic point of view this double-scaling regime is deeply connected with a second family of correlators which we analyse. 
Using the results for the maximal-trace operators, we derive an exact expression for a new integrated correlator involving two coherent-state operators, defined via an exponential generating function of multi-graviton states. At large-$N$ this correlator admits a holographic dual description in terms of a back-reacted geometry known as the AdS bubble. First, we show that the leading supergravity contribution to the integrated correlator agrees with a direct explicit integration of the correlator itself. Secondly, we derive predictions for the integrated version of the Virasoro--Shapiro amplitude evaluated on the AdS bubble background. Lastly, we demonstrate that the large-$N$ non-perturbative contributions to this integrated correlator emerge from giant-magnon configurations in the dual AdS bubble.

\vspace*{0.3cm}

\end{abstract}
\end{titlepage}
\setcounter{tocdepth}{2}
\tableofcontents

\section{Introduction}
\label{sec:intro}

A central challenge in theoretical physics is understanding how gauge 
theories behave at strong coupling. A promising strategy in this direction 
is the development of non-perturbative methods that leverage the power of 
dualities. In this work, we study
$\mathcal{N}=4$ supersymmetric Yang–Mills (SYM) theory with gauge group $SU(N)$, 
a paradigmatic model that serves as an ideal testing ground for such methods. This theory is the maximally supersymmetric gauge theory in four spacetime dimensions, it is 
invariant under electric–magnetic duality \cite{Montonen:1977sn,Witten:1978mh,Osborn:1979tq}, 
and remains (super)conformal at the quantum level. Furthermore, thanks to 
localisation \cite{Pestun:2016zxk} and integrability \cite{Beisert:2010jr}, 
a wide class of
observables in $\mathcal{N}=4$ SYM can be computed exactly. 
Finally, $\mathcal{N}=4$ SYM admits a dual description as type IIB string 
theory on ${\rm AdS}_5 \times {\rm S}^5$ through the celebrated AdS/CFT correspondence \cite{Maldacena:1997re}.

In our programme, four-point correlation functions of half-BPS operators 
play a distinguished role. The shortening condition ensures that 
the scaling dimension \(\Delta\) of a half-BPS operator is fixed and 
equal to the $R-$charge $p$ of the operator. Moreover, four-point correlators are the 
simplest observables that capture non-trivial dynamical information, 
in contrast to two- and three-point functions, which are protected 
and therefore independent of the gauge coupling. A wealth of results 
for four-point correlators with external half-BPS operators is available 
at both weak coupling \cite{Eden:2012tu,Drummond:2013nda,Chicherin:2015edu,Bourjaily:2016evz,Coronado:2018cxj,Caron-Huot:2021usw,Bargheer:2022sfd,He:2024cej,Bourjaily:2025iad} and strong coupling \cite{Goncalves:2014ffa,Rastelli:2016nze,Alday:2017xua,Caron-Huot:2018kta,Alday:2019nin,Goncalves:2019znr,Chester:2019pvm,Drummond:2020dwr,
Abl:2020dbx,Aprile:2020mus,Green:2020eyj,Huang:2021xws,Drummond:2022dxw,
Goncalves:2023oyx,Alday:2023mvu,Aprile:2024lwy,Fernandes:2025eqe}. 
More recently, significant progress has been made in the computation 
of so-called integrated correlators, which will be the focus of this work.

As the name suggests, integrated correlators are a new type of observable
defined by taking a four-point function and integrating the spacetime 
dependence of the inserted operators against particular choices of measures. 
The resulting functions only depend on the complexified Yang Mills coupling, 
\begin{equation}
\tau \coloneqq\tau_1 + i \tau_2= \frac{\theta}{2\pi} + \frac{4\pi i }{\gym^2}\,,\label{eq:tau}
\end{equation}
the number of colours $N$, and the parameters defining the 
inserted operators, for example the $R$-charges of the external states. What makes the study of integrated correlators 
so compelling is the observation that, by combining different 
non-perturbative techniques, it is possible to compute these 
observables exactly and thus predict new results even in absence 
of detailed knowledge about the correlator itself.

The usefulness of integrated correlators was first manifested 
in the pioneering works \cite{Binder:2019jwn, Chester:2020dja}.
There, the authors considered the four-point function $\langle {\cal O}_2(x_1) {\cal O}_2(x_2){\cal O}_2(x_3){\cal O}_2(x_4)\rangle$, of the superconformal primary operator in the stress-tensor multiplet, ${\cal O}_2$,
integrated against a certain measure $\dd \mu(\!\{x_i\}\!)$,\footnote{
It was shown in \cite{Wen:2022oky} that this measure is simply given by $\dd \mu(\!\{x_i\}\!)=\prod_{i=1}^4\dd^4x_i/{\rm vol}(SO(2,4))$.}
and derived a matrix-model representation for this particular 
integrated correlator by exploiting the work of Pestun \cite{Pestun:2007rz} on supersymmetric localisation. 
The strategy is to consider a mass deformation,  $m$, of the 
original ${\cal N}=4$ theory, known as  $\mathcal{N}=2^*$ SYM, 
and  observe that the aforementioned integrated correlator
can also be computed by acting with the four derivatives operator 
$\tau_2^2 \partial_m^2 \partial_\tau \partial_{\bar{\tau}}$ on 
the partition function of $\mathcal{N}=2^*$ SYM on S$^4$, 
eventually setting the mass deformation parameter  $m$ to zero. 
Thanks to supersymmetric localisation \cite{Pestun:2007rz}, the ${\rm S}^4$ 
partition function of $\mathcal{N}=2^*$ SYM can be 
reduced to a matrix model and, as a consequence, so does the integrated correlator itself. 
In this language, the integration measure $\dd \mu$ 
emerges from supersymmetry.

Crucially, even if at the present time a closed form expression for 
the four-point function $\langle \mathcal{O}_2 \mathcal{O}_2 \mathcal{O}_2 \mathcal{O}_2 \rangle$ is not known, by solving 
the matrix model integral it is possible to explore the emergence 
of important non-perturbative effects in the correlator, as well as 
exploring the role played by S-duality in the ${\cal N}=4$ SYM theory 
\cite{Chester:2019jas, Chester:2020vyz, Dorigoni:2021bvj, Dorigoni:2021guq,
Chester:2021aun, Alday:2021vfb, Dorigoni:2022zcr, Dorigoni:2022iem,
Collier:2022emf, Dorigoni:2022cua, Dorigoni:2023ezg, Chester:2023ehi, Alday:2023pet}.

In this paper we shall continue exploring the landscape of 
correlation functions in $\mathcal{N}=4$ SYM, and their integrated 
versions, by considering insertions of more general external operators.
Beyond the lightest half-BPS primary in the stress-tensor multiplet, ${\cal O}_2$, 
there exist half-BPS primary operators with arbitrary $R$-charge $p$. 
For a given charge $p\ge 4$ the space of operators is degenerate.
Amongst the degenerate half-BPS operators with dimension $\Delta=p$
we can construct a special operator $\Omax{p}$, called the \textit{maximal-trace 
operator} in \cite{Paul:2022piq}. For $p$ even, $\Omax{p}$ is defined by taking a suitable power of the lightest operator ${\cal O}_2$,  that is
\begin{equation}\label{intro_maximaltrace}
\Omax{p}= [\mathcal{O}_2]^{\frac{p}{2}}\,,\quad\qquad {\rm with}\quad p\in 2\mathbb{N}\,.
\end{equation}
Closely related to the maximal-trace  operator, we can consider a second special 
operator uniquely built out of the lightest ${\cal O}_2$ operator. 
This is the \emph{coherent-state operator} \cite{Aprile:2025hlt} 
defined by taking an exponential power series \footnote{The constraint 
on the modulus of the parameter $\alpha$ will come from requiring a finite positive 
norm. We refer the reader to Section \ref{sec:RevIntCorr} for the 
precise definitions of the operators here briefly presented. }
\begin{equation}\label{intro_coherent}
\expOO = 
\exp\left(\frac{\alpha}{\sqrt{2}}\, \mathcal{O}_2\right)=  \sum_{n=0}^{\infty} \frac{1}{n!}\left(\frac{ {\alpha} }{\sqrt{2}}\right)^n {\cal O}_2^n \,,\qquad\quad {\rm with}\quad {|\alpha|^2}\in[ 0, \tfrac{1}{2})\,.
\end{equation}
From a quantum-mechanical point of view, the state generated by the operator $\mathcal{O}_H$ is precisely a coherent state of gravitons.

In the present work we focus on four-point correlators that involve the 
two types of operators schematically introduced in \eqref{intro_maximaltrace} and \eqref{intro_coherent} 
and the light ${\cal O}_2$ operators, namely we shall consider
\begin{equation}\label{intro_correlators}
\langle\Omax{p} \Omax{p} {\cal O}_2{\cal O}_2\rangle\,,\qquad \qquad 
\langle \expOO \expOO {\cal O}_2{\cal O}_2\rangle\,.
\end{equation}
In  \cite{Paul:2022piq}  
it was shown that the maximal-trace integrated correlators, averaged 
over the conformally invariant measure $\dd\mu$ mentioned above, can 
again be computed explicitly without detailed knowledge of the 
correlators themselves. Further important structures were then discovered in \cite{Brown:2023cpz,Brown:2023why,Paul:2023rka}. Building on these works, 
and using methods from modular resurgence analysis \cite{Dorigoni:2024dhy}, 
our first result in this paper is to provide modular 
invariant, non-perturbative transseries expansions 
for the maximal-trace integrated correlator in various \emph{scaling 
regimes} where $N$ and/or the $R$-charge $p$ are taken to be large. 
The same techniques allow us to 
derive a modular invariant, non-perturbative large-$N$ transseries expansion
for the heavy-heavy-light-light integrated correlator $\langle \expOO \expOO {\cal O}_2{\cal O}_2\rangle$.

Before discussing in more detail the scaling regimes we shall be 
interested in, we want to comment on the holographic interpretation 
of the integrated correlators here considered. From the point of view of the dual 
string theory description in ${\rm AdS}_5 \times {\rm S}^5$,
it is natural to start from the regime of classical 
(super)gravity, which corresponds to having the central charge 
$c\sim N^2$ as the largest parameter, while at the same time taking the 
't Hooft coupling $\lambda = N \gym^2 $ fixed and large.  
Moving away from  the strict classical gravity limit,~i.e. $N^{-1}=0$ 
and $\lambda^{-1}=0$, any observable in AdS receives 
both $1/N$ quantum gravity  corrections and $1/\lambda$ stringy corrections. 
This is true in particular for the bulk scattering amplitudes that are 
dual to the correlators shown in \eqref{intro_correlators}.
Therefore, the computation of integrated correlators as exact 
functions of the coupling constant provides \emph{new} 
information about scattering processes in AdS,
including quantum gravity corrections and stringy corrections.

Similarly to the CFT side, a preferred set of gravitational observables 
in ${\rm AdS}$ are the four-point amplitudes of half-BPS single-particle 
operators, usually denoted by ${\cal O}_p$, which are dual to ${\rm S}^5$ Kaluza-Klein modes of the ten-dimensional graviton. The  
stress-tensor multiplet is dual to the ${\rm AdS}_5$ graviton multiplet, and the superconformal primary operator 
${\cal O}_2$ introduced above is the lightest of these single-particle operators. 
Due to the degeneracy in the energy spectrum, we find that the
single particle operators form only a subset of all half-BPS operators.
Although their energy can become parametrically heavy, single-particle 
operators can be at most as heavy as the giant gravitons \cite{Aprile:2020uxk}, 
i.e.~D-brane operators of dimension $\Delta\sim N$ \cite{Witten:1998xy,McGreevy:2000cw,Corley:2001zk, Balasubramanian:2001nh,Berenstein:2002ke,Balasubramanian:2002sa,Berenstein:2003ah, Lin:2004nb}.
When inserted in a correlation function, these D-brane operators can be 
treated within the framework of defects \cite{Chen:2025yxg} but otherwise 
they do not alter the underlying ${\rm AdS}_5 \times {\rm S}^5$ geometry.

A different scenario emerges when we consider operators with scaling 
dimension $\Delta\sim N^2$, i.e.~as heavy as the central charge. 
The insertion of any such `huge' operator  at the boundary of ${\rm AdS}$ 
induces a back-reaction which deforms the ${\rm AdS}_5 \times {\rm S}^5$ vacuum. 
For two-point functions, the holographic description of the corresponding 
geometries has been understood in \cite{Abajian:2023jye}. In particular, 
all geometries corresponding to the insertion of two 
`huge' half-BPS operators can be mapped to the beautiful classification work \cite{Lin:2004nb} of 
Lin-Lunin-Maldacena (LLM).

It is interesting to ask how the dynamics underlying the 
geometric transition between light Kaluza-Klein modes and 
back-reacted geometries emerges in the CFT, and vice versa. 
Two observables that probe these dynamics are precisely 
the correlators of {maximal-trace} operators in \eqref{intro_maximaltrace} 
and {coherent-state operators} in \eqref{intro_coherent}, 
studied respectively as function of $p$ and $\alpha$ 
in various scaling regimes. However, 
neither of these correlators is known in closed form and 
therefore
the fact that their corresponding integrated versions can be computed exactly will
prove once again the usefulness and importance of integrated correlators. 

 \vspace{-0.35cm}

\subsection*{Outline and Key Results}
\vspace{-0.2cm}

 The paper is organised as follows.
 In Section \ref{sec:RevIntCorr} we review 
useful properties of half-BPS operators in $\mathcal{N}=4$ SYM and
describe in detail the maximal-trace operators schematically presented in \eqref{intro_maximaltrace}
and the coherent-state operators in \eqref{intro_coherent}. Specifically, we shall
discuss the structure of their two- and four-point functions 
and define the integrated four-point functions of interest.

In Section \ref{sec:ModRes}, after reviewing the basics of the 
${\rm SL}(2, \mathbb{Z})$ spectral representation for 
the maximal-trace integrated correlator found in \cite{Paul:2022piq,Paul:2023rka}, 
we introduce modular resurgence analysis \cite{Dorigoni:2024dhy} 
as a tool to derive modular invariant, non-perturbative expansions 
for the integrated correlators when expanded in a large-parameter regime, 
i.e. large-$N$ and/or large-charge. Along the way we find a landscape 
of physically interesting regimes that we summarise below.

In Section \ref{sec:limits}, we start by discussing the maximal-trace integrated correlator, $\langle\Omax{p} \Omax{p} {\cal O}_2{\cal O}_2\rangle$,
in the regime of large-quantum numbers  
where the charge $p$ is taken to be large, while $N$ is kept finite.
In particular, we clarify that the peculiar nature of the $1/p$ expansion originally discovered in \cite{Paul:2023rka,Brown:2023why}, 
is an example of {\em Cheshire cat resurgence}~\cite{Dunne:2016jsr,Kozcaz:2016wvy}: the interesting phenomenon in which resurgence analysis retrieves
an infinite tower of non-perturbative corrections from a terminating perturbative 
expansion with finitely many terms.
We then focus our attention to various double-scaling regimes:
\vspace{-0.15cm}
\begin{itemize}
\item Section \ref{sec:gammaL2}: 
Large $N$ and large charge $p$ with $p/N^\gamma$ fixed and $0<\gamma<2$.
\item Section \ref{sec:N2}: 
Large $N$ and large charge $p$ with $p/N^2$ fixed.
\item Section \ref{sec:gamma3}: 
Large $N$ and large charge $p$ with $p/N^\gamma$ fixed and $\gamma>2$.
\end{itemize}
\vspace{-0.1cm}
In all these regimes the maximal-trace operators become heavy. However, 
the modular invariant transseries expansions that we derive in the various cases present different types of perturbative and non-perturbative contributions. 
Importantly, the physical interpretation of the non-perturbative sectors 
varies from case to case. We present here some preliminary comments. 

The regime of large $N$ with $p/N^\gamma$ fixed and  $0<\gamma<2$ contains in particular the case where the inserted
heavy operators have dimensions $\Delta\sim N$. However, we claim that these operators do not 
correspond to D-branes. Our analysis shows that the integrated correlator 
of maximal-trace operators are quite different from the case of 
giant gravitons, i.e.~D-brane operators. Integrated correlators of 
giant gravitons have been studied recently in \cite{Brown:2024tru,Brown:2025huy} 
and display a more intricate variety of non-perturbative corrections 
than the case considered here.

The regime of large $N$ with $p/N^2$ fixed is the most significant from a
holographic view-point. As explained earlier, since the dimension of the 
operators considered now scales like the central charge, we expect that 
in the dual gravity description the correlator must be described by a bulk geometry which is no longer ${\rm AdS}_5\times$S$^5$. Our 
non-perturbative analysis for the integrated correlator does provide 
quantitative information about the corresponding gravitational back-reaction. 

This regime is also deeply connected with the second correlator introduced 
in \eqref{intro_correlators} and which involves two coherent-state operators.
By construction, the integrated correlator $\langle \expOO \expOO {\cal O}_2{\cal O}_2\rangle$ 
is built as a series in maximal-trace integrated correlators, 
$\langle\Omax{p} \Omax{p} {\cal O}_2{\cal O}_2\rangle$, so clearly 
the two integrated correlators must be somehow intertwined. Crucially, the insertion 
of two coherent-state operators has a well understood gravity dual that 
we shall use to make quantitative statements. As shown in \cite{Aprile:2025hlt,Giusto:2024trt} this geometry is a special case of an LLM geometry, 
sometimes called ${\rm AdS}$ bubble \cite{Chong:2004ce,Liu:2007xj}, 
and which belongs to the $SO(6)$ consistent supergravity truncation considered in \cite{Cvetic:2000nc}.

In Section \ref{sec:LLM} we give results 
for the perturbative and the non-perturbative sectors
of the integrated correlator 
$\langle \expOO \expOO {\cal O}_2{\cal O}_2\rangle$ valid at large $N$ 
and fixed $\tau$. From these expressions we shall derive the large-$N$ 
genus expansion at fixed 't Hooft coupling $\lambda$, thus generalising 
the analysis carried out  in \cite{Dorigoni:2021guq} for 
$\langle {\cal O}_2{\cal O}_2{\cal O}_2{\cal O}_2\rangle$. Our main 
findings can be summarised as follows:
\begin{itemize}
\item We compute the genus expansion of the $\langle \expOO \expOO {\cal O}_2{\cal O}_2\rangle$ 
integrated correlator, providing data for a new amplitude: the Virasoro-Shapiro 
amplitude on the ${\rm AdS}$ bubble. We show that the 
tree-level pure supergravity contribution that we predict perfectly matches with 
the results we obtain by integrating over the 
insertion points the correlator recently computed in \cite{Aprile:2025hlt}.
\item We compute the non-perturbative completion of the aforementioned 
Virasoro-Shapiro amplitude and identify
a hierarchy of three different types of non-perturbative corrections at large-$\lambda$, namely
\begin{align}
\exp\left(- 2  L_+^* \sqrt{\lambda} \right)\ll\exp\left(- 2 \sqrt{\lambda}\right) \ll \exp\left(- 2 L_-^*  \sqrt{\lambda}\right)\,.
\end{align}
Crucially, the scales $L^*_{\pm}$ are non trivial functions of the parameter $\alpha$ which enters in the definition \eqref{intro_coherent} for the heavy operator $\expOO$. We shall prove that
\begin{equation}
L^*_+=\sqrt{\frac{1+\sqrt{2}\alpha}{1-\sqrt{2}\alpha}}\,,\qquad\qquad
L^*_-=\sqrt{\frac{1-\sqrt{2}\alpha}{1+\sqrt{2}\alpha}}\,.\label{eq:IntroScales}
\end{equation}

\item We reproduce exactly the same scales \eqref{eq:IntroScales}
by analysing the spectrum of giant-magnons on the ${\rm AdS}$ bubble. In this geometric picture, 
we explain that $L^*_+$ and $L^*_-$ correspond precisely 
to the major and minor axis of the elliptical droplet in the LLM description. 
\end{itemize}
Along the way we also show that the genus-zero integrated
Virasoro-Shapiro amplitude on the ${\rm AdS}$ bubble is related to the 
genus-zero integrated correlator of the maximal-trace correlator 
via the mapping in parameters
\begin{equation}
   \textrm{genus-zero:}\qquad \frac{p}{N^2} \leftrightarrow \frac{2\alpha^2}{1-2\alpha^2}\,.
\end{equation}
We provide evidences suggesting that this property holds at genus-zero for all CFT quantities that are \emph{diagonal} with respect to the insertion of two identical maximal-trace operators.  

Lastly, in Section \ref{sec:Conclusions} we comment on possible future directions originating from our results. Our paper contains four appendices in which we collect various technical details.

\section{Correlators of maximal-trace \& coherent-state operators}
\label{sec:RevIntCorr}

In this section we review useful properties of half-BPS operators in $\mathcal{N}=4$ SYM, and
define the half-BPS maximal-trace operators and coherent-state operators mentioned in the introduction. We discuss the structure of their two- and four-point functions and 
then proceed to briefly summarise how to derive from supersymmetric localisation 
the integrated four-point correlation functions of interest for this paper.
In many ways the four-point integrated correlators we shall focus on are the simplest non-trivial observables one can study in $\mathcal{N}=4$ SYM.

\subsection{Maximal-trace operators}

Half-BPS superconformal primary operators in $\mathcal{N}=4$ SYM transform 
in the representation $[0,p,0]$ of the $SU(4)$ $R$-symmetry group and have protected 
scaling dimension $\Delta=p$. A basis for these operators can be constructed explicitly
in the microscopic theory by taking gauge invariant combinations built out of the six scalar 
fields $\Phi_I$, with $I=1,\ldots, 6$, of the $\mathcal{N}=4$ multiplet. 

Adopting the conventions of \cite{Paul:2022piq}, we start by defining \emph{normalised} single trace operators
\begin{equation}\label{eq:singletrace}
    T_p(x,Y) \coloneqq \frac{1}{p}Y_{I_1}\cdots\,{Y_{I_p}}{\rm Tr}\big[ \Phi^{I_1}(x)\,\cdots\, \Phi^{I_p}(x)\big] = \frac{1}{p} {\rm Tr}\big[ (Y\cdot \Phi(x))^p \big]\,,
\end{equation}
where $x$ denotes a spacetime point and $Y_I$ a complex null polarisation vector for the $SO(6)$ $R$-symmetry indices. 
The null condition, $Y \cdot Y=0$, automatically projects the operator \eqref{eq:singletrace} onto the symmetric traceless representation $[0,p,0]$.  
From single-trace operators we construct the multi-trace operator
\begin{equation}\label{eq:multitrace}
    T_{p_1,...,p_n}(x,Y) \coloneqq \frac{p_1 \cdots \,p_n}{p}T_{p_1}(x,Y)\,\cdots\,T_{p_n}(x,Y)\,.
\end{equation}
The total $R$-symmetry charge $p=p_1+...+p_n$ is equal to the scaling dimension 
$\Delta=p$ of the operator. It is important to stress that for fixed charge $p$ 
there is a degeneracy in the spectrum of half-BPS operators. 
For example, at $p=4$ we have two different dimension-four 
half-BPS operators $T_{2,2}$ and $T_4$. 

In this paper we restrict our attention to the so called \textit{maximal-trace} operators, 
denoted hereafter as $\Omax{p}(x,Y)$. These are the unique operators with the maximal 
number of traces for a given charge $p$. More explicitly, 
\begin{equation}\label{eq:Omax}
    \Omax{p}(x,Y) \coloneqq \left\lbrace \begin{matrix} \!\!\!\!\!\!\!\!\!\!\!\!  \!\!\!\! \!\!\!\!\! \! T_{(2,...,2)}(x,Y) = \frac{ \big[2\,T_2(x,Y)\big]^{\frac{p}{2}}}{p}\,,   &\quad   p=2r \equiv 0 \,{\rm mod}\,2\,, \\
    T_{(3,2,...,2)}(x,Y) =\frac{3T_3(x,Y) \big[2\,T_2(x,Y) \big]^{\frac{p-3}{2}}}{p} \,, & \quad\quad\,\,\, p =2r+3\equiv 1 \,{\rm mod}\,2\,.
    \end{matrix}\right.
\end{equation}
These operators were first introduced in \cite{Paul:2022piq}.\footnote{We  
note that $\Omax{p}(x,Y)$ with $p$ odd exists only for $N>2$, since $T_3=0$ 
in a theory with gauge group $SU(2)$.} 

We emphasise that the operator $\Omax{p}(x,Y)$ is in general \emph{not} 
a single-particle operator \cite{Aprile:2020uxk}, often denoted by the symbol $\mathcal{O}_p(x,Y)$.
Single-particle operators correspond to to single-particle Kaluza-Klein fields 
in the dual ${\rm AdS}_5\times {\rm S}^5$ holographic theory. 
In particular, the lightest single-particle operator $\mathcal{O}_2(x,Y)$ 
is the scalar superconformal primary operator in the stress-tensor multiplet 
and it is dual to a single graviton state on ${\rm AdS}_5\times {\rm S}^5$.
The single-particle operators are uniquely defined by requiring that they 
must be orthogonal to all multi-trace operators. We thus conclude that 
the single-particle operator ${\cal O}_p(x,Y)$ equals $T_{p}(x,Y)$ plus 
a particular admixture of multi-trace operators $T_{\vec{p}}(x,Y)$, which 
can be determined via a Gram-Schmidt orthogonalisation procedure with 
respect to the inner-product induced by the two-point functions. 

For $p=2$ and $p=3$ there are no multi-trace operators in $SU(N)$ and 
we easily see from \eqref{eq:Omax} that the maximal-trace operators 
coincide with the single particle ones, i.e. we have
\begin{align}
\Omax{2}(x,Y) &= \mathcal{O}_2(x,Y) = T_2(x,Y)\,,\\
\Omax{3}(x,Y) &= \mathcal{O}_3(x,Y)= T_3(x,Y)\,.
\end{align}
In this work we only discuss maximal-trace operators  $\Omax{p}(x,Y)$,  hence these should not be confused with the single particle operators which instead will not be part of the present discussion.

We note that when the gauge group is $SU(N)$, the space of single-trace operators is exhausted by the set $\{T_2,...,T_N\}$, 
since a simple linear algebra argument based on the Cayley-Hamilton theorem implies that $T_p$ with $p>N$ must belong to the algebra generated by $\{T_2,...,T_N\}$.
The same argument shows that single-particle operators exist 
only for $p\leq N$, since $T_p$ with $p>N$ must necessarily 
be a multi-particle operator. Nevertheless, it is worth mentioning 
that beautiful structures emerge \cite{Yang:2021kot,Holguin:2022zii,Abajian:2023jye} 
when one considers large $N$. In particular, we can construct half-BPS gauge 
invariant operators of dimension $\Delta \propto N$ which
become heavy operators in the large $N$ limit and thus correspond 
to giant-graviton solutions in the dual type IIB superstring theory. 
Giant gravitons are D$3$-brane solutions which wrap an ${\rm S}^3$ 
inside either the ${\rm S}^5$ or the ${\rm AdS}_5$ factor of the 
dual ${\rm AdS}_5\times {\rm S}^5$ geometry, while rotating at the 
speed of light along one of the ${\rm S}^5$ directions \cite{McGreevy:2000cw}.

\subsection{Correlation functions of maximal-trace operators}

Correlation functions of half-BPS operators are severely constrained by superconformal invariance. Importantly, in the interacting theory  two- and three-point functions of all half-BPS operators are protected observables \cite{Lee:1998bxa}, i.e.~they do 
not depend on the gauge coupling, and therefore can be computed directly in the free theory.

To this end, we recall the free theory propagator of the elementary field $Y\!\cdot\Phi(x)$,
\begin{equation}
\langle Y_1\!\cdot\Phi^m_n(x_1)\,\, Y_2\!\cdot\Phi^r_s(x_2)\rangle= \left( \delta^m_s\delta^r_n -\frac{1}{N}\delta^{m}_n\delta^r_s\right) d_{12}\,,\qquad\qquad {\rm with}\quad 
d_{ij}\coloneqq \frac{Y_{ij}}{  x_{ij}^2 }\,,
\end{equation}
where $x_{ij} \coloneqq x_i-x_j$ and $Y_{ij}\coloneqq Y_i\cdot Y_j$ and $m,n,r,s$ are matrix indices of $SU(N)$.
The two-point function of maximal-trace operators \eqref{eq:Omax} is then obtained by Wick contractions and takes the form,
\begin{equation}\label{2_point_maxtrace}
\average{\Omax{p}(x_1,Y_1)\,\Omax{q}(x_2,Y_2)} = \delta_{p,q}\, R_p(N)\,d_{12}^p\,,
\end{equation} 
The two-point function coefficient $R_p(N)$ is function only of the number of colours $N$ and of the $R$-charge $p$ \cite{Gerchkovitz:2016gxx},
\begin{equation}\label{eq:2ptcoef}
    R_{2r}(N) \coloneqq \frac{4^{r}r!}{(2r)^2}\Big(\frac{N^2-1}{2}\Big)_r\,,\qquad R_{2r+3}(N) \coloneqq \frac{3(N^2-1)(N^2-4)4^{r}r!}{N(3+2r)^2}\Big(\frac{N^2+5}{2}\Big)_r\,,
\end{equation}
where $(a)_n \coloneqq \prod_{k=0}^{n-1}(a+k)$ denotes a Pochhammer symbol. 

Since two- and three-point correlators are independent of the Yang-Mills coupling constant, relatively little dynamical information can be extracted from these observables. 
From four-point and higher correlation functions the story changes completely since these observables are only partially non-renormalised \cite{Eden:2000bk} and do depend non-trivially on the dynamics of the theory.

In this paper we are interested in four-point correlation functions of two superconformal primary operators of the stress-tensor multiplet, $\mathcal{O}_2$, and two identical maximal-trace operators, $\Omax{p}$, of equal charge $p\geq 2$,
\begin{equation}\label{4pt_first_time}
 \average{\mathcal{O}_2(x_1,Y_1)\mathcal{O}_2(x_2,Y_2)\Omax{p}(x_3,Y_3)\Omax{p}(x_4,Y_4)}\,.
\end{equation}
Using superconformal Ward identities and the partial non-renormalisation theorem, 
this four point function can be written as \cite{Eden:2000bk,Heslop:2002hp,Nirschl:2004pa},
\begin{equation}\label{Ward_identity_soln}
    \average{\mathcal{O}_2(x_1,Y_1)\mathcal{O}_2(x_2,Y_2)\Omax{p}(x_3,Y_3)\Omax{p}(x_4,Y_4)} = \mathcal{G}_{p,{\rm free}}+d_{12}^2 d_{34}^p \mathcal{I}(z,\bar{z},y,\bar{y}) \mathcal{H}_p(u,v;N,\tau)\,.
\end{equation} 
In the above equation the complex variables $z,\bar{z}$ parametrise the spacetime cross ratios $u,v$, while the variables $y,\bar{y}$ parametrise the $R$-symmetry cross ratios  $\tilde{u}, \tilde{v}$,
\allowdisplaybreaks{
\begin{align}
    u &= \label{eq:uv}z\bar{z}\coloneqq \frac{x_{12}^2x_{34}^2}{x_{13}^2x_{24}^2}\,, \qquad\, \,\,v=(1-z)(1-\bar{z})\coloneqq\frac{x_{14}^2x_{23}^2}{x_{13}^2x_{24}^2}\,,\\*
   \label{eq:munu} \tilde{u} &= y\bar{y} \coloneqq \frac{Y_{12}^2Y_{34}^2}{Y_{13}^2Y_{24}^2}\,,\qquad \tilde{v} = (1-y)(1-\bar{y})\coloneqq \frac{Y_{14}^2Y_{23}^2}{Y_{13}^2Y_{24}^2}\,.
\end{align}}

The first term in \eqref{Ward_identity_soln}, denoted as $\mathcal{G}_{p,{\rm free}}$, 
corresponds to the free part of the four-point function, which can be easily computed 
in the free theory via Wick contractions thus consequently coupling constant independent.
The second term in \eqref{Ward_identity_soln} is the dynamical contribution. 
The prefactor $d_{12}^2 d_{34}^p$ ensures the correct transformation properties 
under conformal transformations, so that the remaining part of the correlator can be written just in term of cross ratios.  
The additional factor $\mathcal{I}$ is purely due to ${\cal N}=4$ kinematics \cite{Eden:2000bk,Heslop:2002hp,Nirschl:2004pa} 
and is conveniently expressed using the variables defined in \eqref{eq:uv}-\eqref{eq:munu},
\begin{equation}\label{eq:Iuv}
    \mathcal{I}(z,\bar{z},y,\bar{y})\coloneqq \frac{(z-y)(z-\bar{y})(\bar{z}-y)(\bar{z}-\bar{y})}{(y\bar{y})^2}\,.
\end{equation}
By stripping off both kinematical prefactors $d_{12}^2 d_{34}^p$ and $\mathcal{I}$, 
all of the dynamical information about the four-point correlator \eqref{4pt_first_time} is then 
contained in the function $\mathcal{H}_p(u,v;N,\tau)$ usually referred to as the \textit{reduced correlator}.

 The reduced correlator is a non-trivial function of the spacetime cross-ratios $u,v$, 
 of the number of colours, $N$, 
 and crucially it is the only part of the full correlator that depends on the complexified Yang-Mills coupling constant
\begin{equation}\label{eq:tau_def}
    \tau:= \tau_1 + i\, \tau_2 = \frac{\theta}{2\pi} +i \frac{4\pi }{\gym^2}\,.
\end{equation}
While the full spacetime and coupling dependence of the reduced correlator is expected to 
be complicated and unlikely to be expressible in closed form, important information can be extracted 
using the method of supersymmetric localisation for a particular averaging of the reduced correlator over a specific measure of the cross-ratios \cite{Binder:2019jwn, Chester:2020dja}.

 The focus of this paper is to analyse a certain spacetime average of the reduced four-point correlator $\mathcal{H}_p(u,v;N,\tau)$ defined as the integral \cite{Binder:2019jwn, Chester:2020dja},
\begin{equation}\label{Integrated_corr_def}
    \mathcal{C}_{p,N}(\tau) \coloneqq -\frac{2}{\pi }\int_0^\infty \dd r\int_0^\pi \dd \theta \frac{r^3\sin^2(\theta)}{u^2}\frac{\mathcal{H}_p(u,v;N,\tau)}{ R_p(N)} \Bigg\vert_{u=1-2r\cos\theta+r^2,v=r^2}\,.
\end{equation}
 The factor $R_p(N)$ appearing in the denominator is defined in \eqref{eq:2ptcoef} and inserted in order to consider the reduced four-point correlator \eqref{4pt_first_time} normalised with respect to the two-point function of the two maximal-trace operators. Building on the results derived in the key works \cite{Paul:2022piq,Paul:2023rka}, we will derive analytic expressions for \eqref{Integrated_corr_def} at large $p$ and large $N$ while keeping an \emph{exact} dependence on the coupling constant $\tau$.

The choice of integration measure in \eqref{Integrated_corr_def} is dictated by 
supersymmetry, albeit in a rather indirect manner, as we now briefly review.
In $\cN=4$ SYM we can introduce a mass parameter $m$ while preserving $\cN=2$ supersymmetry, leading to what is usually known as $\cN=2^*$ SYM. The integrated correlator defined in \eqref{Integrated_corr_def} is then derived from the ${\rm S}^4$ partition function of $\mathcal{N}=2^*$ SYM by taking a specific combination of four derivatives with respect to theory parameters which, upon use of superconformal Ward identities, generates the integration measure in \eqref{Integrated_corr_def}.
We refer to the original papers \cite{Binder:2019jwn,Chester:2020dja} for more details on this derivation.
Importantly, the ${\rm S}^4$ partition function of $\cN=2^*$ SYM can be computed exactly via supersymmetric localisation, reducing the path integral to a matrix model integral \cite{Pestun:2007rz}.
As a consequence, we find that for arbitrary coupling $\tau$, number of colours $N$, and charge $p$ the same integrated correlator \eqref{Integrated_corr_def} can also be evaluated  directly from the localised matrix model formulation,
thus giving direct access to non-perturbative features of the four-point functions \eqref{4pt_first_time}. 
 
At first \cite{Chester:2019jas} considered the integrated correlator $\mathcal{C}_{p=2,N}(\tau)$, related to the four-point function $\langle{\cal O}_2{\cal O}_2{\cal O}_2{\cal O}_2\rangle$, and used the localised matrix model formulation to derive the large-$N$ expansion of $\mathcal{C}_{p=2,N}(\tau)$ at fixed coupling constant $\tau$.
However, it was soon realised that electro-magnetic duality of $\mathcal{N}=4$ SYM has extremely powerful consequences and forces $\mathcal{C}_{{p},N}(\tau)$ to be a real-analytic and modular invariant function of the complexified Yang-Mills coupling $\tau$ for any value of $N$ and charge $p$. This property led to a conjectural expression \cite{Dorigoni:2021bvj,Dorigoni:2021guq}, then proven in \cite{Dorigoni:2022cua}, for the exact integrated correlator $\mathcal{C}_{2,N}(\tau)$ with an arbitrary number of colours $N$. Crucially, the modular invariant coupling constant dependence of $\mathcal{C}_{2,N}(\tau)$ is entirely captured by a particularly simple lattice sum representation.\footnote{Building on \cite{Dorigoni:2022cua} lattice sum representations for integrated correlators of $\langle{\cal O}_2{\cal O}_2{\cal O}_2{\cal O}_2\rangle$ have also been derived in the case of $\mathcal{N}=4$ SYM where the gauge group is a general classical group \cite{Dorigoni:2022zcr} as well as an exceptional Lie group \cite{Dorigoni:2023ezg}.} While the lattice sum representation is convenient for certain calculations, in this work we will make use of a different representation \cite{Collier:2022emf} based on ${\rm SL}(2,\mathbb{Z})$ spectral theory, which will be reviewed in Section \ref{sec:spec}.

Importantly, it is also possible to derive exact expressions for integrated four-point functions involving more complicated external operators. The maximal-trace integrated correlator ${\cal C}_{p,N}$ defined in \eqref{Integrated_corr_def} corresponds to a particular instance of the more general analysis carried out in
\cite{Paul:2022piq}, see also \cite{Brown:2023cpz,Brown:2023why,Paul:2023rka}.
In these references, the authors studied integrated four point correlation functions of two operators $\mathcal{O}_2$ and two general multi-trace half-BPS operators with charge $p>2$. Thanks to these results, it is then possible to compute exactly integrated correlation functions of two operators $\mathcal{O}_2$ and two single-particle states $\mathcal{O}_p$.  
In turn, this computation provided exact non-perturbative data for the scattering processes between gravitons ($p=2$) and higher Kaluza-Klein modes ($p>2$) in ${\rm AdS}_5\times {\rm S}^5$.
Employing similar matrix model techniques \cite{Brown:2024tru}
computed Heavy-Heavy-Light-Light (HHLL) integrated correlation functions where the two heavy operators are determinant operator with conformal dimensions proportional to the number of colours $N$ while the two light operators are ${\cal O}_2$.
The leading large-$N$ expansion of such HHLL integrated correlators was recently derived \cite{Brown:2025huy} for the case in which the heavy operators are either two {sphere giant gravitons}, dual to D3-branes wrapping an ${\rm S}^3$ inside the ${\rm S}^5$, or two ${\rm AdS}$ (or {dual}) {giant gravitons}, dual to D3-branes wrapping an ${\rm S}^3$ inside the ${\rm AdS}_5$ part of the background geometry.

It appears clear that even by restricting our attention to four-point 
functions of half-BPS operators in $\mathcal{N}=4$ SYM, 
there are many options and a huge landscape of regimes that 
one can explore by studying appropriate large-$N$ and/or large quantum numbers limits, 
for different classes of external operators. In particular, coming back 
to the case of present interest, we will focus our discussion on maximal-trace four-point 
correlators $ \average{\mathcal{O}_2\mathcal{O}_2\Omax{p}\Omax{p}}$ with $p$ even 
and analyse their integrated version ${\cal C}_{p,N}(\tau)$ 
defined in \eqref{Integrated_corr_def} in different 
(double-)scaling regimes $N,p\to \infty$:
\vspace{-0.2cm}
\begin{itemize}
\item Section \ref{sec:Large_charge_fixed_N}: Large charge $p$, finite $N$.
\item Section \ref{sec:gammaL2}: Large $N$ and large charge $p=\alpha N^\gamma$ with $0<\gamma<2$ and $\alpha>0$ fixed.
\item Section \ref{sec:N2}: Large $N$ and large charge $p=\alpha N^2$ with $\alpha>0$ fixed.
\item Section \ref{sec:gamma3}: Large $N$ and large charge $p=\alpha N^\gamma$ with $\gamma>2$ and $\alpha>0$ fixed.
\end{itemize}
\vspace{-0.3cm}
Although more details about the physical interpretations behind the different regimes are given in the corresponding sections, we want to emphasise a few key points here.
Firstly, our results of Section \ref{sec:gammaL2} do include the case where the maximal-trace operator $\Omax{p}$ is in the same heavy regime as the giant gravitons operators,~i.e. $\Delta=p = \alpha N$ with $\alpha>0$ fixed as $N\to \infty$. However, we see that the maximal-trace integrated correlator behaves quite differently from the case of the giant gravitons studied in \cite{Brown:2024tru,Brown:2025huy}.
Furthermore, following \cite{Paul:2023rka} we discuss in Section \ref{sec:N2} the `gravity regime' defined by considering the maximal-trace operator $\Omax{p}$ in the regime where $\Delta=p=\alpha N^2$ with $\alpha>0$ fixed as $N\to \infty$. In this regime the dimension of the two $\Omax{p}$ appearing in the correlation function has the same large-$N$ scaling as the central charge of the theory. As a consequence, we see that holographically the insertion of such heavy operators deforms the bulk of the underlying ${\rm AdS}_5\times {\rm S}^5$ geometry to a different spacetime.
This second double-scaling regime is deeply connect to a different family of half-BPS operators, which we now introduce.

\subsection{Coherent-state operators and their correlation functions} \label{sec:coherent}

 In the regime of large $N$ and large 't Hooft coupling $\lambda \coloneqq N \gym^2$, 
 correlation functions of half-BPS operators 
 admit a holographic description as scattering 
 amplitudes in ${\rm AdS}_5\times$S$^5$. 
 We already mentioned that single-particle operators 
 are dual to Kaluza-Klein modes on the ${\rm S}^5$, however 
 the spectrum of the theory contains many more half-BPS states 
 besides the single-particle ones. In fact, both multi-trace operators 
 defined in \eqref{eq:multitrace} and maximal 
 trace operators $\Omax{p}$  in \eqref{eq:Omax} are examples 
 of such operators. We also mentioned that single-particle operators 
 of a given charge $p$ exist as long as $p\leq N$, 
 hence at large $N$ the heaviest of such states can have 
 at most $\Delta = p \sim N$ (where they become giant gravitons 
 \cite{Aprile:2020uxk}). Therefore, in order to probe gravitational backreaction 
 in AdS$_5$, we have to consider operators whose dimension can become heavier than $N$.
 
On the gravity side, the space of half-BPS operators whose dimension scales as $\Delta \sim N^2$ is dual to the space of 
deformations of the ${\rm AdS}_5\times {\rm S}^5$ geometry which preserve the same amount of supercharges. 
From this point of view, \emph{all} half-BPS excitations of ${\rm AdS}_5\times {\rm S}^5$ have been 
classified in a beautiful work by Lin, Lunin, and Maldacena (LLM) \cite{Lin:2004nb}. These LLM spaces 
in general are not described by a consistent truncation {of $10$-dimensional IIB supergravity}.
However, there is one special solution, sometimes called ${\rm AdS}$ 
bubble \cite{Chong:2004ce,Liu:2007xj,Giusto:2024trt},
that is {also} a solution of the $SO(6)$ reduction of type IIB 
supergravity on ${\rm S}^5$  \cite{Cvetic:2000nc}.

This rather well-known consistent truncation retains only fields 
from the graviton multiplet which in the dual CFT correspond to 
components of the stress-tensor multiplet. By construction then, 
a half-BPS solution in this truncation is necessarily dual to an 
operator constructed from the graviton ${\cal O}_2$ and multi-graviton 
operators  ${\cal O}_2^n$, since mixing with higher charge single-particle 
operators would mean introducing higher-charge Kaluza-Klein modes 
on ${\rm S}^5$ and thus would spoil the consistency of the truncation \cite{DHoker:2000xhf}.

The fact that the holographic description of certain
half-BPS operators undergoes a geometric transition 
as their scaling dimension becomes parametrically 
heavy is a fascinating non-perturbative phenomenon, which can 
be explored at the quantitive level by studying certain integrated correlators. 
 The operator we wish to consider for this purpose is a particular 
 one-parameter coherent-state \cite{Aprile:2025hlt}
operator built out of multi-graviton states, 
such that it is dual to a smooth LLM geometry, 
denoted hereafter by $\expOO(\alpha)$. 

More concretely, using the state-operator correspondence, we first introduce the coherent-state as follows,
\begin{equation}\label{eq:OH}
   |Y; \alpha\rangle \coloneqq \sum_{n=0}^{\infty} \frac{1}{n!}\left(\frac{ {\alpha} }{\sqrt{2}}\right)^n |Y;n\rangle,\qquad  |Y;n\rangle 
   \coloneqq\lim_{x\rightarrow 0}\left[ {\cal O}^n_2(x,Y)|0\rangle\right]\,.
\end{equation}
From a state point of view, $|Y; \alpha\rangle$ is precisely a quantum-mechanical coherent state constructed from multi-graviton states.
In terms of operators, we define the corresponding operator
\begin{equation}\label{eq:OHdef}
\expOO(x=0,Y; \alpha) \coloneqq 
\exp\left(\frac{{\alpha}}{\sqrt{2}} \mathcal{O}_2(x=0,Y)\right)\,,
\end{equation}
constructed such that $|Y; \alpha\rangle=
\expOO(0,Y; \alpha)|0\rangle$.
We note that it is not necessary to consider the operator $\expOO$ in radial quantisation, however we find it convenient to do so.\footnote{When 
considering the more general operator $\expOO(x,Y;{\tilde \alpha})=\exp\left(\frac{{\tilde \alpha}}{\sqrt{2}} \mathcal{O}_2(x,Y)\right)$ it is important 
to keep in mind that ${\tilde \alpha}$ transforms under rescaling $x\rightarrow \lambda x$ since $\mathcal{O}_2$ has scaling dimension $\Delta =2$. }

The hermitian conjugate of the coherent state is defined as usual\footnote{We recall that the elementary scalars of the ${\cal N}=4$ multiplet are hermitian matrices, i.e. $(\Phi_I)^\dagger=\Phi_I$.}
\begin{equation}
\langle Y;\alpha| =  \sum_{n=0}^{\infty} \frac{1}{n!}\left(\frac{ \bar{\alpha} }{\sqrt{2}}\right)^n \lim_{x\rightarrow \infty}\left[ |x|^{4n} \langle 0 | {\cal O}^n_2(x,\bar{Y})\right]\,.\label{eq:bra}
\end{equation}
We can then use \eqref{2_point_maxtrace}-\eqref{eq:2ptcoef} to readily evaluate the two point function norm,
\begin{equation}
 \lim_{\substack{ x_1\rightarrow 0 \\ x_2\rightarrow \infty}} \langle \mathcal{O}_H(x_1,Y;\alpha)\mathcal{O}_H(x_2,Y;\alpha) \rangle =   \langle Y;{\alpha}|Y;{\alpha}\rangle =  \left(1-\frac{|Y|^4|\alpha|^2 }{2}\right)^{-\frac{N^2-1}{2}} \,.\label{eq:norm}
\end{equation}
Note that finiteness and positivity of the norm requires $\displaystyle |\alpha|^2 \in[0, {2}/ |Y|^4)$.
The same computation above generalises to 
$\langle Y_1;{\alpha}_1|Y_2;{\alpha}_2\rangle$, where the result  depends only on the combination $(\bar{Y}_1.Y_2)^2(\bar{\alpha}_1\alpha_2)$ which then identifies the physical parameters. We also note that the $\alpha_i$ could be understood as the modulus of the polarisation vector $Y_i$, which are usually taken to have fixed normalisation \cite{Arutyunov:2002fh}. In the following we will consider $\alpha$ to be a real parameter.

We now wish to exploit the structure of 
the four-point function with the maximal-trace operators, in \eqref{Ward_identity_soln}, to derive an expression for the 
correlator between two coherent-state operators $\expOO$ and two light operators $\mathcal{O}_2$, i.e. we consider
\begin{equation}
 \average{\mathcal{O}_2(x_1,Y_1)\mathcal{O}_2(x_2,Y_2)\expOO(0,Y_3;\alpha)\expOO(\infty,Y_4;\alpha)} =  {\langle{Y_4;\alpha}|O_2(x_1,Y_1)\mathcal{O}_2(x_2,Y_2)|{Y_3;\alpha}\rangle}\,.
\end{equation}
Given that the $R$-symmetry dependence in \eqref{Ward_identity_soln} is fully factorised, we 
can,  without loss of generality, restrict ourselves to a particular polarisation vector, which we take to be 
\begin{equation}\label{Ybullet}
Y_{\bullet}\coloneqq (1,i,0,0,0,0)\,.
\end{equation}
With this particular choice of $R$-symmetry polarisation, the operators we are interested in become
\begin{equation}\label{eq:OL}
    O_L(x) \coloneqq \mathcal{O}_2(x,Y_\bullet)  = \frac{1}{2}{\rm Tr}(\Phi_1(x)+i\Phi_2(x))^2\,,\qquad\qquad \expO(x=0;{\alpha}) = \exp\Big(\frac{{\alpha}}{\sqrt{2}}O_L(0)\Big)\,.
\end{equation}
In the following we will focus on the four-point function 
\begin{equation}\label{eq:HHLLini}
\langle  O_L(x_1) \bar{O}_L(x_2) O_H(0;\alpha) \bar{O}_H(\infty;\alpha) \rangle =  {\langle{\alpha}|O_L(x_1)\bar{O}_L(x_2)|{\alpha}\rangle}\,,
\end{equation}
where we use the short-hand notation $|\alpha\rangle$  
to denote the state $|\alpha\rangle \coloneqq |Y_\bullet ;\alpha\rangle = \expO(0;\alpha)|0\rangle$, and $\langle \alpha|$ for its hermitian conjugate \eqref{eq:bra}.
The finite norm condition \eqref{eq:norm} on the coherent-state operator now implies that $\alpha$ lies in the range $0\leq{|\alpha|}<\frac{1}{\sqrt{2}}$. 

Following \cite{Aprile:2025hlt}, throughout this work we shall refer to the four-point 
functions \eqref{eq:HHLLini}, and integrated versions thereof, 
as \textit{Heavy-Heavy-Light-Light} (HHLL) correlators. This will simplify the nomenclature when
discussing in parallel the integrated correlator of maximal-trace operators 
in \eqref{Integrated_corr_def}. 
However, we stress that the operator $\expOO$ defined in \eqref{eq:OHdef} or \eqref{eq:OL} 
is not an eigenvector of the dilation operator, and we emphasise once 
again that the maximal-trace operators $\Omax{p}$ do admit various 
heavy regimes depending on whether $p\sim N$, or $p\sim N^2$, or even larger, 
as explained in the previous section. 

We now proceed to define the  HHLL integrated correlator $\mathcal{C}_{\HHLL}(N,{\alpha};\tau)$. 
To this end, let us observe that the \HHLL~four-point correlation function \eqref{eq:HHLLini} can be expressed as an infinite sum of maximal-trace correlators, 
\begin{equation}\label{eq:HHLL}
    {\langle{\alpha}|O_L(x_1)\bar{O}_L(x_2)|{\alpha}\rangle}= 
    \sum_{r,s=0}^\infty \frac{1}{r!s!}\Big(\frac{{\alpha}}{\sqrt{2}}\Big)^{r+s} \lim_{\substack{ x_3\rightarrow 0 \\ x_4\rightarrow \infty}} \left[|x_4|^{4r} \average{O_L(x_1) \overline{O}_L(x_2)O_L^s(x_3) \overline{O}\,^r_{\!L}(x_4)}\right]\,,
\end{equation}
We note that each four point functions appearing on the right-hand side of \eqref{eq:HHLL} admits the decomposition presented in \eqref{Ward_identity_soln} in terms of free and reduced correlator.
However, it is crucial to stress that only the four-point functions $\average{O_L \overline{O}_L O_L^s \overline{O}\,^r_{\!L}}$ with $r=s$ have a non vanishing reduced correlators. We call these \emph{diagonal} contributions. 

The fact that only diagonal terms in \eqref{eq:HHLL} contribute 
to the reduced correlator is a consequence of superconformal kinematics. 
To see this, we simply have to count the number of allowed $R$-symmetry 
structures of the reduced correlator, which can be readily done by introducing the degree 
of extremality $d$ of a correlator. For a four-point correlator of operators with $R$-charges $\{p_1,...,p_4\}$ the degree of extremality is 
\begin{equation}\label{eq:dExt}
d= \sum_{j=2}^4 \min\left(\tfrac{p_1+p_j}{2},S-\tfrac{p_1+p_j}{2}\right)-S\,,\qquad \qquad {\rm with}\quad S\coloneqq \frac{p_1+p_2+p_3+p_4}{2}\,.
\end{equation}
The number of allowed $R$-symmetry structures of the reduced correlator is then $d(d-1)/2$.

For the four-point function of interest,  $\average{O_L(x_1) \overline{O}_L(x_2)O_L^s(x_3) \overline{O}\,^r_{\!L}(x_4)}$, 
 let us assume without loss of generality that $r\geq s$ so that 
 the operator at position $x_4$ is the heaviest. From \eqref{eq:dExt} we see that the degree of extremality
 of these correlators equals $d=2-\frac{r-s}{2}\leq 2$. Therefore we conclude that:
 when $d<0$ the correlators vanish identically, when $d=0$ or $d=1$ 
 the correlators are protected \cite{Erdmenger:1999pz,Eden:2000gg}
  and thus coincide with their free theory calculations, and only when 
 $d=2$, that is for $r=s$, the correlator is partially renormalised and 
has a non-trivial reduced part. 

From the above considerations, taking into account the 
normalisations $O_L^r(x) = \frac{2r}{2^{r}}\, \Omax{2r}(x,Y_\bullet)$,
and using the decomposition in \eqref{Ward_identity_soln}, we obtain
\begin{equation}\label{eq:HHLL_more0}
  {\langle{\alpha}|O_L(x_1)\bar{O}_L(x_2)|{\alpha}\rangle}=  
  {\cal G}_{{\alpha},{\rm free}} + d^2_{12} \sum_{{r=1}}^\infty  \frac{1}{(r!)^2}
  \lim_{\substack{ x_3\rightarrow 0 \\ x_4\rightarrow \infty}} \left[ \Big(\frac{{\alpha}\, |x_4|^{2}d_{34}}{\sqrt{2}}\Big)^{2r}  \left(\frac{2r}{2^{r}}\right)^2  \mathcal{H}_{2r}(u,v;N,\tau) \right]\,.
\end{equation}
Here we have used that the $R$-symmetry factor defined in \eqref{eq:Iuv} 
reduces to $\mathcal{I}(z,\bar{z},y,\bar{y})\rightarrow 1$, given 
our choice of polarisations \eqref{Ybullet} for which $y,\bar{y}\to \infty$ 
since $Y_{13}=Y_{24}=0$. The numerical factor multiplying $\mathcal{H}_{2r}$ 
in \eqref{eq:HHLL_more0} originates precisely from the proportionality 
constant relating $O_L^r$ to $\Omax{r}$.

In analogy with equation \eqref{Ward_identity_soln} we define the reduced part 
of \HHLL~correlator by removing the free part, and stripping away the kinematic prefactor $d_{12}^2$. 
As a result we define
\begin{equation}\label{eq:Hseries1}
    \mathcal{H}_{{\alpha}}(u,v;N,\tau) \coloneqq
    \sum_{{r=1}}^\infty  \frac{1}{(r!)^2}
  \lim_{\substack{ x_3\rightarrow 0 \\ x_4\rightarrow \infty}} \left[ \Big(\frac{{\alpha}\, |x_4|^{2}d_{34}}{\sqrt{2}}\Big)^{2r}  \left(\frac{2r}{2^{r}}\right)^2  \mathcal{H}_{2r}(u,v;N,\tau) \right]\,.
\end{equation}
All of the dynamical information contained in the \HHLL~correlation 
function are fully captured by the reduced correlator just defined.
The difference between \eqref{eq:HHLL_more0} and \eqref{Ward_identity_soln} is that in \eqref{eq:HHLL_more0} the kinematic prefactor $d_{34}$ cannot be taken out of the sum and hence cannot be factored out.
However, we can easily take the limit over the insertion points to derive the final neat result
\begin{equation}\label{eq:Hseries}
    \mathcal{H}_{{\alpha}}(u,v;N,\tau) = 4\sum_{r=1}^\infty \frac{r^2}{(r!)^2}\Big(\frac{{\alpha}}{\sqrt{2}}\Big)^{2r}\mathcal{H}_{2r}(u,v;N,\tau)\,,
\end{equation}
having used the fact that in this limit the cross-ratios \eqref{eq:uv} reduce to 
$u = \frac{x_{12}^2}{x_1^2}$ and $v = \frac{x_2^2}{x_1^2}$.

We now integrate the reduced correlator in \eqref{eq:Hseries} over the variables $u,v$ with 
respect to the same supersymmetric measure appearing in \eqref{Integrated_corr_def}, 
and thus define the HHLL integrated correlator 
\begin{equation}\label{Integrated_corr_HHLL_def}
    \mathcal{C}_{\HHLL}(N,{\alpha};\tau) \coloneq -\frac{2}{\pi}\int_0^\infty \dd r\int_0^\pi \,\dd \theta \frac{r^3 \sin^2(\theta)}{u^2}
   \frac{ \mathcal{H}_{{\alpha}}(u,v;N,\tau) }{ \langle Y_\bullet;\alpha|Y_\bullet;\alpha\rangle} \Bigg\vert_{u=1-2r\cos\theta+r^2,v=r^2}\,.
\end{equation}
Note that while the integrated correlator for the maximal-trace operators in \eqref{Integrated_corr_def} 
was normalised with respect to the two point function of latter, 
the \HHLL~integrated correlator is normalised with respect to the 
heavy-heavy two-point function \eqref{eq:norm}.

Since $\mathcal{H}_{\alpha}(u,v;N,\tau)$  is expressible as an infinite sum of 
reduced maximal-trace correlators \eqref{eq:Hseries}, we can then make 
use of the definition \eqref{Integrated_corr_def} for the maximal-trace 
integrated correlator ${\cal C}_{2r,N}$ to arrive at
\begin{equation}\label{eq:HHLLIntCorr}
    \mathcal{C}_{\HHLL}(N,{\alpha};\tau) = 4(1-2{\alpha}^2)^{\frac{N^2-1}{2}}\sum_{r=1}^\infty \frac{r^2}{(r!)^2}\Big(\frac{{\alpha}}{\sqrt{2}}\Big)^{2r}R_{2r}(N)\mathcal{C}_{2r,N}(\tau)\,.
\end{equation}

Although an exact closed form expression for $\mathcal{H}_{\alpha}(u,v;N,\tau)$ is 
presently out of reach, we shall see that we have complete analytic control over the integrated 
\HHLL~correlator \eqref{Integrated_corr_HHLL_def} as a function of the number 
of colours $N$, the coherent-state parameter $\alpha$ and the complexified Yang-Mills 
coupling $\tau$. This analysis is carried out in Section \ref{sec:LLM}.
In particular, thanks to our results it is now possible to extract the large-$N$ 
genus expansion at fixed 't Hooft coupling for the integrated \HHLL~correlator 
which carries important non-perturbative information about the dual gravity 
theory, showing in a quantitative manner how the geometric transition happens.

One of the main results of our paper is the derivation of the exact 
large-$N$ fixed-$\tau$ transseries expansion for the \HHLL~integrated 
correlator \eqref{eq:HHLLIntCorr}. Crucially, we shall prove that at 
leading order in the perturbative expansion the integrated correlator 
does in fact match with the supergravity results 
obtained in \cite{Aprile:2025hlt} from the dual LLM geometry. 
Furthermore, we will compute the 
non-perturbative large-$N$ corrections to the \HHLL~integrated correlator 
and show that these effects are related to non-trivial string world-sheet 
giant-magnons configuration on the dual LLM space.

We now move to study the integrated correlator \eqref{Integrated_corr_def} 
of two maximal-trace operators, $\Omax{p}$, and two ${\cal O}_2$ in different 
large-$N$ and/or large-$p$ regimes. With these results, we will 
then derive an exact expression for the \HHLL~integrated correlator \eqref{eq:HHLLIntCorr} 
whose large-$N$ fixed-$\tau$ expansion will display a beautiful 
underlying holographic picture.

\section{Modular invariance and non-perturbative corrections}
\label{sec:ModRes}

In this section we briefly review a modular invariant spectral 
decomposition for the integrated correlator \eqref{Integrated_corr_def} 
with two maximal-trace operators and two 
superconformal primary operators in the stress-tensor multiplet, 
originally derived in \cite{Paul:2022piq}.
We then present some important results from \cite{Dorigoni:2024dhy}  where 
by combining a spectral decomposition approach with methods from resurgence 
analysis we can derive non-perturbative, modular invariant corrections to 
the integrated correlators when expanded in a large-parameter regime, 
i.e. large-$N$ and/or large-charge in later sections.

\subsection{Spectral decomposition for maximal-trace integrated correlators}
\label{sec:spec}

As previously mentioned, the integrated correlator defined in  \eqref{Integrated_corr_def}, 
which we repeat below for convenience, 
 \begin{equation}
    \mathcal{C}_{p,N}(\tau) \coloneqq -\frac{2}{\pi }\int_0^\infty \dd r\int_0^\pi \dd \theta \frac{r^3\sin^2(\theta)}{u^2}\frac{\mathcal{H}_p(u,v;N,\tau)}{ R_p(N)} \Bigg\vert_{u=1-2r\cos\theta+r^2,v=r^2}\,,
\end{equation}
can be computed without needing any knowledge about the reduced correlator itself $\mathcal{H}_p(u,v;N,\tau)$, by exploiting Pestun's matrix model formulation for the $\cN=2^*$ 
SYM partition function on ${\rm S}^4$. 

Regardless of the particular method employed to compute 
$\mathcal{C}_{p,N}(\tau)$, $\mathcal{N}=4$ SYM electro-magnetic duality 
imposes a very strong constraint on all physical observables. Crucially, the integrated correlator~$\mathcal{C}_{p,N}(\tau)$ must be a non-holomorphic, modular invariant function of the complexified 
Yang-Mills coupling $\tau$, i.e. we must have
\begin{align}
&\mathcal{C}_{p,N}(\gamma\cdot\tau) = \mathcal{C}_{p,N}(\tau)\,,\qquad \forall\, \gamma =\left(\begin{matrix} a & b \\ c &d \end{matrix}\right)\in {\rm SL}(2,\mathbb{Z})\,,\\
&\notag {\rm with}\quad \gamma\cdot\tau \coloneqq \frac{a\tau+b}{c\tau+d}\,.
\end{align}

Building on \cite{Paul:2022piq}, with a clever use of ${\rm SL}(2,\mathbb{Z})$ spectral analysis the authors of \cite{Paul:2023rka} evaluated the matrix model representation of $\mathcal{C}_{p,N}(\tau)$ to derive a manifestly modular invariant spectral decomposition (sometimes called Rankin-Selberg decomposition)
\begin{equation}\label{overlap_equation_def}
    \mathcal{C}_{p,N}(\tau) = \average{\mathcal{C}_{p,N}}+ \int_{{\rm Re}(s)=\frac{1}{2}} c_{p,N}(s) \eisen{s}\frac{\dd s}{2\pi i} \,.
\end{equation}
We refer to \cite{Paul:2022piq,Brown:2023cpz,Brown:2023why,Paul:2023rka} for the precise matrix model formulation of the integrated correlator \eqref{Integrated_corr_def} and more general single-particle integrated correlators, while \cite{Collier:2022emf} presents a clear and concise discussion on ${\rm SL}(2,\mathbb{Z})$ spectral methods for integrated correlators.

For this work we can simply understand the role played by ${\rm SL}(2,\mathbb{Z})$ spectral analysis as a way of writing  a real-analytic modular invariant function of $\tau$ in terms of a preferred basis of $L^2$-normalisable eigenfunctions of the Laplace-Beltrami operator $\Delta\coloneqq 4\tau_2^2\partial_\tau\partial_{\bar{\tau}}$.
The spectrum of the Laplacian can be divided into three distinct eigenspaces. The constant function $f(\tau)=1$ is clearly a modular invariant eigenfunction of the Laplace operator with vanishing eigenvalue. The constant term $\average{\mathcal{C}_{p,N}}$ in  \eqref{overlap_equation_def} encodes the contribution from the constant eigenfunction.
The discrete part of the spectrum is rather unruly and it is spanned by a set of modular invariant functions called non-holomorphic Maass cuspforms. Thankfully the such functions do not appear in the spectral decomposition \eqref{overlap_equation_def} and hence will not be presented here.
We stress however the Maass cuspforms may play extremely important roles in the study of CFT partition functions, see for example the beautiful recent work \cite{Perlmutter:2025ngj}.

Lastly, we have the continuous part of the spectrum, which is spanned by the so-called non-holomorphic (or real-analytic) Eisenstein series $\eisen{s}$, with ${\rm Re}(s)=\frac{1}{2}$. The non-holomorphic Eisenstein series $E^*(s;\tau)$ is a modular invariant eigenfunction of the Laplacian with eigenvalue $s(s-1)$ which can be defined as
\allowdisplaybreaks{
\begin{align}
  &\notag E^*(s;\tau) \coloneqq \frac{\Gamma(s)}{2}\sum_{(m,n)\neq (0,0)} \Ymn^{-s} = \frac{1}{2}\sum_{(m,n)\neq (0,0)}\int_0^\infty e^{-t\, \Ymn}\,t^{s-1}\,\dd t\,\\*
&\label{Eisen_series_definitions} =\xi(2s)\tau_2^s  +  \xi(2s-1)\, \tau_2^{1-s}+ \sum_{   k\neq0}e^{2\pi i k\tau_1}    2\sqrt{\tau_2}\,  |k|^{s-\half}\sigma_{1-2s} (k)     \, K_{s-\half}(2\pi |k| \tau_2) \,,
\end{align}}
where $\xi(s) \coloneqq \pi^{-s/2}\Gamma(\frac{s}{2})\zeta(s)=\xi(1-s)$ denotes the completed zeta function, a meromorphic function of $s\in \mathbb{C}$ with simple poles at $s=0,1$.
We also used the short-hand notation for the lattice factor
\begin{equation}
\Ymn \coloneqq \pi \frac{|n\tau+m|^2}{\tau_2}\,.
\end{equation}
While the lattice sum representation on the first line of \eqref{Eisen_series_definitions} is only valid for ${\rm Re}(s)>1$, the Fourier series decomposition with respect to $\tau_1$ on the second line provides a meromorphic continuation to $s \in \mathbb{C}$. 
From \eqref{Eisen_series_definitions} we note that in the present convention the non-holomorphic Eisenstein series satisfies the functional equation
\begin{equation}
E^*(s;\tau) = E^*(1-s;\tau)\,.
\end{equation}

The contour integral in \eqref{overlap_equation_def} corresponds precisely to the non-holomorphic Eisenstein series contribution to the spectral decomposition of the integrated correlator.
The function $c_{p,N}(s)$ appearing as part of the integrand in the spectral decomposition \eqref{overlap_equation_def} is called the \textit{spectral overlap} and it is a meromorphic function of $s$ with exponential decay as $|{\rm Im}(s)|\to \infty$. 
The spectral overlap $c_{p,N}(s)$ can be understood as a certain inner product, called Petersson inner product, between the integrated correlator $ \mathcal{C}_{p,N}(\tau)$ and the non-holomorphic Eisenstein series $E^*(s;\tau)$. Since $E^*(s;\tau) = E^*(1-s;\tau)$ and given that the contour of integration is invariant under $s \leftrightarrow 1-s$, we deduce that the spectral overlap must also satisfy $c_{p,N}(s)= c_{p,N}(1-s)$.
Similarly, the constant term $\average{\mathcal{C}_{p,N}}$ in \eqref{overlap_equation_def} can be understood as the Petersson inner product between $ \mathcal{C}_{p,N}(\tau)$ and the constant function.
Importantly, we have
\begin{equation} 
\average{\mathcal{C}_{p,N}} = \lim_{s\to 1} \big[ c_{p,N}(s) \big]\,.\label{eq:Spec1}
\end{equation}

Given the integral representation \eqref{overlap_equation_def}, one might be tempted to push the contour of integration to infinity and pick up the residues from the poles of the spectral overlap $c_{p,N}(s)$.
However, this procedure yields in general a formal series in non-holomorphic Eisenstein series, which fails to converge for any value of $\tau$. 
Nonetheless, this asymptotic series can be dealt with the general framework 
of modular resurgence technique developed in \cite{Dorigoni:2024dhy},
which is the subject of the next section. 

\subsection{Modular invariant resurgent resummation}\label{Mod_inv_resurgence_section}

There are some general statements about $\mathcal{C}_{p,N}(\tau)$, which simply follow from
the fact that at fixed $N$ the spectral overlap $c_{p,N}(s)$ 
is a meromorphic function of $s$ with poles at integer locations. 
Since analogous statements will be used for $\mathcal{C}_{\HHLL}(N,{\alpha};\tau)$, 
we find it convenient to review the general setup here and postpone 
the explicit computations with the $c_{p,N}(s)$ to the next section.

Consider $p$ and $N$ finite, then the spectral overlap $c_{p,N}(s)$ can be written as \cite{Paul:2022piq,Paul:2023rka}
\begin{equation}\label{c_pN_poly_expr}
    c_{p,N}(s) = \frac{\pi}{\sin(\pi s)} P_{p,N}(s)\,,
\end{equation}
where $P_{p,N}(s)=P_{p,N}(1-s)$ is a polynomial of degree $2N+p-2$ in $s$ with simple zeroes at $s=0,1$ so that $c_{p,N}(s)$ has a finite limit as $s\to 1$. 
The integral representation \eqref{overlap_equation_def} can then be evaluated by shifting the contour of integration towards ${\rm Re}(s)\to\infty$ and collecting residues from all the poles located at $s\in \mathbb{N}^{\geq 1}$.
While $c_{p,N}(s)$ is regular at $s=1$, we note that the Eisenstein series $\eisen{s}$ has a simple pole there with residue $\text{res}_{s=1}\eisen{s}=\frac{1}{2}$.
Proceeding as just described, we arrive at the modular-invariant series representation
\begin{equation}\label{Cpn_Eisen_expansion}
    \mathcal{C}_{p,N}(\tau) = \frac{1}{2}c_{p,N}(1) + \sum_{k=2}^\infty (-1)^{k-1} P_{p,N}(k) \eisen{k}\,.
\end{equation}
We stress that the right-hand side of the above expression is a formal object given that at fixed $\tau$ the Eisenstein series $\eisen{k}$ grows factorially with $k$, as can be easily seen from \eqref{Eisen_series_definitions}. 

Nevertheless, it is possible to find a Borel-like resummation for the formal series \eqref{Cpn_Eisen_expansion}, which will later prove to be useful in disentangling the non-perturbative effects appearing as $N,p\to \infty$. 
To proceed, we use the lattice-sum integral representation \eqref{Eisen_series_definitions} for the non-holomorphic Eisenstein series, and after exchanging the sum over $k$ in \eqref{Cpn_Eisen_expansion} with the integral we arrive at
\begin{equation}\label{C_pN_lattice_sum_rep}
    \mathcal{C}_{p,N}(\tau) = \frac{1}{2}\sum_{(m,n)\in \mathbb{Z}^2}\int_0^\infty e^{-t\Ymn}B_{p,N}(t)\,\dd t\,,
\end{equation}
where the Borel transform is defined as the analytic continuation of the convergent power series,
\begin{equation}\label{eq:Bold}
 B_{p,N}(t) \coloneqq \sum_{k=1}^\infty (-1)^k P_{p,N}(k+1)\,t^k\,.
\end{equation}
We note that the series defining the Borel transform $B_{p,N}(t)$ only converges for $|t|<1$. However, it can be analytically continued to a rational function with a pole at $t=-1$.
Furthermore, as a consequence of the reflection identity $P_{p,N}(s) = P_{p,N}(1-s)$, we find that the Borel transform satisfies the inversion identity $t^{-1}B_{p,N}(t^{-1})= B_{p,N}(t)$. It is quite impressive and extremely non-obvious from the matrix model formulation that integrated correlators (not necessarily of maximal-trace correlators) admit such a simple lattice sum representation \eqref{C_pN_lattice_sum_rep}, see e.g.  \cite{Brown:2023why}.

Now we would like to consider the regimes of large $N$ and/or large $p$. 
Albeit formal, the series expansion \eqref{Cpn_Eisen_expansion} 
will prove to be rather useful in extracting the non-perturbative, 
modular invariant effects that we are interested in.
We shall arrive at these through modular resurgence analysis. 
For a pragmatic introduction to resurgence analysis
we refer to \cite{Dorigoni:2014hea} 
and to \cite{Dunne:2025mye} for a set of pedagogical, physically-oriented lectures. 

Resurgence analysis is a well-known mathematical 
framework for analysing factorially 
divergent power series and extract from them non-perturbative effects. 
In our case, all functions of the complexified coupling $\tau$ 
that we deal with are actually modular invariant functions, 
therefore we expect the non-perturbative terms, at large $N$ and/or large $p$, to display this property.
Our task will be to modify the conventional resurgence analysis paradigm to derive modular invariant non-perturbative corrections, 
generalising what was already done in \cite{Dorigoni:2024dhy}, where it was shown how to do so for
the integrated correlator $\mathcal{C}_{2,N}(\tau)$, at large $N$, reproducing in particular the results of \cite{Dorigoni:2022cua}. 

In the general setting, we denote by $\Phi(\Lambda;\tau)$ a possible $\mathcal{N}=4$ SYM quantity which admits a spectral representation of the form \eqref{overlap_equation_def} and where $\Lambda$ denotes a large parameter,~e.g. either the number of colours or the $R$-charge. Based on a plethora of examples,  in the limit $\Lambda \to\infty$ we find that order by order in $1/\Lambda$  the spectral overlap becomes proportional to 
$\Lambda^{-s}$ and it does contain infinitely many poles located 
at half-integer values of the spectral parameter $s$.
This is true in particular for the spectral overlap of ${\cal C}_{p,N}$ given in \eqref{c_pN_poly_expr}, 
with either $N$ or $p$ playing the role of $\Lambda$.
Compared to the finite $N$ or finite $p$ case, we find that in this limit there is a different issue in computing the large $\Lambda$ expansion from the spectral decomposition.

Taking the spectral representation of $\Phi(\Lambda;\tau)$ and evaluating the contour integral 
as above yields a formal series over half-integral non-holomorphic Eisenstein 
series $\eisen{k+\tfrac{3}{2}}$ with $k\in \mathbb{N}$, which we denote below with 
$\Phi_P(\Lambda;\tau)$, and whose 
schematic form is
\begin{equation}\label{eq:PhiP}
    \Phi_P(\Lambda;\tau) = \sum_{k=0}^\infty b_k \,\Lambda^{-k-\frac{3}{2}} \eisen{k+\frac{3}{2}}\,.
\end{equation}
Quite crucially the real coefficients $b_k$ now grow factorially fast with $k$, unlike the previously discussed case \eqref{Cpn_Eisen_expansion}, where the coefficients $(-1)^{k}P_{p,N}(k+1)$ grew only polynomially in $k$.
As a consequence, the integral representation used to resum the series \eqref{Eisen_series_definitions} can no longer be used since the corresponding Borel transform \eqref{eq:Bold} would now have zero radius of convergence, being itself a factorially divergent formal series. 
In resurgence language, we are presently dealing with a so-called Gevrey-$2$ series.

 At fixed coupling constant $\tau$, we can think of \eqref{eq:PhiP} as a power series in $\Lambda^{-1}$ with factorially divergent yet modular invariant perturbative coefficients $b_k \eisen{k+3/2}$. In the spirit of resurgence analysis, this simple fact actually hides important non-perturbative information. 
 As we now briefly review, a general method to construct the non-perturbative (in $\Lambda$) and modular invariant (in $\tau$) completion to power series of the form \eqref{eq:PhiP} was discussed in \cite{Dorigoni:2024dhy}. The core idea is to make use of the integral representation
\begin{equation}\label{Eisen_largeNp_intRep}
    \Lambda^{-s}\eisen{s} = \int_0^\infty \NPfn{\sqrt{\Lambda}t}\, \frac{2 \Gamma(s)}{\Gamma(2s)}(4t)^{2s-1} \dd t\,,\qquad \NPfn{t} \coloneqq  \summn e^{-4t\sqrt{\Ymn}} \,.
\end{equation}
We substitute the above identity in \eqref{eq:PhiP} and then exchange the sum with the integral in a way analogous to how we derived the lattice sum representation \eqref{C_pN_lattice_sum_rep}. 
We are then led to define a modified Borel transform $\mathcal{B}[\Phi_P](t)$, which is a function of a complex variable $t\in\mathbb{C}$ given by the convergent power series,
\begin{equation}
    \mathcal{B}[\Phi_P](t) \coloneqq \sum_{k=0}^\infty b_k\,\frac{2\Gamma(k+\frac{3}{2})}{\Gamma(2k+3)}(4t)^{2k+2}\,.\label{eq:BorelDef}
\end{equation}
Notice that since the original coefficients $b_k$ grew factorially fast, the Borel transform just defined has a finite radius of convergence. Additionally, we assume that the function $\mathcal{B}[\Phi_P](t)$ can be analytically continued outside its radius of convergence without any natural boundary of analyticity.

Similar to how we derived \eqref{C_pN_lattice_sum_rep}, we would like to use the integral representation \eqref{Eisen_largeNp_intRep} to define the Borel resummation of the formal power series \eqref{eq:PhiP} as the integral transform
\begin{equation}\label{eq:BorelNaive}
  \Phi_P(\Lambda;\tau) \stackrel{?}{=} \int_0^{\infty} \NPfn{\sqrt{\Lambda}t}  \mathcal{B}[\Phi_P](t)\,\dd t\,.
\end{equation}
However, compared to the previously discussed finite $N$ and finite $p$ expression \eqref{C_pN_lattice_sum_rep}, we now face the problem that the modified Borel transform $\mathcal{B}[\Phi_P](t)$ does contain a branch-cut singularity along the contour of integration $t\in \mathbb{R}_{>0}$, thus making the above equation ill-defined.

To circumvent this problem, we choose in the integral representation \eqref{eq:BorelNaive} a different direction of integration, say ${\arg t}= \theta\in [-\pi,\pi)$, in the complex $t$-plane along which the Borel transform $\mathcal{B}[\Phi_P](t)$ has no singularities. Thus we define the directional Borel resummation
\begin{equation}\label{resummation_def}
    \mathcal{S}_\theta [\Phi_P](\Lambda;\tau) \coloneqq \int_0^{e^{i\theta}\infty} \NPfn{\sqrt{\Lambda}t}  \mathcal{B}[\Phi_P](t)\,\dd t \,.
\end{equation}
 This integral defines an analytic function of $\Lambda$ in the wedge ${\rm Re}(\sqrt{\Lambda} e^{i\theta})>0$, which is also modular invariant with respect to $\tau$, since the integration kernel $\NPfn{t}$ defined in \eqref{Eisen_largeNp_intRep} is manifestly so.
 
 Although we have now succeeded in assigning an analytic function to the asymptotic power series \eqref{eq:PhiP}, we still have to face an ambiguity in our resummation procedure since the Borel transform does contain singularities in the Borel plane, in particular along ${\rm arg}(t)=0$. If we choose two resummation directions $-\pi<\theta_1<0$ and $0<\theta_2<\pi$, we find that the two analytic continuations $\mathcal{S}_{\theta_1} [\Phi_P](\Lambda;\tau)$ and $\mathcal{S}_{\theta_2} [\Phi_P](\Lambda;\tau)$ of the same power series \eqref{eq:PhiP} differ on the common domain of analyticity, and in particular $\mathcal{S}_{\theta_1} [\Phi_P](\Lambda;\tau)\neq \mathcal{S}_{\theta_2} [\Phi_P](\Lambda;\tau)$ for $\Lambda>0$. 
 Furthermore, even if the original formal power series \eqref{eq:PhiP} is manifestly real for $\Lambda>0$ and arbitrary $\tau$, neither of the resummations $\mathcal{S}_{\theta_1} [\Phi_P](\Lambda;\tau)$ and $\mathcal{S}_{\theta_2} [\Phi_P](\Lambda;\tau)$ is.
 
 To cure both problems, we introduce a modular-invariant \textit{transseries} completion to the perturbative power series \eqref{eq:PhiP},
\begin{align}
    &\label{eq:TS} \Phi(\Lambda;\tau) = \Phi_P(\Lambda;\tau) + \sigma\, \Phi_{N\!P}(\Lambda;\tau)\,,
\end{align}
where the non-perturbative terms, fully contained in the second factor $\Phi_{N\!P}(\Lambda;\tau)$, are given by yet another infinite series over a different class of modular invariant functions,
\allowdisplaybreaks{
\begin{align}
&\label{NP_series_def}\Phi_{N\!P}(\Lambda;\tau) = \sum_{k=-M}^\infty d_k \Lambda^{-\frac{k+1}{2}}D_\Lambda\Big(\frac{k+1}{2};\tau\Big)\,,\\*
    &\label{eq:Dfct} D_\Lambda(s;\tau) = \summn \exp\big(-4\sqrt{\Lambda\Ymn}\big)\,\Big(16\Ymn \Big)^{-s}\,.
\end{align}}
The coefficients $d_k$ once again grow factorially fast at large $k$ and are completely encoded in the perturbative coefficients $b_k$ of the starting asymptotic series \eqref{eq:PhiP} as shown below.
Given the exponential suppression factor $\exp\big(-4\sqrt{\Lambda\Ymn}\big)$, it appears manifest that the modular invariant function $D_\Lambda(s;\tau)$ and hence  $\Phi_{N\!P}(\Lambda;\tau)$ are non-peturbative in the large parameter $\Lambda$, which we remind the reader will eventually become either the number of colours $N$ or the charge $p$ for the particular integrated correlator $\mathcal{C}_{p,N}(\tau)$ considered.

 The coefficient $\sigma$ multiplying the non-perturbative terms in the transseries \eqref{eq:TS} is called \textit{transseries parameter}, and it is a piecewise constant function of ${\rm arg}(\Lambda)$ chosen in such a way as to obtain an unambiguous Borel resummation of the transseries \eqref{eq:TS}, which is real for $\Lambda>0$.
In more detail, the transseries representation \eqref{eq:TS} is constructed by first evaluating the so called Stokes automorphism, i.e. the discontinuity in directional Borel resummation above and below the singular direction ${\rm arg}(t)>0$, 
\begin{equation}
\label{eq:Stokes}
 \Big(\mathcal{S}_+-\mathcal{S}_-\Big)[\Phi_P] \coloneqq \lim_{\theta\to 0^+} \left[\mathcal{S}_{+\theta} - \mathcal{S}_{-\theta}\right][\Phi_P] (\Lambda;\tau)  = \int_\gamma \NPfn{\sqrt{\Lambda}t}  \mathcal{B}[\Phi_P](t)\,\dd t\,.
\end{equation}
The Hankel integration contour $\gamma$ comes from $+\infty-i\, 0^+$, circles clock-wise around the origin and then goes back to $+\infty+i\, 0^+$. 

Following \cite{Dorigoni:2024dhy}, we assume for simplicity that the Borel transform has a logarithmic branch-cut along $t\in [1,\infty)$ plus possibly an order $M$ pole at $t=1$.
It is an easy exercise in complex analysis to evaluate the contour integral \eqref{eq:Stokes}, which in turn defines the non-perturbative sector $\Phi_{N\!P}(\Lambda;\tau) $ presented in \eqref{NP_series_def} via
\begin{equation}\label{discont}
    \Big(\mathcal{S}_+-\mathcal{S}_-\Big)[\Phi_P](\Lambda;\tau) = -2i\,\mathcal{S}_0[\Phi_{N\!P}](\Lambda;\tau)\,,
\end{equation}
where we have
\begin{align}
    &\label{resummation_NP_def} \mathcal{S}_\theta[\Phi_{N\!P}](\Lambda;\tau) \coloneqq \sum_{k=-M}^{-1}  d_k\,\Lambda^{-\frac{k+1}{2}} \Dfn{\Lambda}{\tfrac{k+1}{2}}+\int_0^{e^{i\theta}\infty}\NPfn{\sqrt{\Lambda}(t+1)} \,\mathcal{B}[\Phi_{N\!P}](t) \,\dd t\,,\\
    &\mathcal{B}[\Phi_{N\!P}](t) \coloneqq \sum_{k=0}^\infty \frac{d_k}{\Gamma(k+1)}t^{k}.
\end{align}

Lastly, the transseries parameter $\sigma$ appearing in \eqref{eq:TS} is correlated with the choice of resummation $\mathcal{S}_\pm$, which is in turn correlated with ${\mbox{arg}}(\Lambda) \gl 0$.
To define a real resummation for $\Lambda>0$, we perform what is called a \textit{ median resummation} \cite{delabaere1999resurgent}
of the transseries \eqref{eq:TS} and for which the transseries parameter is the piecewise constant function $\sigma=\sigma_\pm = \pm i $ according to  ${\mbox{arg}}(\Lambda) \gl 0$.
Given the discontinuity \eqref{discont}, it is easy to see that the median resummation of the transseries is analytic in a neighbourhood of the positive real line $\Lambda>0$ since we have
\begin{equation}\label{requirement_resummation}
    \mathcal{S}_{+} [\Phi_P + \sigma_+ \Phi_{N\!P}](\Lambda;\tau) =  \mathcal{S}_{-} [\Phi_P+\sigma_- \Phi_{N\!P}](\Lambda;\tau)\,,
\end{equation}
and it is furthermore real for $\Lambda>0$ as it should.

As previously mentioned, \cite{Dorigoni:2024dhy} showed that when the above general construction is applied to the study of the large-$N$ expansion for the integrated correlator $\mathcal{C}_{2,N}(\tau)$, the median resummation does coincide with the exact large-$N$ modular invariant transseries expansion of the same physical quantity obtained in \cite{Dorigoni:2022cua} through a different generating-series approach.

It is important to emphasise that the transseries expansion can also be derived directly from the spectral representation \eqref{overlap_equation_def} by expanding the integrand in a regime of large parameters, being that the number of colours $N$ or the charge $p$.
In particular, if we first expand the spectral overlap and then push the $s$-contour of integration to ${\rm Re}(s)\to \infty$ we find that while the poles contribution does immediately produce a formal perturbative expansion of the general form \eqref{eq:PhiP}, the contribution at infinity is much subtler. In \cite{Dorigoni:2024dhy} it was shown that such a contribution does not vanish and it evaluates precisely to the spectral representation of the transseries non-perturbative completion $\Phi_{N\!P}(\Lambda;\tau)$.
A similar analysis can be performed to study the large-$N$ or large-charge expansion of the spectral decomposition \eqref{overlap_equation_def}, where the non-perturbative terms we shall be interested in can be found by analysing the large-$s$ asymptotics of the expanded integrand. However, we find it more immediate and systematic to derive said non-perturbative corrections using the modular resurgence analysis here briefly described.

\section{Non-perturbative structures for maximal-trace operators}
\label{sec:limits}

In this section we apply the methods outlined above 
and study the non-perturbative corrections to the integrated correlator 
$\mathcal{C}_{2r,N}(\tau)$ of two maximal-trace operators of 
even charge $\Omax{p}$ with $p=2r$ and two superconformal primary 
operators in the stress-tensor multiplet $\mathcal{O}_2$ in different 
regimes of large-$N$ and large-charge.

The regimes we consider will behave rather differently:
\vspace{-0.2cm}
\begin{itemize}
\item Section \ref{sec:Large_charge_fixed_N}: large charge $p$, finite $N$;
\item Section \ref{sec:gammaL2}: large $N$ and large charge $p=\alpha N^\gamma$ with $0<\gamma<2$ and $\alpha>0$ fixed;
\item Section \ref{sec:N2}: large $N$ and large charge $p=\alpha N^2$ with $\alpha>0$ fixed;
\item Section \ref{sec:gamma3}: large $N$ and large charge $p=\alpha N^\gamma$ with $\gamma>2$ and $\alpha>0$ fixed.
\end{itemize}
\vspace{-0.2cm}
The regime where $N$ is large and $p$ is fixed has already been discussed in detail in \cite{Paul:2022piq} and it essentially reduces to the case $p=2$ whose large-$N$ transseries expansion has been already extensively studied in \cite{Dorigoni:2022cua,Dorigoni:2024dhy} and hence will not be discussed here.

Let us recall that our starting point is the spectral decomposition \eqref{overlap_equation_def} for the integrated correlator specialised to the case of two even charge $p=2r$ maximal-trace operators.
The corresponding spectral overlap $c_{2r,N}(s)$ was found in \cite{Paul:2023rka}, where it was expressed as the product of two simpler functions,
\begin{equation}\label{overlap_even}
   c_{2r,N}(s)= M_N(s)F_{r,N}(s)\,.
\end{equation}

The first building block $M_N(s)$ does not depend on the charge of the inserted operator $\Omax{2r}$ and it is equal to the spectral overlap of $\mathcal{C}_{2,N}(\tau)$, which is the integrated correlator of four $\mathcal{O}_2$ operators first computed exactly for all $N$ and $\tau$ in \cite{Dorigoni:2021guq,Dorigoni:2021bvj} and then analysed via spectral methods in \cite{Collier:2022emf}.
This function is explicitly given by
\begin{align}\label{c2_spectral_overlap}
    M_N(s)= &\frac{N(N-1)}{4}\cdot \frac{\pi s(1-s)(2s-1)^2}{\sin{(\pi s)}} {}_3F_2(2-N,s,1-s;3,2\vert 1)\,,
\end{align}
where $_pF_q(a_1,...,a_p;b_1,...,b_q\vert z)$ denotes a generalised hypergeometric function.

The second contribution $F_{r,N}(s)$ computed in \cite{Paul:2023rka} becomes absolutely crucial when discussing higher charge operators. This is given by 
\begin{equation}\label{F_p_formula}
    F_{r,N}(s)= \frac{1}{s(1-s)} \left[1-\tFt{-r}{s}{1-s}{1}{\frac{N^2-1}{2}}{1}\right]\,.
\end{equation}
We notice that both hypergeometric functions $_pF_q(a_1,...,a_p;b_1,...,b_q\vert z)$ appearing in \eqref{c2_spectral_overlap} and \eqref{F_p_formula} have as first parameter entry $a_1$ a negative integer. As a consequence, we conclude that for any number of colours $N\in \mathbb{N}^{\geq 2}$ and for any value of the charge $ 2r \in 2\mathbb{N} $ these hypergometric factors simplify dramatically reducing to polynomials in $s(1-s)$ (as already discussed around \eqref{c_pN_poly_expr}). 

Shortly we shall be interested in deriving the large $N$ and large $p=2r$ expansion of the integrated correlator $\mathcal{C}_{2r,N}(\tau)$. To achieve this goal, we find it convenient to construct a generating series for both spectral overlap building blocks $M_N(s)$ and $F_{r,N}(s)$.
The first generating series we consider concerns the charge-independent piece, which yields
\begin{equation}
    M(z;s) \coloneqq \sum_{N=2}^\infty M_N(s)z^{N-1}=-\frac{\pi s(1-s)(2s-1)^2}{2\sin(\pi s)}\cdot \frac{z}{(z-1)^3}\tFo{s}{1-s}{2}{\frac{z}{z-1}}\,.
\end{equation}
This result can be readily derived by substituting the expression \eqref{c2_spectral_overlap} in the generating series and replacing the hypergeometric function by its Gauss sum representation. Exchanging the generating series with the Gauss summation yields the above result.

The large-$N$ expansion for $M_N(s)$ is then promptly obtained from the generating series by using the Cauchy integral formula,
\begin{equation}
    M_N(s) = \oint_{|z|<1} \frac{M(z;s)}{z^{N}}\frac{\dd z}{2\pi i}\,,
\end{equation}
where the contour is oriented counter-clockwise.
We first perform the change of variable $z\to \frac{1}{z}$ exchanging the interior and exterior of the unit circle and yielding
\begin{equation}
    M_N(s) = \frac{\pi s (1-s)(2s-1)^2}{2\sin(\pi s)}\oint_{|z|>1}\frac{z^N}{(z-1)^3}\tFo{s}{1-s}{2}{\frac{1}{1-z}}\frac{\dd z}{2\pi i}\,,\label{eq:MCauchy}
\end{equation}
where the contour orientation is still counter-clockwise. 
The hypergeometric function appearing at the integrand can be simplified using the reflection formula: 
\begin{equation}\label{Hypergeom_ref1}
    \tFo{s}{1-s}{2}{\frac{1}{1-z}} = -\frac{4^{-s}\Gamma(s-1)\tan(\pi s)}{\sqrt{\pi}\Gamma(s+\frac{1}{2})} (z-1)^s\tFo{s-1}{s}{2s}{1-z} + (s \leftrightarrow 1-s)\,.
\end{equation}
We note that the first term on the right side does break the symmetry under $s \leftrightarrow 1-s$. However, this symmetry is of course restored by adding the second reflected term. 

Notice that while the left side of \eqref{Hypergeom_ref1} has a branch cut only along $z\in [0,1]$, the individual terms on the right side now have a branch-cut along $z\in (-\infty,1]$. 
Therefore, upon substitution of \eqref{Hypergeom_ref1} in \eqref{eq:MCauchy} we must deform the contour of integration to the contour $\tilde{\gamma}$ which comes from $-\infty - i \,0^+$, circles counter-clockwise around the branch point at $z=1$ and then goes back to $-\infty +i\,0^+$. 
Using the identity \eqref{Hypergeom_ref1} in \eqref{eq:MCauchy} and then performing the change of variable $z\to \frac{z}{N}+1$ we arrive at 
\begin{equation}\label{eq:Mngamma}
    M_N(s) =  \frac{4^{-s}(2s-1)\sqrt{\pi}\Gamma(s+1)}{\Gamma(s-\frac{1}{2})\cos(\pi s)} N^{2-s}\int_{\gamma}z^{s-3}\Big(1+\frac{z}{N}\Big)^N \tFo{s-1}{s}{2s}{-\frac{z}{N}}\frac{\dd z}{2\pi i} +(s \leftrightarrow 1-s)\,,  
\end{equation}
where the contour $\gamma$ is a simple shift of the previously described contour $\tilde{\gamma}$, which now surrounds the branch-point $z=0$. 

The large-$N$ expansion for the building block $M_N(s) $ is now within reach, since in this limit the integrand can be expanded as
\begin{equation}
z^{s-3}\Big(1+\frac{z}{N}\Big)^N \tFo{s-1}{s}{2s}{-\frac{z}{N}} = e^z z^{s-3} \left[1 - \frac{z(z+s-1)}{2N}+O(N^{-2})\right]\,.
\end{equation}
The $z$-integral can now be evaluated term by term by making use of the integral representation for the reciprocal gamma function
\begin{equation}\label{Gamma_fn_ident}
    \frac{1}{\Gamma(a)} = \int_{\gamma}e^z z^{-a}\frac{\dd z}{2\pi i}\,,
\end{equation}
thus producing the large-$N$ asymptotic expansion
\begin{equation}\label{large_N_series}
    M_N(s) = \sum_{g=0}^\infty N^{2-2g}\left[ N^{-s}\mathcal{M}^{(g)}(s)+N^{s-1}\mathcal{M}^{(g)}(1-s)\right]\,.
\end{equation}
We note that at large $N$ the building block $ M_N(s) $ is expressed as a genus expansion, containing only even powers $N^{2-2g}$ - a fact highly hidden in the original representation. The first few terms in this expansion are given by
\begin{align}
    \mathcal{M}^{(0)}(s) &\label{eq:M0}= \frac{4^{-s}(2s-1)\Gamma(s-2)\Gamma(s+1)}{\sqrt{\pi}\Gamma(s-\frac{1}{2})}\tan(\pi s)\,,\\
    \mathcal{M}^{(1)}(s) &\label{eq:M1}= -\frac{4^{-s-1}(s+5)(2s-1)\Gamma(s+1)^2}{6\sqrt{\pi}(2s+1)\Gamma(s-\frac{1}{2})}\tan(\pi s)\,.
\end{align}

To analyse the second building block $F_{r,N}(s)$ defined in \eqref{F_p_formula} we construct again a corresponding generating series
\begin{equation}
    F_N(y;s) \coloneqq \sum_{r=1}^\infty F_{r,N}(s)y^r = \frac{1}{s(1-s)(y-1)}\left[ \tFo{s}{1-s}{\frac{N^2-1}{2}}{\frac{y}{y-1}}-1\right]\,.\label{eq:Fgenser}
\end{equation}
At this point we need to be careful on how we proceed since the asymptotic expansion of $F_{r,N}(s)$ will change dramatically depending on the particular large parameter limit we may wish to consider.
In particular, the various regimes outlined at the beginning of this section will produce wildly different asymptotic expansions for $F_{r,N}(s)$. For this reason we shall now discuss the different cases separately.

\subsection{Large charge, fixed $N$}
\label{sec:Large_charge_fixed_N}

We start by considering the limit where the charge $p\to\infty$, while the number of colours $N$ is kept finite. Although the physical meaning of such a limit is not immediately clear, in recent years there has been significant progress in understanding how quantum field theories simplify in the limit of large charge insertions, see e.g. \cite{Hellerman:2015nra,Monin:2016jmo,Alvarez-Gaume:2016vff,Hellerman:2017veg,Hellerman:2017sur,Hellerman:2018xpi,Hellerman:2020sqj,Hellerman:2021yqz,Hellerman:2021duh} as well as the recent review \cite{Gaume:2020bmp}. It appears to be generically true that the large charge sector of a quantum field can be described by some kind of effective field theory whose degrees of freedom might differ from the ones appearing in the original theory. While $\mathcal{N}=4$ SYM at large-$N$ has a well-known dual description, it is not quite clear what effective theory one should use to describe the finite-$N$, large-charge limit here considered.

The finite-$N$, large-charge asymptotics of the maximal-trace integrated correlator \eqref{Integrated_corr_def} has already been discussed in \cite{Paul:2023rka,Brown:2023why}, where it was noticed that this expansion is susceptible to subtle differences between the case of even and odd number of colours $N$. In this Section we show that an interesting resurgence analysis structure dubbed \textit{Cheshire cat resurgence} is at play here.

Firstly, we re-derive the finite-$N$ large-charge expansion of the maximal-trace integrated correlator \eqref{Integrated_corr_def} starting from the corresponding spectral overlap \eqref{overlap_even}. Since we are presently working at finite $N$ and the building block $M_N(s)$ given in equation \eqref{c2_spectral_overlap} does not depend on the charge $p=2r$, we simply need to compute the large-$r$ expansion of the second term $F_{r,N}(s)$ presented in equation~\eqref{F_p_formula}.

This analysis closely follows our derivation of the large-$N$ expansion for $M_N(s)$ from its generating series.
We start from the generating series \eqref{eq:Fgenser} to rewrite $F_{r,N}(s)$ via its Cauchy integral representation
\begin{equation}\label{Cauchy_FrN}
F_{r,N}(s) =  \oint_{|y|<1} \frac{F_N(y;s)}{y^{r+1}}\frac{\dd y}{2\pi i} \,.
\end{equation} 
We then combine the change of integration variables $y\to \frac{1}{y}$ together
with the hypergometric function reflection identity
\allowdisplaybreaks{
\begin{align}
     \tFo{s}{1-s}{\frac{N^2-1}{2}}{\frac{1}{1-y}} =&\frac{4^{-s}\sqrt{\pi}\Gamma(\frac{N^2-1}{2})(y-1)^s}{\Gamma(\frac{N^2-1-2s}{2})\Gamma(s+\frac{1}{2})\cos(\pi s)}\tFo{s}{\frac{3-N^2}{2}+s}{2s}{1-y} \\*
     &\notag +(s \leftrightarrow 1-s)\,.
\end{align}}
Lastly, we perform the same change of variables $y\to \frac{y}{r}+1$ as before in order to arrive at the integral representation 
\begin{align}\label{Fr_largep_rep}
    \notag F_{r,N}(s) = \frac{1}{s(1-s)} + &\left[\frac{\Gamma(\frac{N^2-1}{2})\Gamma(s-1)}{\Gamma(\frac{N^2-1-2s}{2})\Gamma(2s+1)\cos(\pi s)}\times \right. \\
    &\left. r^{-s}\int_\gamma y^{s-1}\big(1+\frac{y}{r}\big)^r \tFo{s}{s-\frac{N^2-3}{2}}{2s}{-\frac{y}{r}}\frac{\dd y}{2\pi i}+(s\leftrightarrow 1-s)\right]\,,
\end{align}
where the contour of integration $\gamma$ is identical to the one discussed in \eqref{eq:Mngamma}.

The integrand can be easily expanded at large $r$,
\begin{equation}
y^{s-1}\Big(1+\frac{y}{r}\Big)^r \tFo{s}{s-\frac{N^2-3}{2}}{2s}{-\frac{y}{r}} = e^y y^{s-1} \left[1 - \frac{y(2s+2y+3-N^2)}{4r} + O\big( r^{-2}\big)\right]\,,
\end{equation}
which allows us again to use the gamma function identity \eqref{Gamma_fn_ident} to perform the $y$-integral term by term, thus arriving at the large-charge fixed-$N$ asymptotic series
\begin{equation}\label{Large_charge_small_N_series}
    F_{r,N}(s) = \frac{1}{s(1-s)} +\sum_{g=0}^\infty r^{-g}\left[ r^{-s}\mathcal{F}^{(g)}_N(s)+r^{s-1}\mathcal{F}^{(g)}_N(1-s)\right]\,,
\end{equation}
where the first few terms are given by
\allowdisplaybreaks{
\begin{align}\label{FNzero}
    \mathcal{F}^{(0)}_N(s) &= \frac{2^{-2s}\Gamma(\frac{N^2-1}{2})\Gamma(s-1)}{\sqrt{\pi}s\Gamma(\frac{N^2-1}{2}-s)\Gamma(s+\frac{1}{2})}\tan(\pi s)\,,\\* \label{FNone}
    \mathcal{F}^{(1)}_N(s) &= \frac{2^{-1-2s}\Gamma(\frac{N^2+1}{2})\Gamma(s-1)}{\sqrt{\pi}\Gamma(\frac{N^2-1}{2}-s)\Gamma(s+\frac{1}{2})}\tan(\pi s)\,.
\end{align}}

As it will shortly be very important, we notice a peculiar behaviour of the expanded spectral overlap $\mathcal{F}^{(g)}_N(s)$ depending on the parity of the number of colours $N$. When $N$ is odd (or for that matter generic) due to the trigonometric factor $\tan(\pi s)$ we find that $\mathcal{F}^{(g)}_N(s)$ has infinitely many poles located at half-integer $s$. However, for even $N$ the factor $\Gamma(\frac{N^2-1}{2}-s)$ at denominator has simple poles for all half-integer $ s = \frac{N^2-1}{2}+n$ with $n\in\mathbb{N}$ thus cancelling for ${\rm Re}(s)>0$ all but a finite set of poles coming from $\tan(\pi s)$. This fact will have important repercussions for the resurgent properties of the large-charge expansion of the integrated correlator $\mathcal{C}_{2r,N}$.

We now possess all the pieces for analysing the finite-$N$, large-charge expansion of the maximal-trace integrated correlator \eqref{Integrated_corr_def}.
We rewrite the spectral overlap \eqref{overlap_even}
using \eqref{Large_charge_small_N_series}, 
\begin{align}
    \mathcal{C}_{2r,N}(\tau) \overset{r \to \infty}{=} &\notag \average{\mathcal{C}_{2r,N}} + \intRes \frac{M_N(s)}{s(1-s)}\eisen{s}\frac{\dd s}{2\pi i} \\
    &\label{eq:largeR1}+ 2\sum_{g=0}^\infty r^{-g}\intRes r^{-s}\mathcal{F}^{(g)}_N(s)M_N(s)\eisen{s}\frac{\dd s}{2\pi i}\,,
\end{align}
where the function $M_N(s)$ is presented in equation \eqref{c2_spectral_overlap}.
Following \cite{Paul:2023rka}, we find it useful to split the above equation into two different parts
\begin{equation}\label{eq:Csplit}
\mathcal{C}_{2r,N}(\tau) =\average{\mathcal{C}_{2r,N}} + H_N(\tau) + 2\sum_{g=0}^\infty r^{-g}\intRes r^{-s}\mathcal{F}^{(g)}_N(s)M_N(s)\eisen{s}\frac{\dd s}{2\pi i}\,,
\end{equation}
where the auxiliary, $r$-independent modular invariant function $H_N(\tau)$ is defined as
\begin{equation}\label{eq:HN}
H_N(\tau)\coloneqq \intRes \frac{M_N(s)}{s(1-s)}\eisen{s}\frac{\dd s}{2\pi i}\,.
\end{equation}

In order to isolate the constant term in $\tau$ from the genuine non-trivial modular invariant part, we still need to push the contour of integration for the last integral  in \eqref{eq:largeR1} past ${\rm Re}(s)>1$.
Furthermore, we also remember that the overlap with the constant function is given by 
\begin{equation}
\average{\mathcal{C}_{2r,N}} = \lim_{s\to 1}c_{2r,N}(s) = \lim_{s\to 1}[M_N(s) F_{r,N}(s)]\,,
\end{equation}
which can be easily computed from \eqref{c2_spectral_overlap}-\eqref{F_p_formula},
\begin{align}
\lim_{s\to 1} M_N(s) &\label{MN_sto1_limit}= \frac{N(N-1)}{4} \,,\\*
    \label{FrN_sto1_limit}\lim_{s\to 1}F_{r,N}(s) & = \psi\left( \tfrac{N^2-1}{2}+r\right)-\psi\left(\tfrac{N^2-1}{2}\right)\,,
\end{align}
with $\psi(x) \coloneqq \Gamma'(x) /\Gamma(x)$ the polygamma function.
We can then rewrite~\eqref{eq:Csplit} in the following form,
\begin{equation}
\label{eq:largeR2}
    \mathcal{C}_{2r,N}(\tau) =  \frac{N(N-1)}{4}\left[ F_{r,N}(1)-\sum_{g=0}^\infty r^{-g-1}\mathcal{F}_N^{(g)}(1)\right]+H_N(\tau) +\sum_{g=0}^\infty r^{-g} \mathcal{C}^{(g)}_N(r;\tau)\,,
\end{equation}
where we defined the `genus'-$r$ large-charge modular invariant contribution as
\begin{equation}\label{large_charge_asymp_def}
    \mathcal{C}^{(g)}_N(r;\tau) \coloneqq 2\int_{{\rm Re}(s)=1+\epsilon}r^{-s}\mathcal{F}^{(g)}_N(s)M_N(s)\eisen{s}\frac{\dd s}{2\pi i}\,.
\end{equation}

The first factor in \eqref{eq:largeR2} is constant in $\tau$ and thus trivially modular invariant, while the second term $H_N(\tau)$ defined in \eqref{eq:HN} is a non-trivial modular invariant function of $\tau$, studied extensively in  \cite{Paul:2023rka}, which is however independent from the charge parameter $r$.
Hence, the focus of our present interest will be analysing the resurgence properties of the `genus'-$g$ large-charge modular invariant contribution $\mathcal{C}^{(g)}_N(r;\tau)$ defined in \eqref{large_charge_asymp_def}.

As already discovered in \cite{Paul:2023rka,Brown:2023why}, the perturbative asymptotic expansion of $\mathcal{C}^{(g)}_N(r;\tau)$ at large $r$ changes dramatically depending on whether $N$ is even or odd. While for $N$ odd the large-$r$ perturbative expansion of $\mathcal{C}^{(g)}_N(r;\tau)$ does produce an asymptotic series in half-integral Eisenstein series $\eisen{s}$ with factorially divergent coefficients, for $N$ even the same perturbative expansion reduces to a polynomial in half-integral Eisenstein series.
Crucially, \cite{Paul:2023rka,Brown:2023why} found that independently from the parity of $N$ at large $r$ the functions $\mathcal{C}^{(g)}_N(r;\tau)$ do contain an infinite series of non-perturbative corrections proportional to the modular invariant function $D_r(s;\tau)$ given in \eqref{eq:Dfct}.
In these reference, the non-perturbative corrections say for $N=2$ were determined using generating series techniques since naively it seems impossible to use the modular resurgence methods outlined in Section \ref{sec:ModRes} to reconstruct the non-perturbative sectors from the terminating perturbative expansion.

We want to show that a subtler version of resurgence analysis is at play here.
The dramatic simplification at even $N$, which affects the perturbative expansion of $\mathcal{C}^{(g)}_N(r;\tau)$, while leaving untouched its non-perturbative sector is an example of a structure that has been referred to as {\em Cheshire cat resurgence}; a phenomenon first observed in quantum mechanical examples~\cite{Dunne:2016jsr,Kozcaz:2016wvy} but also in quantum field theory setups, see e.g.~\cite{Dorigoni:2017smz,Dorigoni:2019kux}.
The reason for the naming stems from the fact that exponentially suppressed corrections do survive even when the corresponding perturbative series terminates after finitely many terms, very much like the lingering grin of the cat from Lewis Carroll's Alice in Wonderland.

Firstly, let us discuss the spectral decomposition \eqref{large_charge_asymp_def} without making any assumption on $N$.
From \eqref{FNzero}-\eqref{FNone}, we see that for generic $N$ the integrand has infinitely many poles located at $s=k+\frac{3}{2}$ with $k\in \mathbb{N}$. 
Focusing for concreteness on the genus-$0$ contribution, we substitute \eqref{c2_spectral_overlap} and \eqref{FNzero} in \eqref{large_charge_asymp_def}, and then push the contour of integration to ${\rm Re}(s) \to +\infty$ collecting residues from the poles at $s=k+\frac{3}{2}$ to derive the perturbative asymptotic expansion,
\begin{align}
&\notag \mathcal{C}^{(0)}_N(r;\tau) \sim \mathcal{C}^{(0)}_{N,P}(r;\tau) \coloneqq (N-1) N\,  \Gamma \left(\frac{N^2-1}{2}\right)  \sin \left(\frac{\pi  N^2}{2}\right) \times \\
&\label{eq:C0N}\sum_{k=0}^\infty \frac{ (k+1) \, _3F_2\left(-k-\frac{1}{2},k+\frac{3}{2},2-N;2,3\vert 1\right)}{4^{k+1} \pi ^{\frac{3}{2}}} \cdot \frac{  \Gamma \left(k+\frac{3}{2}\right)  \Gamma \left(k+3-\frac{N^2}{2}\right)}{\Gamma (k+1)}r^{-k-\frac{3}{2}}\, \eisen{k+\frac{3}{2}}\,.
\end{align}
This expression is an asymptotic series in half-integral Eisenstein series precisely of the form \eqref{eq:PhiP} previously discussed.

We note that for generic $N$, the coefficients multiplying $\eisen{k+\frac{3}{2}}$ grow factorially fast. However, when $N$ is an even integer we must be careful since the overall trigonometric factor  $\sin(\pi N^2/2)$ vanishes while the summands with $0\leq k \leq N^2/2-3$ are singular. By either taking a limit over $N$ or by considering directly at the level of integrand \eqref{FNzero} the case $N$ even integer, we conclude that when $N$ is an even integer the asymptotic expansion \eqref{eq:C0N} reduces to a polynomial in $\eisen{s}$ with $s\in\{ \frac{3}{2},...,\frac{N^2-3}{2}\}$.
This is a consequence of the fact that for $N$ even integer, the integrand $\mathcal{F}^{(g)}_N(s)$ becomes holomorphic for ${\rm Re}(s)$ large enough, see e.g. \eqref{FNzero}-\eqref{FNone}.

For the case where $N$ is generic, it is straightforward to apply the modular resurgence methods outlined in Section \ref{sec:ModRes} to define a modular invariant, non-perturbative resummation of the asymptotic perturbative series \eqref{eq:C0N} and rederive the results of \cite{Paul:2023rka,Brown:2023why} obtained via generating series techniques.

For simplicity, we specialise here to $N=3$ for which \eqref{large_charge_asymp_def} at genus-zero simplifies to
\begin{equation}\label{eq:C3int}
    \mathcal{C}^{(0)}_3(r;\tau) = \intRepsilon \frac{3\times 2^{1-2s}\sqrt{\pi}(2s-1)[s(1-s)-6]\Gamma(s)}{\Gamma(4-s)\Gamma(s-\frac{1}{2})\cos(\pi s)}r^{-s}\eisen{s}\frac{\dd s}{2\pi i}\,.
\end{equation}
As already mentioned, this function has an infinite number of poles in the half-plane ${\rm Re}(s)>1$ coming from the term $\cos(\pi s)$ in the denominator. By closing the contour of integration in the right half-plane, or equivalently by substituting $N=3$ into the general expression \eqref{eq:C0N}, we derive the large-$r$ perturbative expansion
\begin{equation}
    \mathcal{C}^{(0)}_{3}(r;\tau) \sim \mathcal{C}^{(0)}_{3,P}(r;\tau) = \sum_{k=0}^\infty \frac{3(k+1)[27+4k(k+2)]}{2^{2k+3} \pi^{\frac{3}{2}}}\cdot \frac{\Gamma(k-\frac{3}{2})\Gamma(k+\frac{3}{2})}{\Gamma(1+k)}\, r^{-k-\frac{3}{2}}\, \eisen{k+\tfrac{3}{2}}\,,
\end{equation}
where we note that the coefficients multiplying the Eisenstein series are factorially divergent.

The Borel transform defined in \eqref{eq:BorelDef} yields for the present case,
\begin{equation}
    \mathcal{B}[\mathcal{C}^{(0)}_{3,P}](t) = 18 t^2\Big[3\tFo{-\tfrac{3}{2}}{\tfrac{3}{2}}{1}{t^2}-t^2\tFo{-\tfrac{1}{2}}{\tfrac{5}{2}}{1}{t^2}-2t^2\tFo{-\tfrac{1}{2}}{\tfrac{5}{2}}{2}{t^2}\Big]\,,
\end{equation}
and, as anticipated, we see that the Borel transform presents a branch-cut singularity along the direction $t^2\in [1,\infty)$.
From the integral representation for the hypergometric function, we find that singularity structure of the Borel transform near the singular point $t=1$ has the form
\begin{equation}\label{eq:C3P}
    \mathcal{B}[\mathcal{C}_{3,P}^{(0)}](t) \sim -\frac{(-6)}{\pi(1-t)} + \mathcal{B}[\mathcal{C}_{3,N\!P}^{(0)}](t-1)\frac{\log(1-t)}{\pi}+\text{reg}(t-1)\,,
\end{equation}
with $\text{reg}(t-1)$ denoting regular terms as $t\to 1$ and
where $\mathcal{B}[\mathcal{C}_{3,N\!P}^{(0)}](t)$ is given by
\begin{equation}
\mathcal{B}[\mathcal{C}_{3,N\!P}^{(0)}](t) = -3(t+1)^2 \left[8\tFo{-\tfrac{3}{2}}{\tfrac{3}{2}}{1}{-t(t+2)}+9\big(1+4t(t+2)\big)\tFo{-\tfrac{1}{2}}{\tfrac{5}{2}}{2}{-t(t+2)}\right]\,.
\end{equation}

Substituting the above expression in the general formula \eqref{resummation_NP_def}, we derive the large-$r$ non-perturbative corrections to $ \mathcal{C}^{(0)}_{3}(r;\tau)$,
\begin{align}
\label{eq:C3NP}
\mathcal{C}_{3,N\!P}^{(0)}(r;\tau) = \sum_{k=-1}^\infty d_{3,k}^{(0)}r^{-\frac{k+1}{2}}\Dfn{r}{\frac{k+1}{2}} \,,
\end{align}
where the first term $d_{3,-1}^{(0)} = -6$ originates from the polar of \eqref{eq:C3P} while higher order terms are captured by
\begin{equation}
 \mathcal{B}[\mathcal{C}^{(0)}_{3,N\!P}](t) = \sum_{k=0}^\infty \frac{d_{3,k}^{(0)}}{k!}t^k =-51-\frac{1839}{4}t-\frac{20211}{8}\frac{t^2}{2}+...\,,
\end{equation}
thus reproducing via a modular resurgent approach the results of \cite{Paul:2023rka,Brown:2023why}.

We now want to understand how to reconstruct in a similar fashion the large-$r$ non-perturbative corrections to $\mathcal{C}_{N}^{(0)}(r;\tau)$ for the case where $N$ is an even integer.
For simplicity we consider $N=2$ where we see an immediate problem: the perturbative sector vanishes identically. 
When $N=2$ the spectral representation
\eqref{large_charge_asymp_def} simplifies to
\begin{equation}\label{eq:C2intrep}
    \mathcal{C}_2^{(0)}(r;\tau)=  \intRepsilon (4r)^{-s}(2s-1)\Gamma(s)\eisen{s}\frac{\dd s}{2\pi i}\,,
\end{equation}
where we note that the integrand is holomorphic in the right half-plane ${\rm Re}(s)>1$.
By closing the contour of integration to the right we encounter no poles; hence the perturbative expansion of $ \mathcal{C}_2^{(0)}(r;\tau)$ is vanishing, i.e. at large-$r$ $\mathcal{C}_2^{(0)}(r;\tau)\sim\mathcal{C}_{2,P}^{(0)}(r;\tau)=0$. Alternatively, we can reach the same conclusion by considering the limit $N\to 2$ of the general perturbative expansion \eqref{eq:C0N}.
Nevertheless, the function \eqref{eq:C2intrep} is obviously non-zero and consequently it must be purely non-perturbative at large~$r$. 

The modular resurgence methods of Section \ref{sec:ModRes} cannot possibly be applied to the vanishing perturbative expansion $\mathcal{C}_{2,P}^{(0)}(r;\tau)$ as to derive the non-perturbative effects $\mathcal{C}_{2,N\!P}^{(0)}(r;\tau)$. Nevertheless, in a variety of circumstances \cite{Dunne:2016jsr,Kozcaz:2016wvy,Dorigoni:2017smz,Dorigoni:2019kux,Dorigoni:2019yoq,Dorigoni:2020oon,Dorigoni:2022bcx,Broadhurst:2025iab} it has been shown how to ``deconstruct zero'',~i.e. how to deform a terminating or even vanishing perturbative expansion as to retrieve the non-perturbative sectors.
Inspired by the Alice in Wonderland creature with a vanishing body and a lingering grin, Cheshire cat resurgence is a method that allows us to reconstruct the non-perturbative effects even when a factorially divergent perturbative expansion is not manifestly present. 

The key point behind this structure is that whenever we encounter a terminating perturbative expansion while still expecting the presence of non-perturbative effects, we should introduce a deformation parameter that reinstates the full resurgence analysis body.
For the present case such deformation is obvious. Since the factorially divergent perturbative expansion \eqref{eq:C0N} terminates after finitely many terms whenever $N$ is an even integer, we should consider its analytic continuation in $N$, and only specialise to $N$ even integer \textit{after} having exploited the resurgence analysis machinery to retrieve the non-perturbative sectors.

Therefore, we are led to consider a deformation of \eqref{eq:C2intrep} where $N=2+\epsilon$,
\begin{equation}\label{eq:C2int}
    \mathcal{C}^{(0)}_{2+\epsilon}(r;\tau)=\intRepsilon \frac{\pi s(1-s)(2s-1)^2}{\sin(\pi s)} r^{-s}\mathcal{F}^{(0)}_{2+\epsilon}(s)\eisen{s}\frac{\dd s}{2\pi i}\,.
\end{equation}
 Keeping only the leading terms in the limit $\epsilon\to 0$, we find the perturbative asymptotic expansion
\begin{equation}\label{eq:C0defP}
   \mathcal{C}^{(0)}_{2+\epsilon}(r;\tau)\sim \mathcal{C}^{(0)}_{2+\epsilon,P}(r;\tau) = \frac{2}{\pi}\sin(2\pi \epsilon)\sum_{k=0}^\infty 2^{-3-2k}(k+1)\Gamma\big(k+\tfrac{3}{2}\big)r^{-k-\frac{3}{2}}\eisen{k+\tfrac{3}{2}}\,.
\end{equation}
Notice that the whole expression is multiplied by $\sin(2\pi\epsilon)$ so that the whole series vanishes when $\epsilon \to 0$. However, at finite $\epsilon$ this is a factorially divergent asymptotic series which we can analyse via the modular resurgence methods of Section \ref{sec:ModRes}.

The Borel transform defined in \eqref{eq:BorelDef} is now given by
\begin{equation}
    \mathcal{B}[\mathcal{C}_{2+\epsilon,P}^{(0)}](t) =\sin(2\pi\epsilon)\, t^2(1-t^2)^{-\frac{3}{2}}\,,
\end{equation}
which has a branch-cut along the direction $t^2\in[1,\infty)$.
It is a little exercise to compute the discontinuity of the Borel transform across the branch cut in the ${\arg}(t)=0$ direction, as in \eqref{eq:Stokes},
\begin{align}
    (\mathcal{S}_+-\mathcal{S}_-)\big[\mathcal{C}^{(0)}_{2+\epsilon,P}\big](r;\tau) &\notag = i \sin(2\pi\epsilon)\int_1^\infty t^2(t^2-1)^{-\frac{3}{2}}\mathcal{E}(\sqrt{r}t;\tau)\dd t\\
    &\label{eq:CNPeps}= i \sin(2\pi\epsilon)\, \, \mathcal{S}_0\left[\mathcal{C}^{(0)}_{2+\epsilon,N\!P}\right](r;\tau)\,.
\end{align}
The requirement that the resummation of \eqref{eq:C0defP} must be real for $r,\epsilon>0$ leads us to consider the transseries representation \eqref{requirement_resummation},
\begin{equation}\label{eq:TSeps}
    \mathcal{C}_{2+\epsilon}^{(0)}(r;\tau) = \mathcal{C}_{2+\epsilon,P}^{(0)}(r;\tau) + \sigma(\epsilon) \,\mathcal{C}_{2+\epsilon,N\!P}^{(0)}(r;\tau)\,,
\end{equation}
where the transseries parameter must satisfy $\rm {Im} \,\sigma(\epsilon) = \pm \sin(2\pi\epsilon)$ according to  ${\mbox{arg}}(r) \gl 0$.

Assuming an analytic dependence on the parameter $\epsilon$, it  seems then natural to {\em conjecture} that the full transseries parameter must be given by 
\begin{equation}
\sigma (\epsilon)= e^{\pm 2\,i\, \pi \epsilon}\,,
\label{eq:TSexp}
\end{equation}
where again the choice in sign is dictated by the lateral Borel resummation employed to resum the perturbative series \eqref{eq:C0defP} which is furthermore correlated with ${\rm arg}(r) \gl 0$.
We do not have a first-principles proof for \eqref{eq:TSexp}, but we notice that such a phenomenon where the full transseries parameter exponentiates as a function of the deformation parameters (in this case $\epsilon$) seems to be ubiquitous when discussing functions with a Cheshire cat structure, see e.g.~\cite{Dorigoni:2020oon, Broadhurst:2025iab}.

Finally, we specialise to the case we were originally interested in, that is we set $\epsilon=0$. While we saw that in this limit the perturbative expansion \eqref{eq:C0defP} vanishes identically, we nevertheless find that surprisingly the non-perturbative terms \eqref{eq:CNPeps} linger on and the transseries parameter simplifies to $\sigma(0)= 1$ so that the deformed transseries \eqref{eq:TSeps} reduces to
\allowdisplaybreaks{
\begin{align}
    \mathcal{C}_2^{(0)}(r;\tau) &\notag =\int_1^\infty t^2(t^2-1)^{-\frac{3}{2}}\mathcal{E}(\sqrt{r}t;\tau)\dd t = \int_0^\infty (t+1)^2 [t(t+2)]^{-\frac{3}{2}}\mathcal{E}(\sqrt{r}(t+1);\tau)\dd t \\*
    &= \sqrt{\frac{\pi}{2}}r^{\frac{1}{4}}D_r\big(-\tfrac{1}{4};\tau \big)-\frac{5}{8}\sqrt{\frac{\pi}{2}}r^{-\frac{1}{4}}D_r\big(\tfrac{1}{4};\tau\big)+...\,,\label{eq:C2NP}
\end{align}}
thus reproducing and extending the results of \cite{Paul:2023rka,Brown:2023why}.

We conclude this section by noting that the large-$r$ non-perturbative sector of $\mathcal{C}_N^{(g)}(r;\tau) $, derived here via resurgence and Cheshire resurgence methods and in \cite{Paul:2023rka,Brown:2023why} via generating series techniques, can in fact be obtained directly from the spectral representations \eqref{eq:C3int}-\eqref{eq:C2int} or in general \eqref{large_charge_asymp_def}.
As shown in \cite{Dorigoni:2024dhy} in the study of the large-$N$ non-perturbative corrections to $\mathcal{C}_{2,N}(\tau)$, while the polar part of the spectral overlap is responsible for the factorially divergent series \eqref{eq:C0N} or the terminating perturbative expansion \eqref{eq:C2int}, we have that the contribution to the spectral integral coming from the contour at infinity ${\rm Re}(s)\to +\infty$ does not vanish and it is purely non-perturbative. This contribution can be evaluated via either saddle point analysis or by exploiting the spectral decomposition \cite{Paul:2023rka} for the modular invariant $D_N(s;\tau)$ functions \eqref{eq:Dfct}, and it can be shown to coincide precisely with the non-perturbative completions \eqref{eq:C3NP}-\eqref{eq:C2NP}.

Although the non-perturbative effects derived in this section could have also been obtained from such a Mellin integral analysis, we emphasise that our present goal was to show that the crucial difference between the transseries expansion of $\mathcal{C}^{(g)}_{N}(r;\tau)$ for even and odd $N$ lies precisely in having Cheshire versus standard resurgence structures at play.

\subsection{Large $N$, large charge with $p = \alpha N^\gamma$ and $\gamma<2$}
\label{sec:gammaL2}

To derive the large-charge expansion of the maximal-trace integrated correlator \eqref{Integrated_corr_def} in the limit where $N\gg1$ and $p =2r= \alpha N^\gamma$ with $\alpha>0$ and $\gamma<2$ fixed parameters, we start again from the spectral overlap \eqref{overlap_even}.

Since in this regime $N$ is taken to be large we can immediately use the previously computed expansion \eqref{large_N_series} for the building block $M_N(s)$. For the second term $F_{r,N}(s)$ presented in equation~\eqref{F_p_formula} we use once more the Cauchy contour integral representation \eqref{eq:Fgenser}.
Given the generating series $F_N(y;s)$ in \eqref{eq:Fgenser}, we simplify the Cauchy integral by rewriting the hypergeometric function appearing in $F_N(y;s)$ in terms of its Gauss series representation and then by deforming the contour of integration from a circle surrounding the pole at $y=0$ to a circle around the pole at $y=1$, yielding
\begin{equation}
    F_{r,N}(s)= -\sum_{k=1}^\infty \frac{(s)_k(1-s)_k}{(\frac{N^2-1}{2})_k k!s(1-s)} \oint_{|y-1|<1}\frac{y^{k-r-1}}{(y-1)^{k+1}}\frac{\dd y}{2\pi i}\,.
\end{equation}

The residue at $y=1$ can be easily evaluated to produce the expansion
\begin{align}
    F_{r,N}(s) &\label{Fr_intermediate_expansion}=\sum_{k=1}^\infty  \frac{(-r)_k  (2-s)_{k-1} (s+1)_{k-1}}{(k!)^2\,\left(\frac{a}{2}\right)_k }\\
    &\notag =\frac{2r}{a}+\frac{(r-1) r (s-2) (s+1)}{a (a+2)}+\frac{2(r-2) (r-1) r (s-3) (s-2) (s+1) (s+2)}{9 a(a+2) (a+4)}+...\,,
\end{align}
where $a\coloneqq N^2-1$. Note that in the present regime where $N\gg 1$ while the charge $p=2r = \alpha N^\gamma$ and $\gamma<2$ it is immediate to deduce that subsequent terms in the above expansion are suppressed further and further in $1/N$. Therefore the final answer can be obtained simply by multiplying the large $N$ expansion for $M_N(s)$ given in \eqref{large_N_series} with that of $F_{r,N}(s)$ readily obtained from \eqref{Fr_intermediate_expansion}.

From the expansion \eqref{Fr_intermediate_expansion}, it is easy to see that at large $N$ with  $p=2r = \alpha N^\gamma$ and $\gamma<2$ the only effect of $F_{r,N}(s)$ is to multiply the different genus $g$ contributions $\mathcal{M}^{(g)}(s)$ given in equation \eqref{large_N_series} by simple polynomials in $s(1-s)$ .
We conclude that in this particular large-$N$ regime the non-perturbative effects to the maximal-trace integrated correlator $\mathcal{C}_{2r,N}(\tau)$ are essentially the same as the ones obtained in the limit $N\to \infty$ with $p$ fixed, that is they are given by linear combinations of the non-perturbative modular invariant functions $D_N(s;\tau)$ defined in \eqref{eq:Dfct}.

It is worth elaborating more on this point for the particular case where $p,N\to \infty$ with $\frac{p}{N}=\alpha$ fixed. 
In this limit we can easily rearrange \eqref{Fr_intermediate_expansion} to obtain the expansion
\begin{equation}
    F_{\frac{\alpha N}{2},N}(s) = \frac{\alpha}{N} + \frac{\alpha^2 (s+1)(s-2)}{4N^2}+\frac{[\alpha ^3 (s-3) (s-2) (s+1) (s+2)-18 \alpha  (s^2-s-4)]}{36 N^3}+ O(N^{-3})\,.
\end{equation}
Consequently, the complete spectral overlap \eqref{overlap_even} can be expanded as
\begin{equation}
    c_{\alpha N,N}(s) = \sum_{g=0}^\infty N^{1-g} \big[ N^{-s}\mathcal{K}^{(g)}(s;\alpha)+N^{s-1}\mathcal{K}^{(g)}(1-s;\alpha)\big]\,,\label{eq:specN}
\end{equation}
with the first few coefficients given by
\allowdisplaybreaks{\begin{align}
    &\mathcal{K}^{(0)}(s;\alpha) =\alpha\mathcal{M}^{(0)}(s)\,,\\*
    &\mathcal{K}^{(1)}(s;\alpha) = \frac{(s-2)(s+1)\alpha^2}{4}\mathcal{M}^{(0)}(s)\,,\\*
    &\mathcal{K}^{(2)}(s;\alpha)= \alpha \mathcal{M}^{(1)}(s) +\frac{[\alpha ^3 (s-3) (s-2) (s+1) (s+2)-18 \alpha  (s^2-s-4)]}{36 }\mathcal{M}^{(0)}(s)\,.
\end{align}}

As already shown in \cite{Paul:2023rka}, from the expansion of the spectral overlap \eqref{eq:specN} it is straightforward to derive the large-$N$ strong 't Hooft coupling expansion $\lambda \coloneqq N g_{_{YM}}^2\gg1$ for the integrated correlator involving two maximal-trace operators $\Omax{\alpha N}$. While the maximal-trace operator here discussed does become heavy at large-$N$, having dimension $\Delta = \alpha N$, it does not quite correspond to a giant graviton operator. 

From the  point of view of $\mathcal{N}=4$ SYM, giant graviton operators are constructed by considering linear combinations of multi-trace operators \eqref{eq:multitrace} given by sub-determinant operators \cite{Balasubramanian:2001nh} and their dual symmetric Schur polynomial counterparts \cite{Corley:2001zk}, both of which are more complicated objects than the maximal-trace operators \eqref{eq:Omax}.
At large-$N$ giant graviton operators become heavy as well, with dimension $\Delta =\alpha N$ and $\alpha$ fixed. Holographically they correspond to D3-brane solutions called respectively \textit{ giant} and \textit{dual giant gravitons}, which wrap an ${\rm S}^3$ inside the ${\rm S}^5$ part of the background ${\rm AdS}_5\times {\rm S}^5$ geometry \cite{McGreevy:2000cw,Balasubramanian:2001nh,Berenstein:2002ke,deMelloKoch:2004crq}, or an ${\rm S}^3$ inside the ${\rm AdS}_5$ part \cite{Hashimoto:2000zp, Grisaru:2000zn}.

Integrated heavy-heavy-light-light correlation functions akin to \eqref{Integrated_corr_def} have been studied in \cite{Brown:2024tru,Brown:2025huy} for the case where the two heavy operators are given by general giant or dual giant gravitons while the light operators are two superconformal primary $\mathcal{O}_2$. In particular, \cite{Brown:2025huy} showed that at the leading order in the $1/N$ expansion, the strong 't Hooft coupling expansion of the giant graviton integrated correlators displays an intricate hierarchy of non-perturbative, exponentially suppressed corrections which depend very non-trivially on the giant graviton parameter $\alpha$ controlling both the scaling dimension $\Delta =\alpha N$ and, holographically, the geometry of the D3-brane solution.

From \eqref{eq:specN}, we see that the maximal-trace integrated correlator $\mathcal{C}_{\alpha N,N}(\tau)$ expanded at large-$N$ cannot possibly produce non-perturbative scales with non-trivial dependence on $\alpha$.  Although the dimension of $\Omax{\alpha N}$ is indeed $\Delta = \alpha N$, it appears clear from its definition \eqref{eq:Omax} that this operator does not involve any single-trace states $T_p(x,Y)$ given in equation \eqref{eq:singletrace} with $p\propto N$. Giant graviton operators most definitely do contain such heavy single-trace operators, thus explaining a richer set of non-perturbative effects than what was observed in this section.

\subsection{Large $N$, large charge with $p = \alpha N^2$}
\label{sec:N2}

We now analyse the large charge expansion of the maximal-trace integrated correlator \eqref{Integrated_corr_def} in the limit $N\gg1$ with $p=\alpha N^2$ and $\alpha >0$. Since the maximal-trace operators are very heavy, having scaling dimension $\Delta = p \sim N^2$, the correlator is of the type Heavy-Heavy-Light-Light. In this context it is always convenient to think about the correlator as the two-point function of the light operators in the background created by the heavy operators. At leading order this background is described by a new saddle of the path-integral which we wish to explore quantitatively.

To proceed with our analysis, we must go back to the Cauchy integral representation \eqref{Cauchy_FrN} for the spectral overlap building block $F_{r,N}(s)$ and consider the case where the charge $2r= \alpha N^2$ with $\alpha$ fixed.
Firstly, we change integration variables to $y\to \frac{r}{y+r}$, leaving us with the expression
\begin{equation}\label{F_rN_int_rep_Nsq}
    F_{r,N}(s) = \frac{1}{s(1-s)}\oint_{|y|>r} \frac{(\frac{y}{r}+1)^r}{y}\left[1-\tFo{s}{1-s}{\tfrac{N^2-1}{2}}{-\tfrac{r}{y}}\right]\frac{\dd y}{2\pi i}\,.
\end{equation}
For the first term in parenthesis it is straightforward to evaluate the $y$-integral by picking up the residue at $y=0$. 
However, for the second term the story is more complicated.

We would now like to expand the integrand at large $r$. However, this task is far from straightforward since we note that the hypergeometric function does not have a manifest expansion in this regime due to the competition between the term $\frac{N^2-1}{2}$ as its third parameter and $-\frac{r}{y}$ as its argument, both of which are scaling as $N^2$. 
To get around this hurdle, we pass to the Mellin-Barnes representation of the hypergeometric function appearing at the integrand of \eqref{F_rN_int_rep_Nsq}, which upon substituting $r = \alpha N^2/2$ takes the form,
\begin{equation}
    \tFo{s}{1-s}{\tfrac{N^2-1}{2}}{-\tfrac{\alpha N^2}{2y}} = \frac{\Gamma(\frac{N^2-1}{2})}{\Gamma(s)\Gamma(1-s)}\int_{-i\infty}^{i\infty} \frac{\Gamma(s+t)\Gamma(1-s+t)\Gamma(-t)}{\Gamma(\frac{N^2-1}{2}+t)}\Big(\frac{\alpha N^2}{2y}\Big)^t\frac{\dd t}{2\pi i}\,.
\end{equation}
Plugging the above expression in \eqref{F_rN_int_rep_Nsq} and exchanging the $y$-integral  with the Mellin integral we arrive at
\begin{align}
  &\label{eq:Finter}   F_{\frac{\alpha N^2}{2},N}(s) =\\
  &\notag \frac{1}{s(1-s)}\Bigg[1-\int_{-i\infty}^{i\infty} \frac{\Gamma(s+t)\Gamma(1-s+t)\Gamma(-t)}{\Gamma(s)\Gamma(1-s)}\cdot \frac{\Gamma(\frac{N^2-1}{2})}{\Gamma(\frac{N^2-1}{2}+t)} \left(\frac{\alpha N^2}{2}\right)^t  \left(\int_{\gamma}\frac{(1+\frac{2y}{\alpha N^2})^\frac{\alpha N^2}{2}}{y^{t+1}}\frac{\dd y}{2\pi i}\right)\frac{\dd t}{2\pi i}\Bigg]\,.
\end{align}
Note that in exchanging the two integrals, we must deform the $y$-integration contour to exactly the same integration contour $\gamma$ previously discussed, that is $\gamma$ comes from $y\to -\infty - i \,0^+$, circles counter-clockwise the branch point at $y=0$ and then goes back to $y\to-\infty +i \,0 ^{+}$.

We are now in a position where we can compute the large-$N$ limit in the double-scaling regime.
Firstly, the integral over $y$ can be performed by expanding its integrand at large $N$, 
\begin{equation}
\frac{(1+\frac{2y}{\alpha N^2})^\frac{\alpha N^2}{2}}{y^{t+1}} = e^{y} y^{-t-1}\left[1-\frac{y^2}{\alpha N^2} +\frac{y^3(3y+8)}{6 \alpha^2 N^4}  +O(N^{-6})\right]\,,
\end{equation}
and then evaluating the contour integral using the identity \eqref{Gamma_fn_ident} for the reciprocal of the gamma function.
Secondly, the ratio of $N$-dependent gamma functions appearing in \eqref{eq:Finter} admits the large-$N$ perturbative expansion
\begin{equation}
    \frac{\Gamma(\frac{N^2-1}{2})}{\Gamma(\frac{N^2-1}{2}+t)}\frac{N^{2t}}{2^t} = 1+\frac{(2-t)t}{N^2}+O(N^{-4})\,.
\end{equation}

When the dust settles, we are left with an expansion in inverse even powers of $N$ with coefficients given by simple Mellin-Barnes integrals over the variable $t$ which can be readily evaluated order by order in $1/N$ thus yielding the asymptotic genus expansion
\begin{equation}
    F_{\frac{\alpha N^2}{2},N}(s) = \sum_{g=0}^\infty N^{-2g}\mathcal{G}^{(g)}(s;\alpha)\,.\label{eq:FN2}
\end{equation}
The first few terms in the above expansion are given by
\begin{align}
    \mathcal{G}^{(0)}(s;\alpha) &\label{eq:G0}= \frac{1-\tFo{s}{1-s}{1}{-\alpha}}{s(1-s)}\,,\\
   \mathcal{G}^{(1)}(s;\alpha) &\label{eq:G1}=(2 \alpha +1) \, _2F_1(2-s,s+1;2;-\alpha ) -  (\alpha +1) \, _2F_1(2-s,s+1;1;-\alpha )\,.
\end{align}
Note that the genus-$g$ spectral overlap satisfies the reflection symmetry $\mathcal{G}^{(g)}(s;\alpha)=\mathcal{G}^{(g)}(1-s;\alpha)$.

Lastly, we combine \eqref{eq:FN2} with the analogous expression \eqref{large_N_series} for the building block $M_N(s)$
as to derive  the expansion for the complete spectral overlap  $c_{2r,N}(s)= M_N(s)F_{r,N}(s)$
in the double-scaling limit $N\gg1$ with $\alpha = 2r/N^2$ fixed
\begin{equation}\label{eq:cDouble}
    c_{\alpha N^2,N}(s) = \sum_{g=0}^\infty N^{2-2g}\big[ N^{-s}\mathcal{A}^{(g)}(s;\alpha)+N^{s-1}\mathcal{A}^{(g)}(1-s;\alpha)\big]\,,
\end{equation}
where the spectral overlap $\mathcal{A}^{(g)}(s;\alpha)$ is given by the linear combination
\begin{equation}
\mathcal{A}^{(g)}(s;\alpha) \label{eq:Ag} = \sum_{g_1=\,0}^g \mathcal{M}^{(g_1)}(s)\,\mathcal{G}^{(g-g_1)}(s;\alpha)\,.
\end{equation}
In particular, the first two terms reduce to the simple linear combinations,
\begin{align}
    \mathcal{A}^{(0)}(s;\alpha) &\label{eq:A0}= \mathcal{M}^{(0)}(s)\mathcal{G}^{(0)}(s;\alpha) \,,\\
    \mathcal{A}^{(1)}(s;\alpha) &\label{eq:A1}= \mathcal{M}^{(0)}(s)\mathcal{G}^{(1)}(s;\alpha)+\mathcal{M}^{(1)}(s)\mathcal{G}^{(0)}(s;\alpha)\,,
\end{align}
which can be made fully explicit using \eqref{eq:M0}-\eqref{eq:M1} and \eqref{eq:G0}-\eqref{eq:G1}.
For example, for the leading correction, which will be analysed shortly, we find
\begin{equation}
 \mathcal{A}^{(0)}(s;\alpha)\label{eq:A0spec}=  \frac{ (2 s-1)  \Gamma (s-2) \Gamma (s-1)}{2^{2s}\sqrt{\pi } \Gamma \left(s-\frac{1}{2}\right)}\tan (\pi  s)[ \, _2F_1(s,1-s;1;-\alpha )-1]\,.
\end{equation}

From the double-scaling expansion \eqref{eq:cDouble} for the spectral overlap it is immediate to write the large-$N$ asymptotic expansion of the integrated correlator,
\begin{equation}\label{eq:CN2}
    \mathcal{C}_{\alpha N^2,N}(\tau) =\average{\mathcal{C}_{\alpha N^2,N}} +\sum_{g=0}^\infty N^{2-2g} \,2\intRes \mathcal{A}^{(g)}(s;\alpha)N^{-s}\eisen{s}\frac{\dd s}{2\pi i}\,.
\end{equation}
We rewrite the above expression by moving the $s$-contour of integration to $s=1+\epsilon$ with $0<\epsilon<1/2$ at the price of picking up the residue from $\eisen{s}$ at $s=1$, thus yielding
\begin{equation}
    \mathcal{C}_{\alpha N^2,N}(\tau) =\average{\mathcal{C}_{\alpha N^2,N}}-\sum_{g=0}^\infty \frac{\mathcal{A}^{(g)}(1;\alpha)}{N^{2g-1}}+\sum_{g=0}^\infty N^{2-2g}\mathcal{C}^{(g)}_\alpha(N;\tau)\,,\label{eq:Calpha}
\end{equation}
where we separated the $\tau$-independent part from the non-trivial genus-$g$ modular invariant contributions
\begin{equation}\label{eq:Cg}
    \mathcal{C}^{(g)}_{\alpha}(N;\tau) \coloneqq 2\intRepsilon \mathcal{A}^{(g)}(s;\alpha)N^{-s}\eisen{s}\frac{\dd s}{2\pi i}\,.
\end{equation}

Before analysing the properties of \eqref{eq:Cg}, it is important to discuss the $\tau$-independent part of the large-$N$ expansion given by the first two terms in \eqref{eq:Calpha}. Firstly, we consider $\average{\mathcal{C}_{\alpha N^2,N}}$ which is simply the spectral overlap of the integrated correlator with the constant function and whose large-$N$ expansion can be determined from \eqref{MN_sto1_limit}-\eqref{FrN_sto1_limit},
\begin{align}
    \average{\mathcal{C}_{\alpha N^2,N}} &\label{eq:Cavg1}=\lim_{s\to 1} \big[ \mathcal{M}_N(s) F_{\frac{\alpha N^2}{2},N}(s) \big]=\frac{N(N-1)}{4}\left[\psi\left( \tfrac{N^2(\alpha+1)-1}{2}\right)-\psi\left(\tfrac{N^2-1}{2}\right)\right]\\
    &\notag \sim \frac{N(N-1)}{4}\log(1+\alpha)+\frac{(N-1)}{2N}\frac{\alpha}{(1+\alpha)}+O(N^{-2})\,.
\end{align}
We notice that the above perturbative expansion does contain both even and odd powers of $N^{-1}$.

However, the second $\tau$-independent term of \eqref{eq:Calpha} cancels precisely these unwanted powers,
\begin{align}
\sum_{g=0}^\infty \frac{\mathcal{A}^{(g)}(1;\alpha)}{N^{2g-1}} &\notag = \mathcal{M}^{(0)}(1) \sum_{g=0}^\infty \frac{\mathcal{G}^{(g)}(1;\alpha)}{N^{2g-1}} = N \mathcal{M}^{(0)}(1) F_{\frac{\alpha N^2}{2},N}(1)\\
&\label{eq:Cavg2}=-\frac{N}{4} \left[\psi\left( \tfrac{N^2(\alpha+1)-1}{2}\right)-\psi\left(\tfrac{N^2-1}{2}\right)\right]\,,
\end{align}
where in the first line we used the expansions \eqref{eq:Ag}-\eqref{eq:FN2}, combined with the fact that $\mathcal{M}^{(g)}(1) = 0$ for all $g>0$ as seen from \eqref{eq:M0}-\eqref{eq:M1}. The final expression is then derived by direct evaluation of \eqref{eq:M0} and \eqref{FrN_sto1_limit}.
Combining the two $\tau$-independent terms \eqref{eq:Cavg1}-\eqref{eq:Cavg2} we obtain
\begin{align}
    \average{\mathcal{C}_{\alpha N^2,N}}-\sum_{g=0}^\infty \frac{\mathcal{A}^{(g)}(1;\alpha)}{N^{2g-1}}&\notag=\frac{N^2}{4}\left[\psi\left( \tfrac{N^2(\alpha+1)-1}{2}\right)-\psi\left(\tfrac{N^2-1}{2}\right)\right]\\
    &\sim \frac{N^2}{4}\log(\alpha+1) + \frac{\alpha}{2(\alpha+1)}+O(N^{-2})\,,\label{eq:largeNSugra}
\end{align}
which is an expansion in inverse powers of $N^2$ as expected from holography.
We will come back to these $\tau$-independent terms in Section \ref{sec:LLM} since they directly relate to the supergravity approximation of the integrated correlator.

We now focus our attention towards the more interesting $\tau$-dependent part \eqref{eq:Cg}. 
 We proceed as in the previous sections and evaluate its large-$N$ perturbative expansion by closing the contour of integration in~\eqref{eq:Cg} towards the right half-plane ${\rm Re}(s)\to \infty$ and collecting residues from the poles of the spectral overlap located at $s=k+3/2$ with $k\in \mathbb{N}$,~see e.g.~\eqref{eq:A0spec}.
 This process produces a formal asymptotic series in $N^{-1}$ whose coefficients are half-integral non-holomorphic Eisenstein series. For the genus-zero contribution this asymptotic expansion is explicitly given by
 \allowdisplaybreaks{
\begin{align}
    &\notag \mathcal{C}^{(0)}_{\alpha}(N;\tau)\sim \mathcal{C}^{(0)}_{\alpha}(N;\tau)_P =\\*
    &\label{eq:C0P}\sum_{k=0}^\infty \frac{(k+1) \Gamma(k-\frac{1}{2})\Gamma(k+\frac{1}{2})}{{2^{2k+1}}\pi^{\frac{3}{2}}\Gamma(k+1)} \big[\tFo{k+\tfrac{3}{2}}{-k-\tfrac{1}{2}}{1}{-\alpha}-1\big]\, N^{-k-\frac{3}{2}}\, \eisen{k+\tfrac{3}{2}}\,.
\end{align}}

The particular example above is indicative of the fact that the large-$N$ perturbative expansion of $\mathcal{C}^{(g)}_{\alpha}(N;\tau)$ is once again an asymptotic series in $N^{-\frac{1}{2}}$ with factorially divergent, modular invariant coefficients.
Proceeding as discussed in Section \ref{sec:ModRes} we therefore use our definition \eqref{eq:BorelDef} to first compute the Borel transform of \eqref{eq:C0P},
\begin{equation}\label{eq:B0sum}
    \mathcal{B}[\mathcal{C}_{\alpha,P}^{(0)}](t) = \sum_{k=0}^\infty \frac{{4}\Gamma(k-\frac{1}{2})\Gamma(k+\frac{1}{2})}{\pi \Gamma(k+1)^2}\Big[\tFo{k+\tfrac{3}{2}}{-k-\tfrac{1}{2}}{1}{-\alpha}-1\Big]t^{2k+2}\,.
\end{equation}
Since the summand contains a hypergeometric function, evaluating the series is not a straightforward task. As a result, we cannot directly infer the analytic properties of the Borel transform. The singularity structures are especially important, since they determine the form of the non-perturbative effects: in particular, the positions of the branch-cut singularities set the scale of the non-perturbative contributions we aim to identify.

In Appendix \ref{sec:NP_appendix_p_Nsq} we exploit an integral representation for the hypergeometric function to derive the complete analytic structure of the Borel transform $\mathcal{B}[\mathcal{C}_{\alpha,P}^{(0)}](t)$ in the complex-$t$ plane as a function of $\alpha>0$.
Since the details are rather technical, we discuss here only the physical consequences of the singularity structure of \eqref{eq:B0sum}.

Firstly, it is immediate to resum the part of \eqref{eq:B0sum} which does not involve a hypergeometric function, yielding
\begin{equation}
\sum_{k=0}^\infty \frac{{4}\Gamma(k-\frac{1}{2})\Gamma(k+\frac{1}{2})}{\pi \Gamma(k+1)^2}t^{2k+2} = -\frac{{16} t^2 E\left(t^2\right)}{\pi }\,,
\end{equation}
with $E(z)$ denoting a complete elliptic integral function.
From the well-known analytic properties of $E(z)$, we deduce that for ${\rm Re}(t)>0$ the Borel transform \eqref{eq:B0sum} has a logarithmic branch-cut singularity for $t\in  ( L_{{\rm I}},+\infty)$, where the first singular point is located at
\begin{equation}
    L_{{\rm I}} =1\,.\label{eq:tI}
\end{equation}

 A more involved analysis presented in appendix \ref{sec:NP_appendix_p_Nsq} shows that, upon summing over $k$, the part of the series \eqref{eq:B0sum} proportional to the hypergeometric function produces an analytic function on the Borel plane, $t$, which possesses two more branch-cuts originating from
\begin{align}
    L_{-}
   &\label{eq:tII} =\sqrt{\alpha+1}-\sqrt{\alpha}\,,\\
    L_{+} &\label{eq:tIII}=\frac{1}{ L_-} = \sqrt{\alpha+1}+\sqrt{\alpha}\,,
\end{align}
and again extending to $t\to+\infty$. Although these are not important for the discussion of the exponentially suppressed terms of physical interest, we also note that \eqref{eq:B0sum} has branch-cut singularities on the negative half-line $t<0$ originating respectively at $-L_{{\rm I}},-L_{-}$ and $-L_{+}$ and extending towards $t\to - \infty$.

In Section \ref{Mod_inv_resurgence_section} we have shown that generally when the Borel transform exhibits discontinuities along the positive real $t$-axis, this has significant implications for the resulting non-perturbative effects.
Such non-perturbative effects are related to the discontinuity of the directional Borel resummation
\begin{equation}
    \mathcal{S}_\theta[\mathcal{C}_{\alpha,P}^{(0)}](N;\tau) = \int_0^{e^{i\theta}\infty}\mathcal{E}(\sqrt{N}t;\tau)\mathcal{B}[\mathcal{C}_{\alpha,P}^{(0)}](t)\dd t\,,
\end{equation}
when the contour of integration $\theta$ crosses the singular Stokes direction $\arg(t)=0$. 
That is, the quantity $\mathcal{S}_{0^+}[\mathcal{C}_{\alpha,P}^{(0)}](N;\tau)-\mathcal{S}_{0^+}[\mathcal{C}_{\alpha,P}^{(0)}](N;\tau) \neq 0$ contains all exponentially suppressed terms in the wedge of the complex $N$-plane ${\rm Re}(N)>0$. 
Crucially, the difference between the two lateral resummations above and below the Stokes ray ${\rm arg}(t)=0$ is fully captured by the analytic structure of the Borel transform. 

Given the above discussion, we then expect three different types of non-perturbative corrections, each one of them associated to a different branch-cut singularity for $t\in [L_i,\infty)$ with $L_i\in\{L_{\rm I},L_{+},L_{-}\}$.
Relegating the technical details to Appendix \ref{sec:NP_appendix_p_Nsq}, we have that the non-perturbative terms can be divided as
\begin{align}
     \mathcal{C}_{\alpha,N\!P}^{(0)}(N;\tau)&\label{NP_terms_pNsq}=\mathcal{C}_{\alpha,N\!P}^{(0),{\rm I}}(N;\tau)+\mathcal{C}_{\alpha,N\!P}^{(0),{-}}(N;\tau)+\mathcal{C}_{\alpha,N\!P}^{(0),{+}}(N;\tau)\,.
\end{align}
Each of the above non-perturbative sectors contains the contribution from the discontinuity of the Borel transform from the corresponding branch-cut originating respectively at $L_{\rm I},L_{\pm}$, taking the general structure
\begin{align}    
    \mathcal{C}^{(0),i}_{\alpha,N\!P}(N;\tau)&=\sum_{k=0}^\infty d^i_k (L^2_i N)^{-\frac{k+1}{2}}D_{L_i^2N}\Big(\tfrac{k+1}{2};\tau\Big)\,, \qquad \qquad \text{with}\quad i\in \{{\rm I},-,+\}\,,
\end{align}
for some perturbative, factorially divergent coefficients $d^i_k$.

In the large $N$ limit the three non-perturbative terms in \eqref{NP_terms_pNsq} give rise to very different non-perturbative scales schematically of the form
\begin{align}
\mathcal{C}_{\alpha,N\!P}^{(0),{\rm I}}(N;\tau) & \sim  \exp\left( - 4 \sqrt{\pi N} \frac{|n\tau+m|}{\sqrt{\tau_2}} \right) \,,\\
\mathcal{C}_{\alpha,N\!P}^{(0),{\pm}}(N;\tau) &\label{eq:NPscales} \sim  \exp\left( - 4 \sqrt{\pi N} (\sqrt{\alpha+1}\pm\sqrt{\alpha}) \frac{|n\tau+m|}{\sqrt{\tau_2}} \right) \,,
\end{align}
where the sum over $(n,m)\neq(0,0)$ is left implicit.
For later reference, we note that for $\alpha>0$ we have that the locations of the singularities on the positive real $t$-axis given in \eqref{eq:tI}-\eqref{eq:tII}-\eqref{eq:tIII} are ordered as $0<L_-<L_{{\rm I}}=1<L_{+}$. 
Therefore at large $N$ the non-perturbative scales in \eqref{eq:NPscales} are ordered in magnitude as
\begin{equation}\label{eq:ScalesOrd}
e^{- 4  (\sqrt{\alpha+1}+\sqrt{\alpha}) |n\tau+m| \sqrt{\frac{\pi N}{\tau_2} } } \ll e^{ - 4   |n\tau+m| \sqrt{\frac{\pi N}{\tau_2} }  } \ll e^{- 4  (\sqrt{\alpha+1}-\sqrt{\alpha}) |n\tau+m| \sqrt{\frac{\pi N}{\tau_2} }  }\,,
\end{equation}
that is the non-perturbative effects due to the discontinuity starting at $L_- = \sqrt{\alpha+1}-\sqrt{\alpha}$ yield the dominant corrections.

For future reference, it is useful to analyse the non-perturbative effects \eqref{eq:NPscales}
in the strong-coupling 't Hooft limit where $N\to \infty$ with $\lambda = N\gym^2= 4\pi N/\tau_2 \gg 1$.
In this limit, the dominant contributions to the modular invariant non-perturbative effects \eqref{eq:NPscales} originate from the lattice points $(n,m)=(0,m)$ with $m\neq0$. Hence in the large 't Hooft coupling limit we rewrite \eqref{eq:ScalesOrd} as
\begin{equation}
e^{-2 |m| (\sqrt{\alpha+1}+\sqrt{\alpha}) \sqrt{\lambda}} \ll e^{-2 |m| \sqrt{\lambda}}  \ll e^{-2 |m| (\sqrt{\alpha+1}-\sqrt{\alpha}) \sqrt{\lambda}}\,.
\end{equation}

We note that the presence of the dominant non-perturbative scale $e^{-2 |m| (\sqrt{\alpha+1}-\sqrt{\alpha}) \sqrt{\lambda}}$ had already been argued for in \cite{Paul:2023rka} starting from considerations on the radius of convergence of the weak 't Hooft coupling expansion $\lambda \to 0$.
Here we have explicitly constructed the ${\rm SL}(2,\mathbb{Z})$ completion of such non-perturbative effects, and have found a new scale of non-perturbative corrections originating from the singularity at $L_+$.
We will comment on the possible holographic interpretation of such non-perturbative scales in Section \ref{sec:LLM}.

\subsection{Large $N$, large charge with $p = \alpha N^\gamma$ and $\gamma>2$}
\label{sec:gamma3}

Lastly we consider the maximal-trace integrated correlator in the regime where $N\gg1 $ and $p=\alpha N^\gamma$ with $\gamma>2$ and $\alpha>0$ fixed.  To this end, we observe that the problem has been solved almost entirely in \cite{Paul:2023rka} where the authors considered the limit $p\to \infty$ with $N$ fixed, with some of these results presented and extended here in Section \ref{sec:Large_charge_fixed_N}.
The only difference between these previous discussions and the present case is that now we are also sending $N\to\infty$. 

Hence we must combine \eqref{Large_charge_small_N_series} with the asymptotic expansion of $M_N(s)$ computed in \eqref{large_N_series}.
This yields the expression
\allowdisplaybreaks{
\begin{align}
   & \notag c_{2r,N}(s)=\frac{1}{s(1-s)}\sum_{g=0}^\infty N^{2-2g-s}\mathcal{M}^{(g)}(s)\\*
&+\sum_{g=0}^\infty\sum_{\ell=0}^\infty N^{2-2g}r^{-\ell}\left[ (Nr)^{-s}\mathcal{M}^{(g)}(s)\mathcal{F}^{(\ell)}_N(s)    +N^{s-1}r^{-s}\mathcal{M}^{(g)}(1-s)\mathcal{F}_{N}^{(\ell)}(s)\right]+(s\leftrightarrow 1-s)\,.
\end{align}}
Although the charge $2r = \alpha N^\gamma$ is the largest parameter in the problem, we are still sending $N\to \infty$, hence each of the functions $\mathcal{F}_N^{(\ell)}(s)$ has also to be expanded at large $N$,
\begin{equation}
    \mathcal{F}_N^{(\ell )}(s) = \sum_{k=0}^\infty N^{2
    \ell +2s-2k}\mathcal{F}^{(\ell,k)}(s)\,,
\end{equation}
where the first few terms given by\allowdisplaybreaks{
\begin{align}
    \mathcal{F}^{(0,0)}(s) &\label{eq:F00}= \frac{8^{-s}\Gamma(s-1)}{\sqrt{\pi}s\,\Gamma(s+\frac{1}{2})}\tan(\pi s) \\*
    \mathcal{F}^{(0,1)}(s)&\notag = -s(s+2) \mathcal{F}^{(0,0)}(s)\,,\qquad \mathcal{F}^{(0,1)}(s) =\frac{s}{4} \mathcal{F}^{(0,0)}(s)\,,\qquad \mathcal{F}^{(1,1)}(s) = -\frac{s(s+1)^2}{4} \mathcal{F}^{(0,0)}(s)\,.
\end{align}}

Finally, remembering that we are considering the limit where both the charge $2r$ and $N$ tend to infinity while keeping $\alpha = 2r/N^\gamma$ fixed,  we derive the asymptotic series,
\allowdisplaybreaks{
\begin{align}
  &  \label{LargeGammaEqnExpansion}  c_{2r,N}(s)=\frac{1}{s(1-s)}\sum_{g=0}^\infty N^{2-2g-s}\mathcal{M}^{(g)}(s)\\*
  &\notag +\sum_{g,\ell,k=0}^\infty N^{2-2g-2k-\ell(\gamma-2)}\Big(\frac{\alpha}{2}\Big)^{-\ell}\Big[\Big(\frac{\alpha N^{\gamma-1}}{2}\Big)^{-s}\mathcal{M}^{(g)}(s)\mathcal{F}^{(\ell,k)}(s)
    +\frac{1}{N}\Big(\frac{\alpha N^{\gamma-3}}{2}\Big)^{-s}\mathcal{M}^{(g)}(1-s)\mathcal{F}^{(\ell,k)}(s)\Big]\\*
    &\notag +(s\leftrightarrow 1-s)\,.
\end{align}}
The first term in the above expression contains only the functions $\mathcal{M}^{(g)}(s)$, thus producing contributions which are almost identical to the expansion of $\mathcal{C}_{2,N}(\tau)$ in the large-$N$ limit. The only difference is the very simple rational function of $s$ multiplying this term but this does not result in any fundamental difference neither at the perturbative nor non-perturbative level.

Much more interesting are the terms in parenthesis on the second line.
Here we note that the prefactor $N^{2-2g-2k-\ell(\gamma-2)}$ becomes more and more suppressed for increasing values of $g,k,\ell$ since we are precisely considering the regime $\gamma>2$. Similarly the first term inside the square bracket is multiplied by the factor $\Big(\frac{\alpha N^{\gamma-1}}{2}\Big)^{-s}$ which will result in a suppressing factor once we analyse the spectral decomposition contour integral \eqref{overlap_equation_def} in the right half-plane ${\rm Re}(s)>1$ as previously done. On the other hand, the behaviour of the second term in parenthesis is rather peculiar since it is suppressed  in the right half-plane ${\rm Re}(s)>1$ only for $\gamma >3$. However, this is not a problem since for $\gamma<3$ we simply need to consider its $s$-reflected term proportional to $N^{(1-s)(3-\gamma)}\mathcal{M}^{(g)}(s)\mathcal{F}^{(\ell,k)}(1-s)$, which is suppressed precisely when $\gamma<3$ and ${\rm Re}(s)>1$.

 Since the asymptotic expansion produced by the first term in \eqref{LargeGammaEqnExpansion} is essentially that of the large-$N$ expansion of $\mathcal{C}_{2,N}(\tau)$ we shall ignore it and hence focus our attention towards the novel and more interesting effects originating from the second line of \eqref{LargeGammaEqnExpansion}.
 Considering the first term in parenthesis for the spectral overlap \eqref{LargeGammaEqnExpansion} we are led to define
 \begin{equation}
    \mathcal{C}_1^{(g,\ell,k)}(N,\alpha,\gamma;\tau) = \intRepsilon 2^{3s}\mathcal{M}^{(g)}(s)\mathcal{F}^{(\ell,k)}(s)\Lambda_1^{-s}\eisen{s} \frac{\dd s}{2\pi i}\,,
\end{equation}
where the auxiliary large parameter $\Lambda$ is given by 
\begin{equation}
\Lambda_1 \coloneqq 4\alpha N^{\gamma-1} \gg 1\,,
\end{equation}
and it will set the scale of the non-perturbative effects.

By combining \eqref{eq:M0} with \eqref{eq:F00}, we find the leading contribution,
\begin{equation}\label{eq:C000}
    \mathcal{C}_1^{(0,0,0)}(\Lambda_1;\tau) = \intRepsilon \frac{2^{1-2s}\Gamma(s-2)\Gamma(s-1)\Gamma(s)\tan(\pi s)^2}{\pi \Gamma(s-\frac{1}{2})^2} \Lambda_1^{-s}\eisen{s}\frac{\dd s}{2\pi i}\,.
\end{equation}
We note that 
 the integrand in the above expression still has poles at $s=k+\frac{3}{2}$ with $k\in \mathbb{N}$, but these are now double poles. 
Following a slightly modified version of the resurgence analysis approach described in Section \ref{Mod_inv_resurgence_section} we can obtain the non-perturbative, exponentially suppressed correction from the perturbative expansion of \eqref{eq:C000}.
Since this discussion differs from the cases previously discussed only from a technical point of view, we prefer to present here just the results and refer to Appendix \ref{sec:appNP2} for the precise analysis of the non-perturbative corrections to \eqref{eq:C000}.

We find that the non-perturbative corrections to the large-$N$, or equivalently large-$\Lambda_1$ perturbative expansion of \eqref{eq:C000} are now set by the new non-perturbative scale $\Lambda_1$, i.e. we have
\begin{equation}
\label{eq:NPC1}
 \tilde{\mathcal{C}}_{1,{ N\!P}}^{(0,0,0)}(\Lambda_1;\tau) \sim  \exp\left( - 4 \sqrt{ \Lambda_1 Y_{mn}(\tau)} \right) = \exp\left( - 8 N^{\frac{\gamma-1}{2}} \sqrt{\alpha Y_{mn}(\tau)} \right) \,,
\end{equation}
where again $Y_{mn}(\tau) = \pi|n\tau+m|^2 /\tau_2$ with $(m,n)\in \mathbb{Z}^2 \setminus\{(0,0)\}$.
Although here we presented for concreteness only the leading contribution, we stress that similar effects are found within the sub-leading corrections as well, i.e. for higher values of $(g,\ell,k)$. 

We now move to discuss the second term in parenthesis for the spectral overlap \eqref{LargeGammaEqnExpansion}.
However, to analyse this second contribution we must separately consider the cases $2<\gamma<3$ and $\gamma>3$, given the arguments presented above. 
Let us start with the case $\gamma>3$ for which we see that the expansion \eqref{LargeGammaEqnExpansion} gives rise to a good 
power series in $1/N$ when ${\rm Re}(s)>1$. 

We then wish to discuss 
 \begin{equation}
    \mathcal{C}_2^{(g,\ell,k)}(N,\alpha,\gamma;\tau) = \intRepsilon \mathcal{M}^{(g)}(1-s)\mathcal{F}^{(\ell,k)}(s) \left( \frac{\alpha N^{\gamma-3}}{2}\right)^{-s} \eisen{s} \frac{\dd s}{2\pi i}\,.
\end{equation}
By combining \eqref{eq:M0} with \eqref{eq:F00}, we find for the leading contribution $(g,\ell,k)=(0,0,0)$,
\begin{equation}
 \mathcal{C}_2^{(0,0,0)}(N,\alpha,\gamma;\tau) = \intRepsilon  \frac{(1-2 s) \sec (\pi  s)}{4 s \Gamma (s+2)}  \left( \frac{\alpha N^{\gamma-3}}{2}\right)^{-s} \eisen{s} \frac{\dd s}{2\pi i}\,.\label{eq:C0002}
\end{equation}
Using the integral representation \eqref{Eisen_series_definitions} for the Eisenstein series we can evaluate exactly the above integral to discover that the large-$N$ expansion \eqref{eq:C0002} has finite radius of convergence and does not contain any non-perturbative effect.
Alternatively, one can easily check that for ${\rm Re}(s)\to+\infty$ the integrand of \eqref{eq:C0002} vanishes exponentially fast.
A similar behaviour is also found for the sub-leading corrections, i.e. for higher values of $(g,\ell,k)$.

Surprisingly, the same analysis for the case $2<\gamma<3$ reveals a rather different story.
As previously discussed, when $2<\gamma<3$ we should not consider the second term in parenthesis for the spectral overlap \eqref{LargeGammaEqnExpansion} but rather use its reflected $s\to 1-s$ counterpart.
For $2<\gamma<3$ we then define
 \begin{equation}
    \tilde{\mathcal{C}}_2^{(g,\ell,k)}(N,\alpha,\gamma;\tau) = \intRepsilon 2^{-3s}\mathcal{M}^{(g)}(s)\mathcal{F}^{(\ell,k)}(1-s) \Lambda_2^{-s} \eisen{s} \frac{\dd s}{2\pi i}\,,
\end{equation}
where we defined a second auxiliary large parameter 
\begin{equation}
\Lambda_2 \coloneqq \frac{ N^{3-\gamma}}{4\alpha} \gg 1 \,,
\end{equation}
which will set the scale of new non-perturbative effects.

We use again \eqref{eq:M0} and \eqref{eq:F00} to derive the leading contribution,
\begin{equation}\label{eq:C0003}
    \tilde{\mathcal{C}}_2^{(0,0,0)}(\Lambda_2;\tau) = \intRepsilon \frac{2^{-2 s-3} (2 s-1) \Gamma (s-2) \tan (\pi  s) }{\pi  (s-1)} \Lambda_2^{-s}\eisen{s}\frac{\dd s}{2\pi i}\,.
\end{equation}
The integrand once more has simple poles at $s=k+\frac{3}{2}$ with $k\in \mathbb{N}$, hence we can directly apply the resurgence analysis approach described in Section \ref{Mod_inv_resurgence_section} to first obtain the large-$N$, or equivalently large-$\Lambda_2$, perturbative expansion and from this derive the non-perturbative effects.
Since the present analysis is pretty much identical to the cases previously discussed we simply state the results for the non-perturbative contributions to \eqref{eq:C0003} which are now set by the new non-perturbative scale $\Lambda_2$.
In the regime where $2<\gamma<3$ we find a new type of non-perturbative, exponentially suppressed corrections of the form
\begin{equation}\label{eq:NPC2}
 \tilde{\mathcal{C}}_{2,{ N\! P}}^{(0,0,0)}(\Lambda_2;\tau) \sim  \exp\left( - 4 \sqrt{ \Lambda_2 Y_{mn}(\tau)} \right) = \exp\left( - 2 N^{\frac{3-\gamma}{2}}\sqrt{\frac{Y_{mn}(\tau)}{\alpha} } \right) \,,
\end{equation}
where again $Y_{mn}(\tau) = \pi|n\tau+m|^2 /\tau_2$ with $(m,n)\in \mathbb{Z}^2 \setminus\{(0,0)\}$.

Currently we are not aware of any physical interpretation for the novel non-perturbative 
effects \eqref{eq:NPC1}-\eqref{eq:NPC2}, 
either in the field theory or in the dual holographic setting.
The reason is that we analysed the maximal-trace integrated correlator in the 
regime where $N\gg1 $ and $p=\alpha N^\gamma$ with $\gamma>2$ and $\alpha>0$ fixed, 
in which case the dimension of the corresponding maximal-trace operators is $\Delta \propto N^\gamma$ with $\gamma>2$ 
and thus parametrically larger than the central charge. 
The central charge is usually the largest parameter in holography,
therefore we do not know what the physics of this limit is (if any at all) from the dual string theory side.
Nevertheless, it would be interesting to understand if holography can still be applied in this regime 
and if so, what is the semi-classical interpretation 
for the non-perturbative effects \eqref{eq:NPC1}-\eqref{eq:NPC2}  here discovered.

\section{Coherent-state correlators \& their holographic duals}
\label{sec:LLM}

In Section \ref{sec:coherent} we introduced a particular quantum mechanical coherent-state, constructed from multi-graviton states via the exponential series, 
\begin{equation}\label{repeatdefOH}
   | \alpha\rangle = {O}_H(0;\alpha) |0\rangle =  \sum_{n=0}^{\infty} \frac{1}{n!}\left(\frac{ {\alpha} }{\sqrt{2}}\right)^n \lim_{x\rightarrow 0} \Big[ O_L^n (0)|0\rangle \Big]\,,
\end{equation}
where the light operator $O_L(x)$  denotes the graviton operator $\mathcal{O}_2(x,Y_\bullet) $ for a particular choice of polarisation vector, see \eqref {eq:OL}.

From the definition of the coherent-state operator 
we then rewrote the four-point correlator 
\begin{equation}\label{sec:LLM_HHLL}
\average{O_L(x_1) \bar{O}_L(x_2) O_H(0;\alpha) \bar{{O}}_H(\infty;\alpha)} = {\langle{\alpha}|O_L(x_1)\bar{O}_L(x_2)|{\alpha}\rangle}
\end{equation}
as the series expansion  \eqref{eq:HHLL}
built out from the four-point correlators between the same light 
operators $O_L(x_1)\bar{O}_L(x_2)$ and the maximal-trace operators.
Considering the integrated version of this HHLL correlator defined in \eqref{Integrated_corr_HHLL_def},
we arrived at the result \eqref{eq:HHLLIntCorr} rewritten here for convenience,
\begin{equation}\label{eq:CHHLL}
    \mathcal{C}_{\HHLL}(N,{\alpha};\tau) = 4(1-2{\alpha}^2)^{\frac{N^2-1}{2}}\sum_{r=1}^\infty \frac{r^2}{(r!)^2}\Big(\frac{{\alpha}}{\sqrt{2}}\Big)^{2r}R_{2r}(N)\mathcal{C}_{2r,N}(\tau)\,,
\end{equation}
where $\mathcal{C}_{2r,N}(\tau)$ denotes the maximal-trace integrated correlator with  $R$-charge $2r$ defined in \eqref{Integrated_corr_def}, while $R_{2r}(N)$ is 
the coefficient of the corresponding two-point function given in \eqref{eq:2ptcoef}.

In this section we use the series representation \eqref{eq:CHHLL} to first evaluate 
$\mathcal{C}_{\HHLL}(N,{\alpha};\tau)$ exactly as a function of $N$, $\alpha$ and 
$\tau$ and then analyse its non-perturbative large-$N$ expansion at fixed coupling $\tau$.
With these results it becomes straightforward to compute the large 't Hooft coupling limit, i.e. 
$N\gg1$ with $\lambda = N \gym^2 \gg1$.  Along the way
we find an intriguing parallel between the results for the \HHLL~coherent-state integrated correlator and the similar 
discussion presented in Section \ref{sec:N2} for the 
maximal-trace integrated correlator in the double-scaling 
limit $N\gg1$ with $R$-charge $p = \alpha N^2$ and $\alpha>0$ fixed.

In Section \ref{sec:PertLLM} we 
show that the HHLL coherent-state integrated correlator, computed via supersymmetric 
localisation, in the supergravity regime $N\gg1$ and $\lambda\gg1$ agrees with 
the actual integration of the correlator derived independently in \cite{Aprile:2025hlt}.
However our analysis goes beyond the supergravity regime and poses strong constraints 
on a new AdS amplitude: the Virasoro-Shapiro amplitude 
dual to $\langle{\alpha}|O_L(x_1)\bar{O}_L(x_2)|{\alpha}\rangle$, which at present times
is almost entirely unknown.

In Section  \ref{sec:NPLLM} we compute the large-$N$ non-perturbative contributions 
to the \HHLL~integrated correlator \eqref{eq:CHHLL} and show that they 
admit a beautiful  interpretation in terms of giant magnon solutions 
in the underlying holographic geometry. 

\subsection{Spectral decomposition of the HHLL integrated correlator}
\label{sec:HHLL}

To derive the spectral decomposition of the coherent-state integrated correlator
$\mathcal{C}_{\HHLL}(N,\alpha;\tau)$, 
we start from that of the maximal-trace integrated correlators given in  \eqref{overlap_equation_def} 
and proceed with a similar analysis to the one carried out in Section \ref{sec:N2}.

By construction, $\mathcal{C}_{\HHLL}(N,\alpha;\tau)$ admits the 
${\rm SL}(2,\mathbb{Z})$ spectral decomposition,
\begin{equation}
    \mathcal{C}_{\HHLL}(N,\alpha;\tau) = \average{\mathcal{C}_{\HHLL}}+\intRes c_{\HHLL}(N,\alpha;s)\eisen{s}\frac{\dd s}{2\pi i}\,,\label{eq:CHHLLint}
\end{equation}
where the spectral overlap $c_{\HHLL}(\alpha,N;s)$ is expressed  thanks to \eqref{eq:CHHLL} 
as an infinite series over the maximal-trace spectral overlaps \eqref{overlap_even},
\begin{equation}\label{eq:cHHLLser}
    c_{\HHLL}(N,\alpha;s) = M_N(s)(1-2\alpha^2)^{\frac{N^2-1}{2}}\sum_{r=1}^\infty \frac{2^r\alpha^{2r}}{ r!}\Big(\frac{N^2-1}{2}\Big)_r F_{r,N}(s)\,.
\end{equation}
The function $c_{\HHLL}$ depends on the two building blocks $M_N$ and $F_{r,n}$, defined respectively in  \eqref{c2_spectral_overlap} 
and \eqref{F_p_formula}, that also determine the maximal-trace integrated correlators.

Borrowing results from the previous section, the expression \eqref{eq:cHHLLser} 
can be simplified further. Firstly, we specialise the formula \eqref{F_p_formula} for $F_{r,N}(s)$ to the case where $r\in \mathbb{N}$ as it appears in \eqref{eq:cHHLLser}, 
thus yielding the finite sum 
\begin{equation}
    F_{r,N}(s) = \frac{r!}{s(s-1)}\sum_{k=1}^r \frac{(s)_k(1-s)_k (-1)^k}{\big(\frac{N^2-1}{2}\big)_k (k!)^2 (r-k)!}\,.
\end{equation}
We substitute the above expression in \eqref{eq:cHHLLser} and exchange 
the two summmations over $k$ and $r$ to arrive at the compact expression
\begin{equation}
    c_{\HHLL}(N,\alpha;s)=\frac{M_N(s)}{s(1-s)}\left[1-\tFo{s}{1-s}{1}{\frac{2\alpha^2}{2\alpha^2-1}}\right]\,.\label{eq:cHHHLLspec}
\end{equation}
Quite remarkably, $c_{\HHLL}(N,\alpha;s)$ is fully factorised. 
In particular, the $N$ dependence of the spectral 
overlap \eqref{eq:cHHLLser} comes entirely from the building block $M_N(s)$.

The large-$N$ expansion of $M_N(s)$ has already been organised in \eqref{large_N_series} 
as a genus expansion in the quantities ${\cal M}^{(g)}(s)$, see e.g.~\eqref{eq:M0}-\eqref{eq:M1}. 
Hence, it is immediate to derive the large-$N$ expansion of the spectral overlap \eqref{eq:cHHHLLspec} 
which can then be substituted in the spectral integral \eqref{eq:CHHLLint} to arrive at
\begin{equation}\label{eq:CHHLLgenus}
    \mathcal{C}_{\HHLL}(N,\alpha;\tau) = \average{\mathcal{C}_{\HHLL}}+\log(1-2\alpha^2)\sum_{g=0}^\infty N^{1-2g}\mathcal{M}^{(g)}(1) +\sum_{g=0}^\infty N^{2-2g}\mathcal{C}_{\HHLL}^{(g)}(N,\alpha;\tau)\,.
\end{equation}
The function $\mathcal{C}_{\HHLL}^{(g)}(N,\alpha;\tau)$ is the
$\tau$-dependent genus-$g$ contribution of the HHLL integrated 
correlator, defined as follows,
\begin{equation}
    \mathcal{C}_{\HHLL}^{(g)}(N, \alpha;\tau) \coloneqq 2\intRepsilon \frac{\mathcal{M}^{(g)}(s)}{s(1-s)}\left[ 1-\tFo{s}{1-s}{1}{\tfrac{2\alpha^2}{2\alpha^2-1}}\right]N^{-s}\eisen{s}\frac{\dd s}{2\pi i}\,.\label{eq:Cgenus}
\end{equation}

The $\tau$-independent part of the \HHLL~integrated correlator is given by 
the first two terms in \eqref{eq:CHHLLgenus} and can be simplified further.
The first of such contributions is the spectral overlap with the constant function, 
$\average{\mathcal{C}_{\HHLL}}$. As discussed in \eqref{eq:Spec1}, this average can 
be obtained by considering the limit as $s\to 1$ of the spectral overlap $c_{\HHLL}(N,\alpha;s)$.
Given \eqref{eq:cHHHLLspec} we make use of \eqref{MN_sto1_limit} to derive 
\begin{equation}
    \average{\mathcal{C}_{\HHLL}} = \lim_{s\to 1}\left[c_{\HHLL}(N,\alpha;s)\right] = -\frac{N(N-1)}{4}\log(1-2\alpha^2)\,.\label{eq:Cavg}
\end{equation}
For the second $\tau$-independent factor of \eqref{eq:CHHLLgenus} we use the results 
derived in Section \ref{sec:limits}, see in particular \eqref{eq:M0}-\eqref{eq:M1}, to derive
\begin{equation}
\mathcal{M}^{(0)}(1) = -\frac{1}{4}\,,\qquad \mathcal{M}^{(g\geq 1)}(1) = 0\,.\label{eq:Ms1}
\end{equation}
Substituting \eqref{eq:Cavg} and \eqref{eq:Ms1} in \eqref{eq:CHHLLgenus} we obtain
\begin{equation}
    \mathcal{C}_{\HHLL}(N,\alpha;\tau) = -\frac{N^2}{4}\log(1-2\alpha^2)+\sum_{g=0}^\infty N^{2-2g}\mathcal{C}_{\HHLL}^{(g)}(N,\alpha;\tau)\,.\label{eq:CHHLLgenus2}
\end{equation}
The simplicity of this final expression is remarkable.

We now turn our attention to the $\tau$-dependent terms $\mathcal{C}_{\HHLL}^{(g)}(\alpha;\tau)$ 
in the genus expansion \eqref{eq:CHHLLgenus}. 
Focusing on the leading genus-zero contribution we use the expression for $\mathcal{M}^{(0)}(s)$ derived in \eqref{eq:M0} to arrive at,
\begin{align}
&\notag \mathcal{C}_{\HHLL}^{(0)}(N, \alpha;\tau) =\\*
 &\label{eq:C0exact}\intRepsilon \frac{(2s-1)\Gamma(s-2)\Gamma(s-1)}{2^{2s-1}\sqrt{\pi}\Gamma(s-\frac{1}{2})}\tan(\pi s)\left[\tFo{s}{1-s}{1}{\tfrac{2\alpha^2}{2\alpha^2-1}}-1\right] N^{-s}\eisen{s}\frac{\dd s}{2\pi i}\,.
\end{align}
Surprisingly, we find that modulo a redefinition of the parameters this genus-zero spectral representation 
\eqref{eq:C0exact} is identical to that of the maximal-trace integrated correlators, $\mathcal{C}^{(0)}_{\tilde{\alpha}}(N;\tau)$,
in the double scaling regime, $p =\tilde{\alpha} N^2$. The latter was given in  \eqref{eq:Cg},
and we repeat it here below for convenience, 
\begin{equation}\label{eq:repeatcalAspectral}
    \mathcal{C}^{(0)}_{\tilde \alpha}(N;\tau) \coloneqq 2\intRepsilon\mathcal{A}^{(0)}(s;\tilde \alpha)N^{-s}\eisen{s}\frac{\dd s}{2\pi i}\,,
\end{equation}
where the spectral overlap $\mathcal{A}^{(0)}(s;\tilde{\alpha})= \mathcal{M}^{(0)}(s)\mathcal{G}^{(0)}(s;\alpha)$ takes the explicit form \eqref{eq:A0spec}.
By comparing the two expressions \eqref{eq:C0exact} and \eqref{eq:repeatcalAspectral}, we deduce the identity
\begin{equation}\label{eq:C0HHLLId}
\mathcal{C}_{\HHLL}^{(0)}(N, \alpha;\tau)  = \mathcal{C}^{(0)}_{\tilde{\alpha}}(N;\tau)\quad\qquad {\rm with}\quad  \tilde{\alpha} = \frac{2\alpha^2}{1-2\alpha^2}\,.
\end{equation}
Furthemore, we also recognise that the $\tau$-independent 
logarithmic term  in \eqref{eq:CHHLLgenus2} is identical to that of the maximal-trace 
integrated correlators in \eqref{eq:largeNSugra}, under the same mapping of parameters, 
namely
\begin{equation}
-\frac{N^2}{4}\log(1-2\alpha^2) =  \frac{N^2}{4}\log(\tilde \alpha+1)  \qquad {\rm with}\quad  \tilde{\alpha} = \frac{2\alpha^2}{1-2\alpha^2}\,.
 \end{equation}
By including the logarithmic terms into the genus-zero contribution, we conclude that the full genus-zero contribution to
the  \HHLL~integrated correlator is mapped to the maximal-trace integrated correlator.
This mapping is also consistent with the physical regime of parameters, since the range 
in which the coherent state is well defined, i.e.~when the two-point function norm is finite and positive, $0 \leq \alpha < \frac{1}{\sqrt{2}}$, 
is mapped precisely to $0\leq \tilde{\alpha}<\infty$.

At this point, an obvious question to ask is whether we can 
understand the genus-zero relation between the coherent-state 
and maximal-trace correlators \eqref{eq:C0HHLLId} from a more 
fundamental quantum field theory point of view.
  
One approach to this question is to show that the same relation \eqref{eq:C0HHLLId} 
holds for the three-point functions that appear in the common OPE to both 
four-point correlators. We shall start by considering the exchange of the 
stress-tensor chiral primary half-BPS operator ${\cal O}_2$, or equivalently 
the single trace operator $T_2$ defined in  \eqref{eq:singletrace}, 
for which the three-point functions with  
$ \Omax{p}  \Omax{p} \sim [T_2\ldots T_2][T_2\ldots T_2]$ is 
protected - 
we are using the definition of $\Omax{p}$ in 
\eqref{eq:Omax} with $p=0\,{\rm mod}\,2$.
We will compare afterwards with a similar computation for 
$\langle {\cal O}_H{\cal O}_H {\cal O}_2\rangle$. 

In general, for $\langle \Omax{p}  \Omax{q}  \mathcal{O}_2\rangle$ 
with $ \Omax{p}  \Omax{q} \sim [T_2\ldots T_2][T_2\ldots T_2]$,
there are only two possible topologies depending on the 
number of $T_2$ in each multi-trace operator, i.e.~depending on 
whether $p=q$ or $|p-q|=2$, namely
\begin{equation}\label{topologies}
\begin{array}{ccc}
\begin{tikzpicture}[scale=1.3]  
\def\shift{.35}

\def\latoxuno{-.35}
\def\latoxdue{-.38+1.5}
\def\latoyuno{.7}
\def\latoydue{-.25}

\foreach \x in {0,.07}
\draw (\latoxuno-.02+\x,\latoyuno) -- (\latoxuno-.02+\x,\latoydue);

\foreach \x in {.14,.21,.28}
\draw[fill=black] (\latoxuno+.03+\x,\latoyuno/2+\latoydue/2) circle (.3pt);

\foreach \x in {.42,.49,.56}
\draw (\latoxuno+.02+\x,\latoyuno) --(\latoxuno+.02+\x,\latoydue);

\draw (\latoxuno-.02+.68,\latoyuno) -- (\latoxuno-.02+1.2, \latoyuno/2+\latoydue/2);

\draw (\latoxuno-.02+.68,\latoydue) -- (\latoxuno-.02+1.2, \latoyuno/2+\latoydue/2);

\draw  (\latoxuno+.25,\latoydue-.2) node[scale=.6] {$[T_2... T_2 T_2]$};
\draw  (\latoxuno+.25,\latoyuno+.2) node[scale=.6] {$[T_2... T_2 T_2]$};

\draw (\latoxuno-.02+1.35, \latoyuno/2+\latoydue/2) node[scale=.6] {$T_2$};
\end{tikzpicture} 
 & \rule{2cm}{0pt} &
\begin{tikzpicture}[scale=1.3]  
\def\shift{.35}

\def\latoxuno{-.35}
\def\latoxdue{-.38+1.5}
\def\latoyuno{.7}
\def\latoydue{-.25}

\foreach \x in {0,.07}
\draw (\latoxuno-.02+\x,\latoyuno) -- (\latoxuno-.02+\x,\latoydue);

\foreach \x in {.14,.21,.28}
\draw[fill=black] (\latoxuno+.03+\x,\latoyuno/2+\latoydue/2) circle (.3pt);

\foreach \x in {.42,.49}
\draw (\latoxuno+.02+\x,\latoyuno) --(\latoxuno+.02+\x,\latoydue);

\draw (\latoxuno-.02+.68-.01,\latoyuno+.01) -- (\latoxuno-.02+1.2, \latoyuno/2+\latoydue/2);

\draw (\latoxuno-.02+.68,\latoyuno+.08) -- (\latoxuno-.02+1.2, \latoyuno/2+\latoydue/2+.08);

\draw  (\latoxuno+.25,\latoydue-.2) node[scale=.6] {$[T_2\,\,...\, T_2]$};
\draw  (\latoxuno+.25,\latoyuno+.2) node[scale=.6] {$[T_2... T_2 T_2]$};

\draw (\latoxuno-.02+1.35, \latoyuno/2+\latoydue/2) node[scale=.6] {$T_2$};
\end{tikzpicture} 
\end{array}
\end{equation}
where a straight line denotes a Wick contraction 
between the adjoint scalars $\Phi^{I}(x)$ of $\mathcal{N}=4$ 
which form the microscopic constituents of $T_2$.
The topology on the left is \emph{diagonal} with respect to the 
maximal-trace operators, i.e.~we must have $p=q$, while the 
topology on the right is non-diagonal, $|p-q|=2$, 
and it is equivalent to a two-point function.

If we consider a maximal-trace operator with $p=\tilde\alpha (N^2-1)$, 
we can compute the diagonal topology presented with 
the left diagram in \eqref{topologies}, exactly as function of $N$, 
yielding the result 
\begin{equation}\label{seconda}
\frac{ \langle \Omax{p}(Y_{1},x_1) \Omax{p}(Y_{2},x_2) T_2(Y_3,x_3)\rangle}{ \langle \Omax{p}(Y_{1},x_1) \Omax{p}(Y_{2},x_2)\rangle } 
=    (N^2-1)   \tilde \alpha\, \frac{ d_{13} d_{23} }{d_{12}}\,,
\end{equation}
referring to \eqref{eq:singletrace} for the precise normalisation of the operator $T_2$. As it will shortly be useful, 
we rewrite here for convenience the two point function \eqref{eq:2ptcoef},
\begin{equation}\label{again_twopoint}
\langle \Omax{p}(Y_{1},x_1) \Omax{p}(Y_{2},x_2)\rangle=
 \frac{2^{p}(\frac{p}{2})!}{p^2}\Big(\frac{N^2-1}{2}\Big)_{\frac{p}{2}}\,d_{12}^p \,,
\end{equation}    
and remind the definition  $d_{ij}= Y_i \cdot Y_j /   |x_i-x_j|^2$.

Let us now consider the HHL three-point function $\langle {O}_H  O_H  \mathcal{O}_2\rangle$.  
From the definition of $O_H$ as a series expansion in multi-graviton states, 
see e.g~\eqref{repeatdefOH}, we expand both heavy operators and rewrite the HHL 
three-point function as a series in three-point functions $\langle [T_2\ldots T_2][T_2\ldots T_2]T_2\rangle$, 
for which both diagonal and non-diagonal topologies in \eqref{topologies} do contribute. 
Using the previous expressions \eqref{seconda} and \eqref{again_twopoint} we obtain
\begin{align}
& \frac{ \langle \expOO(Y_1,x_1) \expOO(Y_2,x_2) T_2(Y_3,x_3)\rangle}{\langle \expOO(Y_1,x_1) \expOO(Y_2,x_2) \rangle }\label{terza}=      
  {(N^2-1)}{}\left[\frac{2 \alpha^2 }{ (1{-} 2\alpha^2) }\cdot \frac{ d_{13} d_{23} }{ d_{12} }
+  \frac{1}{\sqrt{2}}\frac{ \alpha}{(1-2\alpha^2)}\cdot \frac{ d_{13}^2+d_{23}^2}{d_{12} }\right] \,.
\end{align}
We note that the above expression is an exact result for all $N$, consequence of 
\eqref{seconda} and \eqref{again_twopoint}. 

In order to compare the maximal-trace three-point function 
\eqref{seconda} against the HHL three-point function \eqref{terza}, we only need 
to consider a spacetime and $R$-symmetry configuration such that the contribution 
to  \eqref{terza} coming from the non-diagonal topology vanishes identically. 
This is achieved by requiring that $d_{13}^2+d_{23}^2=0$. Under this assumption, 
the three-point functions \eqref{seconda} and \eqref{terza} coincide provided 
we use the same map in parameters, $\tilde{\alpha} = \frac{2\alpha^2}{1-2\alpha^2}$, 
as presented before.

We now wish to reconsider the derivation above at large-$N$ to be able to generalise our results to non-protected three-point couplings. 
At large-$N$, the leading order term of the three-point function \eqref{seconda} 
has a straightforward derivation that only uses factorisation into two and three-point 
functions and  some simple combinatorial argument.\footnote{
The combinatorics work as follows. First we note that $\langle T_{2}^{n}  
T_{2}^{n} T_2\rangle = (n!)^2/(n-1)!  \langle T_2 T_2 T_2\rangle 
\langle T_2 T_2\rangle^{n-1} (1+O(N^{-2}))$ and
$\langle T^n_2 T^n_2 \rangle= n! \langle T_2 T_2\rangle^n(1+O(N^{-2}))$.
Then a simple calculation yields $\langle T_2 T_2 T_2\rangle= 8 N^2+O(N^0)$. 
The quoted result hence follows upon setting $n=\frac{p}{2 }= 
\frac{\tilde \alpha}{2}N^2 +O(N^0)$.} In this way, we obtain the result,
\begin{equation}\label{secondaseconda}
\frac{\langle \Omax{p} \Omax{p} T_2\rangle }{ \langle \Omax{p} \Omax{p}\rangle \,} = 
\frac{p}{2}\, \frac{ \langle T_2 T_2  T_2\rangle \,}{ \langle T_2 T_2 \rangle  }\Big(1+O(N^{-2})\Big)\,,
\end{equation}
which coincides with \eqref{seconda} when expanded at leading order. 
We stress that the overall factor of $p$ is entirely due to combinatorics, The only $p$-dependence comes from the two-point function normalisation \eqref{again_twopoint} for 
$\langle \Omax{p} \Omax{p}\rangle \,$, since the quantity $\langle T_2 T_2  T_2\rangle/\langle T_2 T_2 \rangle$ 
is manifestly indepedent from $p$. 
We can then repeat the derivation of the HHL three-point coupling 
for the diagonal contributions directly at large $N$, showing that 
the large-$N$ three-point functions \eqref{seconda} and \eqref{terza} 
coincide provided $\tilde{\alpha} = \frac{2\alpha^2}{1-2\alpha^2}$

This combinatorial argument does not use the fact that $\langle T_2 T_2  T_2\rangle$ 
is protected, but rather it relies crucially on the three-point coupling 
$\langle \Omax{p} \Omax{p} T_2\rangle$ scaling like $p \times\!\!\  \langle \Omax{p} \Omax{p}\rangle$.
Given that this scaling is just a consequence of large-$N$ factorisation, our argument 
applies to all three-point coupling between $\Omax{p}  \Omax{p}$ and 
double-particle operators \cite{Aprile:2018efk}, both protected and 
non-protected, for which we find that in general
\begin{equation}\label{quarta}
\frac{\langle \Omax{p} \Omax{p} [T_{q_1} \partial^\ell\scalebox{1.3}{$\square$}^{\frac{1}{2}(t-q_1-q_2)} T_{q_2}]\rangle }{ \langle \Omax{p} \Omax{p}\rangle \,} = 
\frac{p}{2}\, \frac{ \langle T_2 T_2  [T_{q_1} \partial^\ell\scalebox{1.3}{$\square$}^{\frac{1}{2}(t-q_1-q_2)} T_{q_2}]\rangle \,}{ \langle T_2 T_2 \rangle  }(1+O(N^{-2}))\,.
\end{equation}
Any such three-point coupling has the property that is identical to a 
three-point coupling in which the maximal-trace operators are replaced by 
the coherent-state operators $O_H$ upon mapping $\tilde{\alpha} = \frac{2\alpha^2}{1-2\alpha^2}$, 
provided the large-$N$ scaling  is $p \times\!\!\  \langle \Omax{p} \Omax{p}\rangle$ 
as in \eqref{secondaseconda} or \eqref{quarta}.

Armed with this information, we recall that the reduced HHLL correlator \eqref{eq:Hseries} is expressed as a sum over reduced maximal-trace correlators which are all purely diagonal. 
It appears then that by factorising the HHLL correlator 
in products of three point functions, in the appropriate channels, 
one can gain some understanding about the relation \eqref{eq:C0HHLLId}
directly at the level of the HHLL correlator.
For example, this would be the case for the exchange of the stress-tensor multiplet where the four-point function factorises as
\begin{equation}
\average{\mathcal{O}_2\mathcal{O}_2\expOO\expOO} \sim \langle \mathcal{O}_2\mathcal{O}_2 T_2\rangle \langle T_2 \expOO\expOO\rangle
\end{equation}
A similar argument can be repeated for both the protected and the long twist-four double-particle operators. 
This line of reasoning, combined with our results from the integrated correlator, 
supports the conjecture that the relation \eqref{eq:C0HHLLId} 
can be uplifted to the genus-zero reduced correlators. 
However, we leave the proof of this conjecture for future investigations.

A second question we wish to address regarding the genus-zero relation 
between the coherent-state and maximal-trace correlators \eqref{eq:C0HHLLId} 
is whether this property persists at higher genus. Unfortunately, beyond genus 
zero we see that the HHLL and maximal-trace correlators become genuinely different 
quantities, not related by the simple mapping between $\alpha$ and $\tilde \alpha$, 
that is
\begin{equation}\label{eq:nonequalityCgHHLLCgMT}
\mathcal{C}_{\HHLL}^{(g>0)}(N, \alpha;\tau)  \neq \mathcal{C}^{(g>0)}_{\tilde\alpha}(N;\tau) \quad\qquad {\rm with}\quad  \tilde{\alpha} = \frac{2\alpha^2}{1-2\alpha^2}\,.
\end{equation}

The mathematical reason for this is rather simple. The spectral overlap 
$\mathcal{A}^{(g>0)}(s,\tilde{\alpha})$ for the maximal-trace correlator 
has been computed in \eqref{eq:Ag} as a sum over terms $\mathcal{M}^{(g_1)}(s)\mathcal{G}^{(g_2)}(s,\tilde{\alpha})$ 
with $g_1+g_2=g$ and $g_1,g_2\in \mathbb{N}$. If we use the relation between 
$\tilde\alpha$ and $\alpha$ in \eqref{eq:C0HHLLId} we see that the  single 
contribution with $g_2=0$, which corresponds to $\mathcal{M}^{(g)}(s)\mathcal{G}^{(0)}(s,\tilde{\alpha})$, is again identical to the 
complete genus-$g$ spectral overlap \eqref{eq:Cgenus} for the \HHLL~integrated 
correlator. However, at higher genus we find that the two correlators are 
genuinely different quantities \eqref{eq:nonequalityCgHHLLCgMT} due to the 
presence of the additional contributions $\mathcal{M}^{(g_1)}(s)\mathcal{G}^{(g_2)}(s,\tilde{\alpha})$ with $g_2\neq0$. 
In Appendix \ref{comparison_Nsq_coherent_app} we analyse in detail 
the genus-one spectral overlap for both \HHLL~and double-scaling maximal-trace 
integrated correlators to show that indeed the two observables
are not related by the simple identity \eqref{eq:C0HHLLId}, which therefore 
does not persist to sub-leading orders.

The modular invariant genus expansion of the HHLL integrated correlators $\mathcal{C}_{\HHLL}(N, \alpha;\tau)$ we just derived,
imposes constraints on the correlator itself. Furthermore, since at large $N$ and strong coupling the `un-integrated' correlator
admits a holographic description,  we deduce that the aforementioned constraints do provide valuable information on the 
underlying geometrical description. In the next two sections we will make this interplay very explicit and present constraints for the HHLL correlator both at the perturbative and non-perturbative level in a large 't Hooft coupling genus-expansion.

\subsection{Perturbative tests at strong coupling \& dual geometry
}
\label{sec:PertLLM}

In \cite{Chester:2019pvm,Binder:2019jwn,Chester:2020dja}, see also \cite{Alday:2021vfb}, it was shown that the integrated correlator of the four-point function $\langle {\cal O}_2 {\cal O}_2 {\cal O}_2 {\cal O}_2\rangle$ poses extremely non trivial constraints on the higher-derivative expansion of the dual Virasoro-Shapiro amplitude in ${\rm AdS}$.  Here we generalise that 
analysis to the integrated correlator of maximal-trace operators and 
to the \HHLL~integrated correlator. For the latter we will show that the 
leading order contribution agrees perfectly with the result obtained 
by integrating the supergravity correlator given in \cite{Aprile:2024lwy,Aprile:2025hlt} and 
computed in the ${\rm AdS}$ bubble. Furthermore we will derive new constraints for the higher-derivative corrections to supergravity in the ${\rm AdS}$ bubble.

To proceed, we note that since we have already derived the large-$N$ expansion of $ \mathcal{C}_{\tilde\alpha N^2}$  and ${\cal C}_{\HHLL}$ at fixed string coupling $\tau$, it becomes straightforward to use our results from section \ref{sec:N2} and \ref{sec:HHLL} to derive the large-$N$ genus expansion at fixed 
coupling 't Hooft coupling $\lambda$. 
In order to obtain the fixed  't Hooft coupling genus expansion from the finite-$\tau$ results presented in Section \ref{sec:N2} and \ref{sec:HHLL}  we simply need to remove all non-zero Fourier modes from the real-analytic Eisenstein series
\eqref{Eisen_series_definitions} and perform the replacement
\begin{equation}
E^*(s;\tau) \to \xi(2s) (4\pi N)^s \lambda^{-s}   +  \xi(2s-1)\,(4\pi N)^{1-s} \lambda^{s-1}\,,\label{eq:EisenHooft}
\end{equation}
having express the imaginary part of $\tau$ in favour of the 't Hooft coupling $\lambda$, 
i.e. $\tau_2 = 4\pi N/\lambda$. This replacement is justified by the fact that the non-zero Fourier modes of the real-analytic Eisenstein 
series \eqref{Eisen_series_definitions} correspond to Yang-Mills 
instantons, which are exponentially suppressed in the limit considered here.  

To obtain the large-$N$ genus expansion at fixed  't Hooft coupling of either integrated correlator, $ \mathcal{C}_{\tilde\alpha N^2}$  or ${\cal C}_{\HHLL}$, we start from their modular invariant genus expansion and substitute in the corresponding spectral integral representations the reduced expression \eqref{eq:EisenHooft} for $E^*(s;\tau)$. We then evaluate each contour integral as previously discussed and lastly isolate all terms contributing to a given power $N^{2-2g}$ with $g\in \mathbb{N}$. 

We shall consider first the \HHLL~correlator,
and discuss the analogous results for the maximal-trace operators later on. 
Proceeding as just outlined, we replace $E^*(s;\tau)$ with \eqref{eq:EisenHooft} in  the spectral integrals \eqref{eq:Cgenus},
evaluate the contour integral via the residue theorem and lastly re-arrange the fixed-$\tau$ expansion \eqref{eq:CHHLLgenus}
to arrive at the genus expansion
\begin{align}\label{eq:ThooftHHLL}
  \mathcal{C}_{\HHLL}(N,\alpha;\lambda) 
  \sim 
  -\frac{N^2}{4}\log(1-2\alpha^2)+\sum_{g=0}^\infty N^{2-2g} & \left[\ \,
  \sum_{k=0}^\infty \tilde{\mathcal{B}}^{(g)}(k+\tfrac{3}{2};\alpha) \xi(2k+3) \left(\frac{\lambda}{4\pi}\right)^{-k -\frac{3}{2}}\right.\\
  &\left.\, +\sum_{k=0}^{g-1} \tilde{\mathcal{B}}^{(g-k-1)}(k+\tfrac{3}{2};\alpha) \xi(2k+2) \left(\frac{\lambda}{4\pi}\right)^{k +\frac{1}{2}}  \right]\,,\notag
\end{align}
where $\tilde{\mathcal{B}}^{(g)}(k+\frac{3}{2};\alpha)$ denotes the residue  from the genus $g$ 
contribution  \eqref{eq:Cgenus} evaluated at $s=k+3/2$ with $k\in \mathbb{N}$.
Analogously to the analysis previously carried out for $\langle {\cal O}_2 {\cal O}_2 {\cal O}_2 {\cal O}_2\rangle$, 
we identify three different types of perturbative contributions, each of which has a clear interpretation on the gravity side:
\vspace{-0.3cm}
\begin{itemize}
\item[(i)] The supergravity contribution:  $\displaystyle-\frac{N^2}{4}\log(1-2\alpha^2)$.\vspace{-0.2cm}
\item[(ii)] The genus-$g$ Virasoro-Shaphiro contribution: $\displaystyle \sum_{k=0}^\infty \tilde{\mathcal{B}}^{(g)}(k+\tfrac{3}{2};\alpha) \xi(2k+3) \left(\frac{\lambda}{4\pi}\right)^{-k -\frac{3}{2}}$.\vspace{-0.2cm}
\item[(iii)] The genus-$g$ counterterm contributions: $\displaystyle \sum_{k=0}^{g-1} \tilde{\mathcal{B}}^{(g-k-1)}(k+\tfrac{3}{2};\alpha) \xi(2k+2) \left(\frac{\lambda}{4\pi}\right)^{k +\frac{1}{2}}$.
\end{itemize}

For the genus-zero contribution, corresponding to tree-level and higher-derivatives corrections, 
we derive from \eqref{eq:C0exact} the result
\begin{equation}
\tilde{\mathcal{B}}^{(0)}(k+\tfrac{3}{2};\alpha)=\frac{ (k+1) \Gamma \left(k-\frac{1}{2}\right) \Gamma \left(k+\frac{1}{2}\right) }{2^{2k+1}\,\pi ^{\frac{3}{2}} \Gamma (k+1)} \left[1- 
\tFo{k+\tfrac{3}{2}}{-k-\tfrac{1}{2}}{1}{\tfrac{2\alpha^2}{2\alpha^2-1}}\right]\,.\label{eq:B0VS}
\end{equation}
Here we will not give the explicit form of ${\cal B}^{(g)}(k+\tfrac{3}{2};\alpha)$, which can be derived from \eqref{eq:Cgenus}, however we note that the genus-$g$ contributions to the Virasoro-Shaphiro perturbative series are factorially divergent power series in $1/\sqrt{\lambda}$. This can be checked explicitly in the case of the genus-zero expression given above. We stress that the same  divergent behaviour is also encountered  \cite{Dorigoni:2021guq} in the genus expansion of  $\langle {\cal O}_2 {\cal O}_2 {\cal O}_2 {\cal O}_2\rangle$ at large-$\lambda$.  We will come back to the physical consequences of this fact  in the next section.

As an important consistency check, we now show 
that the $\lambda$ independent, supergravity contribution (i) 
can be reproduced by integrating as in \eqref{Integrated_corr_HHLL_def} the four-point \HHLL~correlator
computed in \cite{Aprile:2025hlt}.
The authors of \cite{Aprile:2025hlt} studied precisely
the leading connected term of the four-point correlator 
$\langle{\alpha}|O_L(x_1)\bar{O}_L(x_2)|{\alpha}\rangle$
showing that it admits a holographic description valid 
as an analytic function of $\alpha$. 
This correlator was obtained by computing the two point function of the 
light operators in the background of the ${\rm AdS}$ bubble, and analytically continuing 
the result in $\alpha$. It was checked that the analytic continuation gives 
the correct result even when $\alpha$ is infinitesimally small. 

The solution technique put forward in \cite{Aprile:2025hlt} was to first solve
the connection problem for the dilaton wave-equation 
in the background of the ${\rm AdS}$ bubble, and then use a Ward identity to relate the dilaton to $O_{L}$, 
thus arriving at $\langle{\alpha}|O_L(x_1)\bar{O}_L(x_2)|{\alpha}\rangle$. 
The validity of the analytic continuation 
in $\alpha$ was checked against non trivial CFT data, 
both protected and non-protected \cite{Aprile:2025hlt}. 
Here we provide yet another independent consistency check. 

The wave equation for the bulk dilaton field is a Heun equation whose connection problem 
has an elegant solution in the language of Seiberg-Witten theory~\cite{Aminov:2020yma}. In our conventions, 
the final result of \cite{Aprile:2025hlt} for the \HHLL~four-point correlator reads
\begin{equation}\label{Psi_introduction}
  \frac{\langle{\alpha}|O_L(1)\bar{O}_L(x, \bar{x})|{\alpha}\rangle}{\langle \alpha| \alpha\rangle}\Bigg|_{\rm tree}\!\!=\ {N^2}{}\sum_{\ell=0}^{\infty} \oint_{}^{}\, 
    ({\ell+1}) (x\bar{x})^{\frac{-\omega-\ell-4}{2}} \left( \frac{x^{\ell+1}-\bar{x}^{\ell+1}}{x-\bar{x}} \right) E_{\rm gravity}(\omega,\ell;\alpha)\,\frac{\dd\omega}{2\pi i}\;,
    \end{equation}
where
\begin{equation}\label{Egravity}
E_{\rm gravity}(\omega,\ell;\alpha)=\sum_{ \kappa\in D }^{ } \left(\frac{1}{\kappa+\frac{\ell}{2}+2+a(\omega,\ell;\alpha)}+\frac{1}{\kappa+\frac{\ell}{2}+2-a(\omega,\ell;\alpha)}\right)\;.
\end{equation}
The function $a(\omega,\ell;\alpha)$ plays a crucial role since it 
computes the mass-spectrum of the exchanged operators in the heavy-light channel.
The integration over $\omega$ in \eqref{Psi_introduction} is performed 
via Cauchy theorem by using the Feynman contour. 
This fixes the summation $\kappa\in D$, left implicit in \eqref{Egravity}.

In the approach of~\cite{Aminov:2020yma},  $a(\omega,\ell;\alpha)$ corresponds to the quantum Seiberg-Witten period of 
the spectral curve associated with the Heun equation in the ${\rm AdS}$ bubble geometry. 
In particular, it is computed by inverting the Matone-relation \cite{Matone:1995rx},
\begin{equation}\label{Matonerel}
u=-\frac{1}{4}-a^2+a_t^2+a_0^2 +t \partial_t F(a, a_0,a_t,a_1,a_\infty, t)\,,
\end{equation}
where $a=a(\omega,\ell;\alpha)$, while the parameters are $a_t=a_{\infty}=1$, $a^2_0=\frac{(\ell+1)^2}{4}$,  
\begin{equation}
a_1^2=\frac{1+2\omega^2-\ell(\ell+2)}{4},\quad u=\frac{4(2\alpha^2+1)-(1-2\alpha^2)(\omega^2-\ell(\ell+2))}{4},\quad t=-\frac{2\alpha^2}{1-2\alpha^2}\,.
\end{equation}
The function $F$ is the classical {Liouville theory} conformal block which can be found in \cite[C.1.4]{Bonelli:2022ten}. 

The HHLL integrated correlator 
has been defined in~\eqref{Integrated_corr_HHLL_def} in terms of the reduced correlator~${ \mathcal{H}_{{\alpha}}(u,v;N,\tau) }$ in~\eqref{eq:HHLL_more0}-\eqref{eq:Hseries}. Here we are interested in the tree-level contribution to the reduced correlator,~i.e.
${ \mathcal{H}_{{\alpha}}(u,v;N,\tau) } \vert_{\rm{tree}}$,
which we have to extract from \eqref{Psi_introduction}.
Thus, following the same type of splitting into free 
and dynamical part, employed in  \eqref{eq:HHLL_more0}, 
we find convenient to introduce the function ${\cal K}_{\alpha}(u,v)$, 
defined as
\begin{equation}
 \left.\frac{\langle{\alpha}|O_L(1)\bar{O}_L(x, \bar{x})|{\alpha}\rangle}{\langle \alpha|\alpha\rangle}\right\vert_{\rm tree}=
  {\langle{0}|O_L(1)\bar{O}_L(x, \bar{x})|{0}\rangle}{}
 +{N^2}{} {\cal K}_{\alpha}(u,v)\,,\label{eq:Ksugra}
\end{equation}
and which is related to  ${ \mathcal{H}_{{\alpha}}(u,v;N,\tau) }\vert_{\rm{tree}}$ through the equation
\begin{equation}
{ \mathcal{H}_{{\alpha}}(u,v;N,\tau) } \vert_{\rm{tree}}=N^2{ \langle \alpha|\alpha\rangle}  \frac{u^2}{4}{\cal K}_{\alpha}(u,v)\,,
\end{equation} 
where we used $d_{12}^2 = 4/u^2$ - since $x_{12}^4= u^2$ for the insertions points arranged as in \eqref{eq:Ksugra}. With these conventions, the tree-level HHLL integrated correlator becomes, 
\begin{equation}
\label{eq:Kintegrated}
    {\cal C}_{\HHLL}\Big|_{\rm tree} =
    -\frac{\,N^2}{2\pi}\int_0^\infty \dd r\int_0^\pi \dd \theta  \,{r^3\sin^2(\theta)}{}{\cal K}_{\alpha}(u,v)\,\Big\vert_{u=1+r^2-2r\cos\theta,v=r^2}\,.
\end{equation}

While the problem of determining the exact $\alpha$ dependence of the correlator in \eqref{Psi_introduction} is hard,
the small $\alpha$ expansion is calculable. The period $a(\omega,\ell;\alpha)$ has the following expansion
\begin{equation}
a(\omega,\ell;\alpha) = \frac{\omega}{2} +\sum_{n= 1}^\infty \alpha^{2n} \gamma_n(\omega,\ell) \,,
\end{equation}
where the various $\gamma_n$ can be computed from the known Young diagram expansion of $F$.
Plugging the equation above in $E_{\rm gravity}$ yields a power series expansion in $\alpha$ 
for the integrand of \eqref{Psi_introduction}.
We can then evaluate the $\omega$ contour integral by collecting residues from the poles 
located at {$\omega=-\ell-4-2\kappa$} thus obtaining an expansion for the correlator \eqref{Psi_introduction} 
as a power series for small $x,\bar{x}$, where we note that $u=(1-x)(1-\bar{x})$ and $v=x \bar{x}$.

The free term in \eqref{eq:Ksugra} 
is the only term contributing at order $O(\alpha^0)$, for which we find
\begin{align}
  {\langle{0}|O_L(1)\bar{O}_L(x, \bar{x})|{0}\rangle}{}= & {N^2}{}
 \sum_{\kappa=0}^{\infty}\sum_{\ell=0}^{\infty} 2 (\ell+1) (x \bar{x})^{\kappa} \left(\frac{ x^{\ell+1 } - \bar{x}^{\ell+1}}{x - \bar{x}}\right)  = {N^2}{} \left( \frac{2}{u^2} \right)\,.
 \end{align}
 which correctly reproduces the two-point function normalisation. 
 For the  dynamical part of \eqref{eq:Ksugra} we shall write
 \begin{equation}
 {\cal K}_\alpha(u,v)=\sum_{n=1}^{\infty} \frac{\alpha^{2n}}{n!} K_{\alpha}^{(n)}(u,v)\,,
 \end{equation}
 where the $n!$ has been introduced to follow the conventions of \cite{Aprile:2025hlt} 
 so that we may borrow their explicit results.
 For reference, let us give the first 
 non trivial order, $O(\alpha^2)$, where we have
\begin{equation}
\gamma_1(\omega,l)=\frac{(\omega-\ell-2)(\omega-\ell)(\omega+\ell)(\omega+\ell+2)}{4(\omega-1)\omega(\omega+1)}\,.
\end{equation}
The contour integral is done by collecting residues from the poles 
located at {$\omega=-\ell-4-2\kappa$}, now with $\kappa\ge -1$, and yields
 \begin{align}\label{def_Kappa}
{K}_{\alpha}^{(1)}(u,v)= & 
-8 u^2\overline{D}_{2422}(u,v)\,,
  \end{align}
 where $\overline{D}_{\Delta_1 \Delta_2 \Delta_3 \Delta_4}(u,v)$ 
is the well known $4$-point contact Witten diagram  (see Appendix D of \cite{Dolan:2001tt}).
This order in the small-$\alpha$ expansion
is obtained by replacing the two heavy operators 
$O_H(\alpha)=e^{\frac{\alpha}{\sqrt{2}} O_L} \rightarrow \alpha/\sqrt{2} \,O_L$  
in the HHLL four-point function \eqref{sec:LLM_HHLL}. The 
expression \eqref{def_Kappa} then matches identically with the expected result 
from the known $\langle O_L O_L O_L O_L\rangle$ correlator.

At higher order in the small-$\alpha$ expansion the story becomes rather laborious. 
However, as shown in \cite{Aprile:2025hlt}, we find that the transcendental 
content of the correlator at order $O(\alpha^{2n})$ consists of
ladder integrals, and their derivatives, up to degree $2n$. 
Results for $n=2$ and $n=3$ can be found in \cite{Aprile:2025hlt}.
We have evaluated numerically the integrals,
\begin{equation}
\label{eq:KIintegrated}
    I_n \coloneqq -\frac{1}{2\pi}\int_0^\infty \dd r\int_0^\pi \dd \theta \, 
     {r^3\sin^2(\theta)}{} {K}_{\alpha}^{(n)}(u,v)\Big\vert_{u=1+r^2-2r\cos\theta,v=r^2}\,.
\end{equation} using the explicit 
results given in \cite{Aprile:2025hlt} and established
\begin{equation}
I_1 = \frac{1}{2} \,, \qquad I_2 = 1 \,, \qquad I_3 = 4 \,.\label{eq:NumInt}
\end{equation}
Our result for $I_1$ agrees with the known result from \cite{Binder:2019jwn,Chester:2020dja} 
(after matching  conventions), while $I_2$ and $I_3$ are new results.\footnote{We also checked 
the integration \cite{Chester:2019pvm} for the one-loop correlator in \cite{Aprile:2017bgs}, 
and found perfect agreement.} 

We can finally compare with the supergravity contribution (i) to \eqref{eq:ThooftHHLL}. 
By rewriting the logarithmic term as
\begin{equation}\label{exactformula}
-\frac{N^2}{4}\log(1-2\alpha^2) = N^2 \sum_{n=1}^{\infty} \frac{\alpha^{2n}}{n!} I_n\,, \qquad \qquad {\rm with}\quad I_n=2^{n-2}(n-1)!\,,
\end{equation}
we find perfect agreement between the analytic result from the integrated correlator 
and the actual numerical integration \eqref{eq:NumInt}. 
It would be interesting to derive the exact supergravity contribution 
(i) to \eqref{eq:ThooftHHLL} from the `un-integrated' correlator of  \cite{Aprile:2025hlt}  
by exploiting the spectral curve approach underlying the Heun connection problem that led to \eqref{Egravity}-\eqref{Matonerel}.

We conclude this section with some comments concerning the large-$N$ genus expansion 
at fixed  't Hooft coupling of the maximal-trace correlator $\mathcal{C}_{\tilde\alpha N^2}$. 
Following the same procedure which led to the genus expansion \eqref{eq:ThooftHHLL} for the \HHLL~correlator, we arrive at  
\allowdisplaybreaks{
\begin{align}
 \mathcal{C}_{\tilde\alpha N^2}(N;\lambda) \sim \frac{N^2}{4}\log(1+\tilde\alpha)+\sum_{g=0}^\infty N^{2-2g} &\notag \left[c_g(\alpha)+ 
  \sum_{k=0}^\infty \tilde{\mathcal{A}}^{(g)}(k+\tfrac{3}{2};\tilde\alpha) \xi(2k+3) \left(\frac{\lambda}{4\pi}\right)^{-k -\frac{3}{2}}\right.\\*
  &\ \ \ \ \ \left.+\sum_{k=0}^{g-1} \tilde{\mathcal{A}}^{(g-k-1)}(k+\tfrac{3}{2};\tilde\alpha) \xi(2k+2) \left(\frac{\lambda}{4\pi}\right)^{k +\frac{1}{2}}\right]\label{eq:ThooftCalpha}\,.
\end{align}}

As discussed in section \ref{sec:HHLL}, we know that the genus-zero contributions of the maximal-trace integrated correlator and the \HHLL~integrated correlator are related via the simple mapping of parameters $\tilde{\alpha} = \frac{2\alpha^2}{1-2\alpha^2}$ in \eqref{eq:C0HHLLId}. Unsurprisingly, we then find that 
\begin{equation}
    \tilde{\cal A}^{(0)}\Big(k+\tfrac{3}{2};\tilde\alpha\Big)= {\tilde{\cal B}}^{(0)}\Big(k+\tfrac{3}{2};\sqrt{\tfrac{\tilde \alpha}{2(1+\tilde\alpha)}}\Big)\,,
\end{equation}
so that also in the 't Hooft limit, the genus-zero HHLL and maximal-trace correlators are related via the simple map in parameters as above, where importantly we also note that $c_0(\alpha)=0$ in \eqref{eq:ThooftCalpha}.
However, as previously mentioned, the higher genus contributions to the two integrated correlators are not related via a simple mapping in parameters. A direct consequence of this fact can be immediately seen by comparing the two 't Hooft genus expansions \eqref{eq:ThooftHHLL}  and \eqref{eq:ThooftCalpha}.
For every genus $g\geq 1$, we see that the maximal-trace correlator $\mathcal{C}_{\tilde\alpha N^2}$ receives a $\lambda$ independent contribution denoted by
 $c_{g}(\alpha)$ in \eqref{eq:ThooftCalpha}. For example, the value $c_1=\frac{\tilde \alpha}{2(1+\tilde \alpha)}$ can be read from \eqref{eq:largeNSugra}.
 Such contributions are completely absent for the \HHLL~integrated correlator \eqref{eq:ThooftHHLL}. 
 
 It would be extremely interesting to understand how to derive such $\lambda$-independent terms from the holographic side to clarify the difference between HHLL and maximal-trace correlators from a geometric point of view. Unfortunately, since the deviation between the two integrated correlators appears at the sub-leading order $O(N^0)$, this makes any holographic calculation much harder.
We find it remarkable that the 't Hooft expansion of the \HHLL~integrated correlator has the simple structure in \eqref{eq:ThooftHHLL}.

\subsection{Non-perturbative effects \& their holographic interpretation}
\label{sec:NPLLM}

In this section we compute the non-perturbative corrections to the large-$N$ expansion of the modular 
invariant genus-zero \HHLL\,integrated correlator \eqref{eq:Cgenus}, derived in the previous section, 
and then discuss their physical interpretation.

Firstly, we stress that the identity \eqref{eq:C0HHLLId} relating
$\mathcal{C}_{\HHLL}^{(0)}(N, \alpha;\tau) $ to $\mathcal{C}^{(0)}_{\tilde{\alpha}}(N;\tau)$ with $\tilde{\alpha} = \frac{2\alpha^2}{1-2\alpha^2}$ has been derived at the level of spectral decompositions and hence it does hold at the non-perturbative level in $N$.
An immediate consequence of this fact is that the large-$N$ perturbative series of the two integrated correlators 
coincide via the same mapping in parameters $\alpha \leftrightarrow \tilde{\alpha}$ as above.
The large-$N$ perturbative expansion of the correlator   $\mathcal{C}^{(0)}_{\HHLL}$, denoted 
hereafter with $\mathcal{C}^{(0)}_{\HHLL,P}$, is\footnote{The same perturbative series can also be obtained 
from \eqref{eq:C0exact} by pushing the contour of integration towards ${\rm Re}(s)\to +\infty$ while collecting 
residues from the simple poles located at $s= k+\frac{3}{2}$ with $k\in \mathbb{N}$.}
\allowdisplaybreaks{
\begin{align}
    &\label{eq:C0asy} \mathcal{C}_{\HHLL}^{(0)}(N, \alpha;\tau)\sim \mathcal{C}^{(0)}_{\HHLL,P}(N,\alpha;\tau) \coloneqq \\*
    &\notag \sum_{k=0}^\infty \frac{(k+1)\Gamma(k-\frac{1}{2})\Gamma(k+\frac{1}{2})}{2^{2k+1}\pi^{\frac{3}{2}}\Gamma(k+1)}  \left[ \tFo{k+\tfrac{3}{2}}{-k-\tfrac{1}{2}}{1}{\tfrac{2\alpha^2}{2\alpha^2-1}}-1\right] N^{-k-\frac{3}{2}} \eisen{k+\tfrac{3}{2}}\,.
\end{align}}
We can then proceed to analyse this asymptotic series via its Borel transform and repeat the steps outlined in Section \ref{Mod_inv_resurgence_section}.
Using  \eqref{eq:BorelDef} we immediately find that the Borel transform is given by 
\begin{align}\label{eq:BorelIdd}
    \mathcal{B}[\mathcal{C}^{(0)}_{\HHLL,P}](\alpha,t) &\notag=\sum_{k=0}^\infty \frac{4\Gamma(k-\frac{1}{2})\Gamma(k+\frac{1}{2})}{\pi \Gamma(k+1)^2}\left[\tFo{k+\tfrac{3}{2}}{-k-\tfrac{1}{2}}{1}{\tfrac{2\alpha^2}{2\alpha^2-1}}-1\right]t^{2k+2}
\end{align}
On the RHS we precisely recognise the same functional form found for the Borel of the genus-zero maximal-trace correlator \eqref{eq:B0sum}, thus we arrive at the identity,
\begin{equation}\label{eq:BorelId}
     \mathcal{B}[\mathcal{C}^{(0)}_{\HHLL,P}](\alpha,t)=\mathcal{B}[\mathcal{C}^{(0)}_{\tilde{\alpha},P}](t)\Big\vert_{\tilde{\alpha} = \frac{2\alpha^2}{1-2\alpha^2}}\,.
\end{equation}
Hence, both the large-$N$ perturbative and non-perturbative corrections 
to the HHLL leading order contribution in \eqref{eq:C0exact} are directly 
related to those of $\mathcal{C}^{(0)}_{\tilde{\alpha}}(N;\tau)$ via the mapping $\tilde{\alpha} = \frac{2\alpha^2}{1-2\alpha^2}$. 

The analytic properties of the Borel transform for the 
maximal-trace integrated correlator on the LHS of \eqref{eq:BorelId} 
have been analysed in detail in Section \ref{sec:N2}.
From the singularity structure of $\mathcal{B}[\mathcal{C}^{(0)}_{\tilde{\alpha},P}](t)$, 
we then derive immediately the  non-perturbative correction 
to the large-$N$ expansion of the \HHLL~integrated correlator. 
Upon substituting the map $\tilde{\alpha} = \frac{2\alpha^2}{1-2\alpha^2}$ in \eqref{eq:tI}-\eqref{eq:tII}-\eqref{eq:tIII}, we deduce that the Borel 
transform $\mathcal{B}[\mathcal{C}^{(0)}_{\HHLL,P}](\alpha,t)$ 
of the genus-zero \HHLL~correlator has three different 
branch-cuts along the positive real $t$-axis starting respectively at,
\begin{equation}\label{eq:geoscalesHHLL}
  L^*_{\rm I} = 1\,,\qquad  L^*_{-} = \sqrt{\frac{1-\sqrt{2}\alpha}{1+\sqrt{2}\alpha}}\,, \qquad L^*_{+}=\frac{1}{L^*_{-}}=\sqrt{\frac{1+\sqrt{2}\alpha}{1-\sqrt{2}\alpha}}\,.
\end{equation}
Correspondingly, the exact large-$N$ transseries expansion of the 
leading  contribution \eqref{eq:C0exact} to the \HHLL~integrated correlator is not
given by the perturbative series \eqref{eq:C0asy} alone, but crucially it must also contain important non-perturbative and modular invariant corrections.
These non-perturbative effects are derived in a straightforward manner from the analogous non-perturbative sectors \eqref{eq:NPscales} for the maximal-trace integrated correlator in the double scaling regime using the map in parameters \eqref{eq:BorelId},
\allowdisplaybreaks{
\begin{align}
\mathcal{C}_{\HHLL,N\!P}^{(0),{\rm I}}(N;\tau) &\notag \sim  \exp\left( - 4 \sqrt{\pi N} \frac{|n\tau+m|}{\sqrt{\tau_2}} \right) \,,\\*
\mathcal{C}_{\HHLL,N\!P}^{(0),{\rm -}}(N;\tau) &\label{eq:NPscalesHHLL} \sim  \exp\left( - 4 \sqrt{\pi N} \sqrt{\frac{1-\sqrt{2}\alpha}{1+\sqrt{2}\alpha}} \frac{|n\tau+m|}{\sqrt{\tau_2}} \right) \,,\\*
\mathcal{C}_{\HHLL,N\!P}^{(0),{\rm +}}(N;\tau) &\notag \sim   \exp\left( - 4 \sqrt{\pi N}\sqrt{\frac{1+\sqrt{2}\alpha}{1-\sqrt{2}\alpha}}\frac{|n\tau+m|}{\sqrt{\tau_2}} \right) \,,
\end{align}}
where the sum over $(n,m)\neq(0,0)$ is left implicit.

We can rewrite the arguments of the exponentials in \eqref{eq:NPscalesHHLL} in the regime of large 't Hooft coupling, i.e. $N\gg1$ with $\lambda=N \gym^2 = \frac{4\pi N}{\tau_2} \gg 1$. The lattice sum is dominated
by the lattice points $(n,m)=(0,m)$ with $m\neq0$, and thus in this limit we obtain the hierarchy of novel non-perturbative scales to the \HHLL~integrated correlator
\begin{align}
\exp\left(- 2|m|\sqrt{\lambda} \sqrt{\frac{1+\sqrt{2}\alpha}{1-\sqrt{2}\alpha}}\right)\ll\exp\left(- 2|m|\sqrt{\lambda}\right) \ll \exp\left(- 2|m|\sqrt{\lambda} \sqrt{\frac{1-\sqrt{2}\alpha}{1+\sqrt{2}\alpha}}\right)\,,
\label{eq:NPscalesHHLLtHooft}
\end{align}
valid for the physical parameter range $0 < \alpha < {1}/{\sqrt{2}}$. We note in 
particular that for $\alpha\to0$ the three non-perturbative scales all reduce to the single quantity $e^{- 2\sqrt{\lambda}}$. 

It is interesting to ask what is the semi-classical origin 
of the non-perturbative scales \eqref{eq:NPscalesHHLL}-\eqref{eq:NPscalesHHLLtHooft} 
in the dual gravity theory, which describes the insertion 
of the coherent-state operators. As in the previous section, 
we set up the discussion starting from the easier case where 
we consider the four-point stress-tensor correlator 
$\langle {\cal O}_2 {\cal O}_2 {\cal O}_2 {\cal O}_2\rangle$, 
and then move to the \HHLL~correlator.

We begin by connecting the perturbative contributions 
to $\langle {\cal O}_2 {\cal O}_2 {\cal O}_2 {\cal O}_2\rangle$ at large $N$ and strong coupling with the low energy expansion around the flat space limit \cite{Alday:2023mvu}, see also 
the discussion in \cite{Caron-Huot:2024tzr}.  
The leading term at genus-zero 
is the well-known flat-space Virasoro-Shapiro amplitude. 
Curvature corrections around the flat-space  limit
can be organised into layers, where each layer is a non-trivial function with its own expansion to 
all orders in $\frac{1}{\sqrt{\lambda}}$. The integrated version of these layers is precisely the genus expansion given in \cite{Dorigoni:2021guq}. From the work of Alday, Hansen, and Silva \cite{Alday:2022uxp,Alday:2022xwz}, 
it is understood that such low-energy expansion is related to the CFT-data of short-string operators. 
These states $O$ have scaling dimensions $\Delta_O \sim \lambda^{\frac{1}{4}}$ and three-point couplings with ${\cal O}_2 {\cal O}_2$
which at 
leading order behave as ${C_{O,{\cal O}_2 {\cal O}_2}} \sim2^{-\Delta}\Delta^{3}/\sin(\frac{\pi}{2}\Delta)$ .\footnote{As shown in \cite[(3.18)]{Alday:2022uxp},  in low-energy expansion theory these short strings 
contribute perturbatively to sum rules. The 
suppression of the three-point couplings is compensated by the $\omega_{\tau,\ell}(s,t;q)$ block.} 
The study of the integrated correlators \cite{Dorigoni:2021guq}
shows that this low energy expansion 
is factorially divergent and thus requires the presence of non-perturbative effects.

Beyond these short-string states, Maldacena and Hofman 
identified big-string states of dimensions $\Delta\sim \lambda^{\frac{1}{2}}$, 
called giant magnons in \cite{Hofman:2006xt}. 
It seems plausible that these big-string states provide a 
link between the perturbative expansion of the correlator
and its non-perturbative completion as we see emerging 
from the study of the integrated correlator.

The analysis of the integrated correlator for $\langle {\cal O}_2 {\cal O}_2 {\cal O}_2{\cal O}_2\rangle$, closely parallels our analysis of 
the \HHLL~integrated correlator.
However, the key advantage in the HHLL correlator is that the parameter 
$\alpha$ enters the non-perturbative scales
$L^*_{+}$ and $L^*_{-}$ in \eqref{eq:geoscalesHHLL}.
This $\alpha$-dependence provides an opportunity 
to test the conjectured link between big-string states and non-perturbative effects
that we mentioned above.   
To justify this further, we now compute the spectrum of giant magnons in 
the gravity dual to the coherent-state operators, 
and show that the quantities $L^*_{\pm}$ in \eqref{eq:geoscalesHHLL} 
precisely control the dimensions of the giant magnons. 
The holographic picture will also manifest that the 
relation $L^*_{+} L^*_{-}=1$ is no coincidence.

We explained in Section \ref{sec:coherent} that the coherent-state 
operators are dual to a back-reacted geometry, the ${\rm AdS}$ bubble, 
which belongs to a consistent truncation of type IIB supergravity in ten spacetime dimensions. 
In the previous section we used precisely this background to
test the results of the integrated correlator against the actual 
integration of the leading large-$N$ correlator computed from 
the gravity side, see \eqref{Psi_introduction}. In that context, 
we only used information about the five-dimensional 
asymptotically ${\rm AdS}$ part of the full ten-dimensional 
background. In order to study the giant magnons, 
we need to consider the internal space as well. The 
full ten-dimensional metric of the ${\rm AdS}$ bubble is discussed in Appendix \ref{app:AdSbubble},  
here we present a streamlined version of that discussion, which is fine-tuned to our purpose.  

To set up the notation, let us consider the ten-dimensional metric 
\begin{equation}\label{metricPOPEtext}
ds^2_{10}=G_{\mu\nu} d X^\mu dX^\nu = \sqrt{ \Delta }\, ds_{5}^2 +  \frac{1}{\sqrt{\Delta}} (T_{ij})^{-1} D\mu^i D\mu^j\,,
\end{equation}
where the variables $\mu^i$ with $i\in\{0,...,5\}$ are defined by 
\begin{equation}\label{spherecoordinates}
\mu^0 + i \mu^3 = \cos\theta \cos\chi \,e^{i \psi_1},\qquad
\mu^1 + i \mu^2 = \cos\theta \sin\chi \,e^{i \psi_2},\qquad 
\mu^4+i \mu^5 = \sin\theta \,e^{i\phi}\,,
\end{equation}
with $(\theta, \chi,\phi,\psi_1,\psi_2)$ coordinates on an ${\rm S}^5$.
The five dimensional metric $ds_{5}^2$ is given by
\begin{equation}\label{5dmetricAdSbubble_text}
ds^2_5= - {\faH}(r)^{-\frac{2}{3}} \left(1+\frac{r^2}{L^2} {\faH}(r)\right) dt^2 + {\faH}(r)^{\frac{1}{3}} \left( \frac{dr^2}{1+\frac{r^2}{L^2}{\faH}(r)} + r^2 d\Omega_3^2 \right)\,.
\end{equation} 
Lastly, we defined $\Delta=T_{ij}\mu^i\mu^j$ and the covariant derivative appearing in \eqref{metricPOPEtext} is 
\begin{equation}
D\mu^i=d \mu^i+A^{ij}\mu^j\,,
\end{equation}
 where $A^{ij}$ is an antisymmetric matrix of one-forms gauge fields. 
The ${\rm AdS}$ bubble is an excitation of the vacuum 
${\rm AdS}_5\times {\rm S}^5$ solution, which is induced by a 
single gauge field in the $4$-$5$ plane, with a profile along the $dt$ component. 
The gauge field sources a certain unimodular scalar matrix $T_{ij}$ 
parametrised by a neutral and a charged scalar field. As a consequence 
of the half-BPS condition satisfied by this geometry, all fields 
can be written in terms of a single function $\faH(r)$ whose general 
form can be found in Appendix  \ref{app:AdSbubble}. 

The ten-dimensional metric $ds^2_{10}$ given in \eqref{metricPOPEtext} 
preserves an $SO(4)\times SO(4)$ symmetry. The first $SO(4)$ acts on 
the $d\Omega_3^2$ factor of the five-dimensional geometry,  as 
manifest in \eqref{5dmetricAdSbubble_text}. 
The second $SO(4)$ acts on the internal space, and in particular on 
the $\mu^i$ variables defined in \eqref{spherecoordinates} where we 
notice that the coordinates $\chi,\psi_1,\psi_2$ parametrise an 
${\rm S}^3$ with standard metric of the form $d\tilde{\Omega}_3^2= d\chi^2 + \cos^2\!\chi\, d\psi_1^2  + \sin^2\!\chi\,d\psi_2^2$. 

A giant magnon state is a solution to  
the equations of motion derived from the Nambu-Goto action. 
Its motion takes place in a $\mathbb{R}\times$S$^2$ sub-manifold 
of the geometry where S$^2\subset$ S$^3$ of the internal space. 
The worldsheet time is 
identified with the global time coordinate, $t\in (-\infty,+\infty)$. 
The string sits 
at the centre of space, i.e.~at $r=0$, and 
extends in global time $t$, thus explaning the $\mathbb{R}$ component of the motion.
The worldsheet space coordinate, $\sigma$, 
is identified with the angle $\phi$ and spans a particular interval  $I$.  
The string sits at $\chi=\psi_1=\psi_2=0$
and moves in the direction $\theta \coloneqq \theta(\sigma)$. 
The S$^2$ part of its motion is then spanned by the coordinates $\mu^0,\mu^4,\mu^5$ in \eqref{spherecoordinates}.

More in detail, let us consider the Nambu-Goto (NG) action, 
\begin{equation}
S=\frac{ \sqrt{\lambda} }{ 2\pi }\int\!\!\dd t\, \dd\sigma\, \sqrt{ - {\rm det}( g_{ab}) }\,,
\qquad\qquad {\rm with} \quad g_{ab}=  G_{\mu\nu}\partial_{a}X^{\mu}\partial_{b}X^{\nu}\,.
\end{equation}
The induced metric and the action can 
be computed explicitly by using formulae in Appendix \ref{app:AdSbubble},
in particular, we refer to Appendix \ref{10dmetricPope} for a derivation 
of the giant magnon solution. They key points of that derivation can be rephrased
by following a suggestion of \cite{Hofman:2006xt}, noticing that
the embedding of the string can be described equivalently
by the coordinate $\sin \theta$ with $\theta\coloneq\theta(\sigma)$.
It is then natural to describe the motion of the string by employing the same combination of variables that define the coordinates $\mu^{4}=\mu^{4}(\sigma)$ and $\mu^5=\mu^5(\sigma)$ in \eqref{spherecoordinates}. 

In these coordinates we can bring the NG action to a very simple form,
\begin{equation}\label{rew_MH_magnons}
S=\frac{ \sqrt{\lambda} }{ 2\pi }\int\!\! \dd t\, \dd\sigma\, \sqrt{ 
L^{*2}_{-} \left(\frac{d\mu^4}{d\sigma}\right)^2 +L^{*2}_{+} \left(\frac{d\mu^5}{d\sigma}\right)^2 }
\,,\qquad \qquad {\rm with}\quad  L^*_{\pm}=\sqrt{\frac{1\pm \sqrt{2}\alpha}{1\mp \sqrt{2}\alpha}}\,,
\end{equation}
which is precisely the action describing a geodesic motion on 
the plane $4$-$5$ with flat metric. However, since the worldsheet 
coordinate $\sigma$ extends over an interval, the solutions correspond 
to straight lines inside an ellipse with major axis $L_+$ and minor 
axis $L_-$. Remarkably, the quantities $L^*_{\pm}$ are exactly the 
same quantity that set the scales for the non-perturbative 
effects that we found in the HHLL integrated correlator from 
resurgence analysis, see  \eqref{eq:geoscalesHHLL}. Here, in 
the holographic picture, they have just appeared from the study 
of the giant magnon solutions.

Moreover, the ellipse emerging in \eqref{rew_MH_magnons} 
is precisely the droplet of the ten-dimensional bubbling geometries 
of Lin-Lunin-Maldacena \cite{Lin:2004nb}. This is quite a nice result given that 
the coordinates used to parametrise the ${\rm AdS}$ bubble 
in \eqref{metricPOPEtext} do not make manifest the droplet of \cite{Lin:2004nb}.
Thanks to our identification between the field theory and geometry parameters, we can also understand the relation $L^*_+ L^*_-=1$ as the geometric condition for which the area of the droplet, in our case the area of 
the ellipse, must be quantised in Planck units. 
Presently, this condition amounts to the fact that the non-perturbative 
scales  \eqref{eq:geoscalesHHLL} do satisfy $L^*_+ L^*_-=1$.

The energy of the giant magnon is computed by the length 
of the solution in the $4$-$5$ plane. In the dual CFT this 
quantity computes the twist, $\Delta-J,$ of the dual operator. 
To understand this point it should be noted that at the boundary $r\rightarrow\infty$, 
the metric becomes ${\rm AdS}_5\times {\rm S}^5$, however 
the metric for the five-sphere is actually given by
$ds_{{\rm S}^5}^2=\sin^2\! \theta\,(d\phi+dt)^2 + d\theta^2 + \cos^2\! \theta\, d\tilde{\Omega}_3^2$. 
Thus, from the boundary point of view, a giant 
magnon spins on the sphere {due to the $ (d\phi+dt)^2 $ term}.

To compute the energy spectrum, it is useful to start from $\alpha=0$, hence first recovering the vacuum solution of Maldacena-Hofman in the circular droplet.
This is the solution where $\mu^4=\sin\theta_0$ and $\mu^5= \sin\theta_0
\tan \sigma$, with $\sigma\in[-\frac{\pi}{2}+\theta_0,+\frac{\pi}{2}-\theta_0]$. 
The motion of the string is constant in the direction $\mu^4$ while interpolating between two opposite points for the $\mu^5$ coordinates. The energy of this solution  
reads 
\cite{Hofman:2006xt}, 
\begin{equation}\label{MHformula}
\Delta-J= \frac{\sqrt{\lambda}}{\pi}\cos\theta_0\,.
\end{equation}
The quantity $\Delta-J$ is maximal at $\theta_0=0$, and vanishes identically  for $\theta_0=\frac{\pi}{2}$ where the straight line degenerates to a point. The latter is a representation of the known half-BPS operator. 

Note that for the ${\rm AdS}_5\times {\rm S}^5$ vacuum, all 
solutions to the NG action equations of motion are equivalent 
to $\mu^4=\sin\theta_0$ and $\mu^5= \sin\theta_0
\tan \sigma$, due to rotational invariance of the circular droplet. 
When $\alpha\neq 0$, the droplet gets deformed to an ellipse 
and rotational invariance is broken. 
In this case we distinguish two fundamental solutions moving 
along either the major or minor axis. These giant magnon 
solutions are respectively given by
\begin{equation}\label{giantM_Lplus}
\left\lbrace \begin{array}{l}
 \mu^4_+=\sin\theta_0 \\
\mu^5_+= \sin\theta_0\tan \sigma
\end{array}
\right.
\qquad\qquad {\rm with}\quad
\Delta-J= \frac{\sqrt{\lambda}}{\pi}L^*_+\cos\theta_0\,,
\end{equation}
where the mangon moves along the major axis of the ellipse, and
\begin{equation}\label{giantM_Lmeno}
\left\lbrace \begin{array}{l}
\mu^4_-=\sin\theta_0\tan \sigma \\
\mu^5_-= \sin\theta_0
\end{array}
\right.
\qquad\qquad  {\rm with}\quad\Delta-J=\frac{\sqrt{\lambda}}{\pi}L^*_-\cos\theta_0\,,
\end{equation}
where the magnon moves along the minor axis.\footnote{Following the work of BMN \cite{Berenstein:2002jq}, the relation \eqref{MHformula} between energy and twist has been understood also in the gauge theory side \cite{Santambrogio:2002sb,Beisert:2005tm}. It would be interesting to generalise these arguments to the ${\rm AdS}$ bubble case as well.}

We emphasise that the giant magnons \eqref{giantM_Lplus}-\eqref{giantM_Lmeno} here constructed are \textit{finite energy} solutions. 
However, the non-perturbative effects we discovered in the study 
of integrated correlators seem to suggest that from a path-integral 
point of view we should be looking for \textit{finite action} 
solutions of the Euclidean equations of motion. 
We believe it is no 
accident that the quantities \eqref{giantM_Lplus}-\eqref{giantM_Lmeno} 
at their maxima for $\theta_0=0$ reduce precisely to the non-perturbative scales \eqref{eq:NPscalesHHLLtHooft}. Hence we expect these finite energy 
solutions  
to be directly related to finite action configurations via Wick 
rotation, although showing how this works 
amounts to a formidable task that we leave for future works. 

We conclude this section by providing some speculations 
regarding the expected properties of the 
finite-action giant magnons. These properties are expected on the 
basis of what we found for the \HHLL~integrated 
correlator, using the resurgence analysis carried out in Section \ref{sec:N2}
and the results presented in Appendix \ref{sec:NP_appendix_p_Nsq}. 
On the one hand, the leading non-perturbative effects associated with the 
scale $L_-^*$ contribute to the \HHLL~correlator with a purely 
imaginary overall factor $i$. On the other hand, the non-perturbative 
effects associated with the scale $L_+^*$ are sub-dominant and provide 
for a purely real contribution to the correlator. This suggests that 
semi-classically the corresponding Euclidean saddles should have one 
and two negative-mode directions respectively.\footnote{Our argument 
is solely based on the reality properties of the one-loop determinant 
around a saddle-point. Of course it might very well be that these saddles 
have a larger number of negative directions. However, we find more 
plausible to believe that the minimal choice is actually the correct one. } 
A naive Wick rotation of the real-time giant-magnon solutions
seems to be supporting this claim.

Considering the solution \eqref{giantM_Lmeno}, 
we expect the string configuration 
responsible for the non-perturbative scale  $L^*_-$ to stretch along 
the minor axis of the ellipse. In field space, small fluctuations of this configuration would then possesses a negative 
direction where the string slips off to one side, thus collapsing 
to a point. Considering instead the solution \eqref{giantM_Lplus}, we  
expect the string configuration responsible for the 
non-perturbative scale $L^*_+$ to stretch along the major axis 
of the ellipse. Such a configuration would then have two negative 
directions: one where the strings caps off to a point as in the 
previous case, while the other where the string rotates towards 
the minor axis solution, which has a lower on-shell action. 
Lastly, when 
$\alpha=0$ the ellipse reduces to a circle and the two non-perturbative 
string configurations are related by rotational invariance, 
as displayed precisely by the non-perturbative effects studied in Section \ref{sec:N2}.

It would be interesting to study the 
Euclidean problem in more detail to provide concrete quantitative 
support for our speculative comments.

\section{Conclusions}
\label{sec:Conclusions}

In this work we used non-perturbative methods to analyse 
the maximal-trace integrated correlator 
$\mathcal{C}_{p,N}(\tau) $ defined in \eqref{Integrated_corr_def} and 
first discussed in \cite{Paul:2022piq}, and the HHLL integrated correlator
$\mathcal{C}_{\HHLL}(N,{\alpha};\tau)$ defined in \eqref{Integrated_corr_HHLL_def}.
Even though  the corresponding `un-integrated' correlators are not known in closed form, 
we were able to derive exact, modular invariant transseries expansions 
for both $\mathcal{C}_{p,N}(\tau)$ and $\mathcal{C}_{\HHLL}(N,{\alpha};\tau)$ 
using techniques from modular resurgence analysis.
We then used our results to study a variety of regimes where the number of colours $N$ and/or the $R$-symmetry charge $p$ are taken to be large. 

Considering the integrated correlator $\mathcal{C}_{p,N}(\tau)$, 
we studied the limit of large charge 
with $N$ fixed, as well as the limit where both $N$ and 
the charge $p$ are taken to be large with $p/N^\gamma$ fixed for different values of 
$\gamma>0$. From a holographic perspective, 
the double scaling limit where $N,p\gg1$ with $p/N^2 = \tilde{\alpha}$ 
fixed, appears to be the most interesting regime. In this limit, we found 
the remarkable result that the genus-zero contribution to 
$\mathcal{C}_{\tilde{\alpha} N^2,N}(\tau)$ is identical to the 
genus-zero contribution to  $\mathcal{C}_{\HHLL}(N,{\alpha};\tau)$ 
under the simple map in parameters $\tilde{\alpha} = {2\alpha^2}/({1-2\alpha^2})$.

We then focussed on the HHLL correlator whose 
holographic description in supergravity was recently understood 
in \cite{Aprile:2025hlt}. The heavy operator inserted is dual to a coherent 
state constructed from multi-graviton states in AdS.
 The corresponding dual back-reacted geometry in type IIB supergravity is
the known AdS bubble - a particular example of an 
LLM space with an elliptic droplet. As we did for $\mathcal{C}_{p,N}(\tau)$, we computed as well  for $\mathcal{C}_{\HHLL}(N,{\alpha};\tau)$ the complete,~i.e.~perturbative and non-perturbative, modular invariant 
large-$N$ transseries expansion.

From this transseries we were immediately able to derive the large-$N$ genus expansion for the HHLL integrated correlator in the strong 't Hooft coupling limit, $\lambda \gg 1$.
At the perturbative level, we showed that the tree-level 
contribution matches identically the integral over the insertion points of the 
`un-integrated' correlator computed in  \cite{Aprile:2025hlt}.
At the non-perturbative level, we used resurgence analysis 
to identify three different types of exponentially suppressed 
corrections in $\sqrt{\lambda}$. Importantly, two of these corrections 
are expressed in terms of the $\alpha$-dependent scales $L^*_{\pm}$ given in \eqref{eq:IntroScales}. 
We were able to reproduce exactly the same scales 
by analysing the spectrum of giant-magnons 
on the ${\rm AdS}$ bubble. In this picture,  we also confirmed that the quantities
$L^*_+$ and $L^*_-$ correspond precisely to the major and minor axis of the elliptical 
droplet in the LLM description. The relation  $L^*_+L^*_-=1$ found surprisingly from resurgence analysis, 
can then be translated to the quantisation condition for the area of the droplet  in Planck units.

This work opens up several future research directions worth investigating. 
Firstly, we find it remarkable that the integrated correlator ${\cal C}_{\HHLL}(N,\alpha;\tau)$ admits an ${\rm SL}(2,\mathbb{Z})$
decomposition for which the spectral overlap neatly factorises as function of $\alpha$ times a function of $N$, 
as shown in \eqref{eq:cHHHLLspec}. This property is most definitely not generally found in other types of integrated correlators. In fact, it is not even shared 
by the maximal-trace integrated correlator ${\cal C}_{p,N}(\tau)$,
which provides the building block for constructing ${\cal C}_{\HHLL}(N,\alpha;\tau)$. It might be possible that this property is manifest in an alternative matrix model description of the HHLL integrated correlator, and it
would be very interesting to understand 
this from a physics view-point.

From a holographic perspective, the genus expansion of the HHLL integrated correlator \eqref{eq:ThooftHHLL} 
parallels that of the stress-tensor correlator as a series of corrections 
to the low-energy expansion of the dual superstring theory, as well 
as curvature corrections around the flat-space limit. 
It would be very interesting to combine arguments from the CFT bootstrap, 
for example those proposed in \cite{Alday:2023mvu}, with our results for the 
HHLL integrated correlator in order to construct the Virasoro-Shapiro 
amplitude on the ${\rm AdS}$ bubble, so far almost completely unexplored.
On a similar note, as discussed in Section \ref{sec:NPLLM}, it would also be extremely important to show
from a proper Euclidean path-integral approach
how the giant magnon solutions, or more general long string solutions, 
contribute to the non-perturbative completion  of the Virasoro-Shapiro amplitude.

Another important issue which deserves further investigation is the relation between the HHLL correlator $\mathcal{C}_{\HHLL}(N,{\alpha};\tau)$ and the maximal-trace correlator $\mathcal{C}_{\tilde{\alpha} N^2,N}(\tau) $ with $p/N^2 =\tilde{\alpha}$ fixed.
While at genus-zero we found that the two correlators are surprisingly related via a map in parameters, $\tilde{\alpha} \leftrightarrow \alpha$, we have also proved that 
this relation does not hold at higher orders.
This fact appears clear when comparing the large-$N$ genus expansion at large 't Hooft coupling $\lambda$ of both integrated correlators given respectively in \eqref{eq:ThooftHHLL} and \eqref{eq:ThooftCalpha}.
A striking 
difference is that the only $\lambda$-independent term in the genus expansion of the HHLL integrated correlator \eqref{eq:ThooftHHLL} appears at leading order, $O(N^2)$, while the 
maximal-trace correlator has a series of $\lambda$-independent corrections at any given order in $1/N$. 

From a holographic point of view, such contributions correspond to the integration of 
supergravity loops. The genus expansion for the HHLL integrated correlator is very reminiscent of that 
for the stress-tensor integrated correlator  \cite{Dorigoni:2022cua} and in both cases we 
find that supergravity-loops integrate to zero beyond one-loop. 
Correspondingly, given that such 
corrections are instead present for the maximal-trace integrated correlator this suggests 
that the maximal-trace correlator and the HHLL correlator correspond to 
the same AdS bubble geometry only at the leading order.

Lastly, in this work we focussed our attention only towards operators constructed 
out of the superconformal primary operator $\mathcal{O}_2$ in the stress-tensor multiplet. 
Firstly, using the results of \cite{Paul:2022piq,Brown:2023cpz,Brown:2023why,Paul:2023rka} 
it is possible to analyse the integrated correlator for the maximal-trace operators 
with odd $R$-charge $p$ given in \eqref{eq:Omax}, and schematically of the 
form $\Omax{p}\sim \mathcal{O}_3 \,[\mathcal{O}_2]^{\frac{p-3}{2}}$.
From our analysis, it appears clear that when $p/N^2$ is kept fixed as 
$N\gg1$, this integrated correlator should now correspond (at least at leading order) 
to a four-point function of two gravitons and two higher Kaluza-Klein modes 
on the dual AdS bubble geometry. We believe it should be possible to combine 
bootstrap ideas and results from \cite{Aprile:2025hlt} to compute this 
four-point function. The odd maximal-trace integrated correlator 
would then provide a consistency check for this calculation as discussed in Section \ref{sec:PertLLM}.

While in this previous setup we would only be inserting two quanta of the higher Kaluza-Klein mode $\mathcal{O}_3$, the present work suggests that interesting new phenomena may appear when we insert instead a coherent state built from $\mathcal{O}_3$.
We stress that the results of \cite{Paul:2022piq,Brown:2023cpz,Brown:2023why,Paul:2023rka} lead to exact expressions for integrated four-point functions with two $\mathcal{O}_2$ and two identical half-BPS operators $\mathcal{O}_p$ not necessarily of maximal-trace type nor single-particle operators.
In particular, we believe it should be possible to push the analysis of \cite{Paul:2022piq,Brown:2023cpz,Brown:2023why,Paul:2023rka}  as to derive exact expressions in both the number of colours $N$ and the Yang-Mills coupling $\tau$, for an integrated version of the four-point function
\begin{equation}\label{eq:HHLLConc}
 \average{\mathcal{O}_2(x_1,Y_1)\mathcal{O}_2(x_2,Y_2)\widetilde{\mathcal{O}}_H(0,Y_3)\widetilde{\mathcal{O}}_H(\infty,Y_4)} \,,
\end{equation}
where the coherent-state `heavy' operator $\widetilde{\mathcal{O}}_H$ is now built out of the single-particle state $\mathcal{O}_3$, i.e.
\begin{equation}
\widetilde{\mathcal{O}}_H(0,Y; \alpha) = 
\exp\left({\alpha}\, \mathcal{O}_3(0,Y)\right)\,.
\end{equation}

Importantly, also in this case the dual geometry is a LLM space whose profile is known from \cite{Vazquez:2006id,Anempodistov:2025maj}. However, unlike for the graviton coherent-state 
discussed in the present work, the space is not a consistent truncation and one needs to use
original LLM description. 
The study of these more general HHLL correlators \eqref{eq:HHLLConc} will help understanding  in
broader scenarios how the properties of the geometry are captured by the integrated correlators.  

Our work suggests a possible way forward in this direction. For the coherent-state operator 
here considered, we have found the factorisation property \eqref{eq:cHHHLLspec} for 
the  spectral overlap, which can be written as the product of the spectral overlap for 
the original  stress-tensor integrated correlator \cite{Binder:2019jwn, Chester:2020dja}  
times a function which captures all the properties of the underlying geometry. 
It is tantalising to conjecture that something similar may happen for a wider class of HHLL correlators, 
such as \eqref{eq:HHLLConc}, and thus, at least in this class, one might hope to make a precise statement about the 
way in which the spectral overlaps depend on the parameters of the geometry. 

\vskip 0.2cm
\noindent {\large {\bf Acknowledgments}}
\vskip 0.2cm
We thank Benjamin Basso, James Drummond, Stefano Giusto, Paul Heslop, Ben Hoare, 
Grisha Korchemsky, Vasilis Niarchos, Hynek Paul, Simon Ross, Rodolfo Russo, Michele Santagata, and Congkao Wen for many useful discussions. We would
also like to thank Rodolfo Russo and Congkao Wen for providing helpful comments on the draft.

FA is supported by
RYC2021-031627-I funded by MCIN/AEI/10.13039/501100011033 and by the NextGeneration EU/PRTR. 
FA also acknowledges support from the
``HeI" staff exchange program \href{https://cordis.europa.eu/project/id/101182937}{DOI 10.3030/101182937} financed by the MSCA.  FA would like to thank the organisers of \href{https://indico.mpp.mpg.de/event/11294/}{Symbology@25} for the hospitality. 
DD is grateful to the RIKEN Center for Interdisciplinary Theoretical and Mathematical Sciences for the hospitality during the final stages of this project. DD is supported by
the Royal Society grants ICA$\backslash$R2$\backslash$242058 and IEC$\backslash$R3$\backslash$243103. RT is supported by the grant $\text{CUP D93C24000490005}$ issued by Università degli Studi di Parma.

\appendix

\section{Non-perturbative effects in the regime $p=\alpha N^2$}
\label{sec:NP_appendix_p_Nsq}

In this appendix we study the analytic structure of the Borel transform \eqref{eq:B0sum} as to understand the large-$N$ non-perturbative contributions to the integrated correlator \eqref{Integrated_corr_def} in the regime where the charge $p$ of the maximal-trace operator inserted is given by $p=\alpha N^2$ with $\alpha>0$ a fixed parameter.

Starting from \eqref{eq:B0sum} we first make use of the hypergeometric functions identity,
\begin{equation}
    \tFo{k+\tfrac{3}{2}}{-k-\tfrac{1}{2}}{1}{-\alpha} = \big(\sqrt{\alpha+1}-\sqrt{\alpha}\big)^{2k+3}\tFo{k+\tfrac{3}{2}}{\tfrac{1}{2}}{1}{\tfrac{4\sqrt{\alpha(\alpha+1)}}{(\sqrt{\alpha}+\sqrt{\alpha+1})^2}}\,,
\end{equation}
and then rewrite the right-hand side of the above equation  using the integral representation
\begin{equation}
    \tFo{k+\tfrac{3}{2}}{\tfrac{1}{2}}{1}{\tfrac{4\sqrt{\alpha(\alpha+1)}}{(\sqrt{\alpha}+\sqrt{\alpha+1})^2}} = \int_0^1 \frac{\Big(1-\frac{4x\sqrt{\alpha(\alpha+1)}}{(\sqrt{\alpha}+\sqrt{\alpha+1})^2}\Big)^{-k-\frac{3}{2}}}{\pi\sqrt{x(1-x)}}\dd x\,.
\end{equation}
By exchanging the sum in \eqref{eq:B0sum} with the $x$-integral we arrive at
\begin{equation}\label{Borel_largeN_largeP}
    \mathcal{B}[\mathcal{C}_{\alpha,P}^{(0)}](t) = \frac{{16}t^2}{\pi}E(t^2)-\frac{{16}t^2}{\pi}\int_0^1 \frac{ v_{\alpha}(x)^{-\frac{3}{2}}}{\pi\sqrt{x(1-x)}} E\big(t^2 v_{\alpha}(x)\big)\dd x \,,
\end{equation}
where $E(z)=\frac{\pi}{2}\tFo{\frac{1}{2}}{-\frac{1}{2}}{1}{z}$ denotes the complete elliptic integral function
and where we defined the auxiliary quantity
\begin{equation}
 \quad v_{\alpha}(x) \coloneqq (\sqrt{\alpha}+\sqrt{\alpha+1})^2-4x\sqrt{\alpha(\alpha+1)}\,.
\end{equation}

From the well-known analytic structure of the elliptic integral $E(z)$, we immediately see that the first term in \eqref{Borel_largeN_largeP} has logarithmic branch-cuts along $t\in (-\infty,- L_{\rm I})\cup (L_{\rm I},+\infty)$ where $L_{{\rm I}}=1$  just as in the cases previously discussed. However, the singularity structures in the complex $t$-plane arising from the $x$-integral are more complicated and interesting.
For sufficiently small values of $|t|$ the integrand remains holomorphic throughout the interval of integration $x\in [0,1]$, but for large enough $|t|$ the integral over $x$ develops branch-cut singularities. 
It is easy to see that a branch-cut must exist since the elliptic integral $E(z)$ has a branch point at $z=1$, hence we expect a first singularity in $t$ to appear when
\begin{equation}
    t^2=(\sqrt{\alpha}+\sqrt{\alpha+1})^2-4x\sqrt{\alpha(\alpha+1)}\label{eq:tsing}
\end{equation}
for some $x$ in the range of integration. 

We now use the integral representation \eqref{Borel_largeN_largeP} to compute the difference in lateral resummations across the Stokes line ${\rm Re}(t)>0$,
\begin{equation}\label{eq:StokesApp}
    (\mathcal{S}_+-\mathcal{S}_-)[\mathcal{C}^{(0)}_{\alpha,P}](N;\tau) = -2i\,\Big[\mathcal{S}_0[\mathcal{C}_{\alpha,N\!P}^{{\rm I},(0)}](N;\tau)+ \mathcal{I}_\alpha(N;\tau)\Big]\,.
\end{equation}
The first term in the above equation simply comes from having computed the discontinuity in lateral resummations for the first term appearing in \eqref{Borel_largeN_largeP} and proportional to the elliptic integral function $E(t^2)$.
This first contribution can be written explicitly as
\allowdisplaybreaks{
\begin{align}
    \mathcal{S}_0[\mathcal{C}_{\alpha,N\!P}^{{\rm I},(0)}](N;\tau)& \notag = \int_0^\infty {4}t(t+2) (t+1)^2\tFo{\tfrac{1}{2}}{\tfrac{3}{2}}{2}{-t(2+t)}\mathcal{E}(\sqrt{N}(t+1);\tau)\,\dd t \\*
 &\label{eq:C1NP}   = \mathcal{C}_{\alpha,N\!P}^{{\rm I},(0)}=\frac{{8}}{N}\Dfn{N}{1}+\frac{{28}}{N^{\frac{3}{2}}}\Dfn{N}{\tfrac{3}{2}}+\frac{{33}}{N^2}\Dfn{N}{2}+...\,,
\end{align}}
where we used the known formula for the discontinuity of the elliptic integral
\begin{equation}\label{eq:discE}
    \lim_{\epsilon \to 0^+}\Big[ E(z+i \epsilon )- E(z-i \epsilon )\big] = \frac{i\pi}{2}(1-z)\tFo{\tfrac{1}{2}}{\tfrac{3}{2}}{2}{1-z} =-2 i \big[E(1-z)-K(1-z)\big]\,,
\end{equation} 
valid for $z>1$ and where $K(z)$ is the complete elliptic integral of the first kind.

The term~$\mathcal{I}_\alpha(N;\tau)$ in~\eqref{eq:StokesApp} comes from the second factor in~\eqref{Borel_largeN_largeP}. To evaluate this contribution we exchange the $x$ and $t$ integrals and then use the discontinuity formula~\eqref{eq:discE} arriving at
\begin{equation}\label{Ialpha_definition}
    \mathcal{I}_{\alpha}(N;\tau)=\int_0^1 \frac{{4}v_{\alpha}(x)^{-\frac{3}{2}}}{\pi \sqrt{x(1-x)}}\int_{v_\alpha(x)^{\frac{1}{2}}}^\infty \frac{ t^2(v_\alpha(x)-t^2)}{ v_{\alpha}(x)}\,\tFo{\tfrac{1}{2}}{\tfrac{3}{2}}{2}{\tfrac{ v_\alpha(x)-t^2}{ v_{\alpha}(x)}}\mathcal{E}(\sqrt{N}t;\tau)\,\dd t\,\dd x\,.
\end{equation}
We note here that the lower extrema of integration $t =v_{\alpha}(x)^{\frac{1}{2}}$ corresponds precisely to the starting point of the branch cut for the elliptic integral function which appeared in \eqref{Borel_largeN_largeP}. 
In particular over the integration domain $x\in[0,1]$ we see that $v_{\alpha}(x)^{\frac{1}{2}}$ takes the minimal value $L_{-}=v_{\alpha}(1)^{\frac{1}{2}}$ at $x=1$ and maximal value $ L_{+}=v_{\alpha}(0)^{\frac{1}{2}}$ at $x=0$ where $L_{+} = L_{-}^{-1} = \sqrt{\alpha+1}-\sqrt{\alpha}$. 

To analyse the asymptotic expansion of \eqref{Ialpha_definition} at large-$N$ we perform the change of variables $t\to t \,{v_\alpha(x)}^{\frac{1}{2}}$ thus yielding
\begin{equation}
    \mathcal{I}_\alpha(N;\tau) = \frac{{4}}{\pi}\int_1^\infty t^2(1-t^2)\tFo{\tfrac{1}{2}}{\tfrac{3}{2}}{2}{1-t^2}\int_0^1\frac{\mathcal{E}(\sqrt{N v_\alpha(x)}t;\tau)}{\sqrt{x(1-x)}}\dd x\,\dd t\,.
\end{equation}
The integral over $t$ can be readily evaluated by remembering that the factor $\mathcal{E}(\sqrt{\Lambda}t;\tau)$ defined in \eqref{Eisen_largeNp_intRep} serves the purpose of being a modular invariant Borel-like integration kernel.
We first shift integration variables $t\to t+1$ and then expand the hypergeometric function in its Gauss series around the origin and perform the $t$-integration term by term to arrive at
\begin{equation}
\mathcal{I}_\alpha(N;\tau) = \sum_{\ell=2}^\infty c_\ell \sum_{(m,n)\neq(0,0)} (N Y_{mn}(\tau))^{-\frac{\ell}{2}}  \int_0^1  \frac{ e^{ -4 \sqrt{  N Y_{mn}(\tau)  v_\alpha(x)}}}{\sqrt{ x(1-x) v_\alpha(x)^\ell}} \dd x\,,\label{eq:appStep}
\end{equation}
where the first few coefficients are
\begin{equation}
c_2 =-\frac{1}{{2}\pi} \,, \qquad c_3 = -\frac{7}{{16}\pi} \,, \qquad c_4 = -\frac{33}{{256} \pi} \,.
\end{equation}

To express \eqref{eq:appStep} as  a transseries at large-$N$ we simply need to change integration variables from $x$ to 
\begin{equation}
t^2 \coloneqq  v_\alpha(x) =  (\sqrt{\alpha}+\sqrt{\alpha+1})^2-4x\sqrt{\alpha(\alpha+1)}\,,
\end{equation}
thus yielding
\begin{equation}
\mathcal{I}_\alpha(N;\tau) = \sum_{\ell=2}^\infty 4c_\ell \sum_{(m,n)\neq(0,0)} (N Y_{mn}(\tau))^{-\frac{\ell}{2}}  \int_{L_-}^{L_+}   e^{ -4 t \sqrt{  N Y_{mn}(\tau) }} t^{-\ell}[ (t-L_-)(L_+-t)]^{-\frac{1}{2}} \dd t\,.\label{eq:appStep2}
\end{equation}
We claim that the integral representation \eqref{eq:appStep2} coincides with the median resummation for the non-perturbative sector which does contain two distinct non-perturbative scales controlled precisely by $L_-$ and $L_+$.

To proceed we note that the integrand of \eqref{eq:appStep2} has a square-root branch cut singularity starting at $t = L_+$ and extending to $t\to \infty$. 
For simplicity we can focus our attention solely towards computing the integral in \eqref{eq:appStep2} and thus write the identity
\allowdisplaybreaks{\begin{align}
&\notag I_{\alpha,\ell}(N;\tau)\coloneqq \int_{L_-}^{L_+}   e^{ -4 t \sqrt{  N Y_{mn}(\tau) }} t^{-\ell}[ (t-L_-)(L_+-t)]^{-\frac{1}{2}} \dd t =\\*
&\label{eq:medianresum}\int_{L_-}^{\infty \pm i \epsilon}   e^{ -4 t \sqrt{  N Y_{mn}(\tau) }} t^{-\ell}[ (t-L_-)(L_+-t)]^{-\frac{1}{2}} \dd t\mp i \int_{L_+}^{\infty }   e^{ -4 t \sqrt{  N Y_{mn}(\tau) }} t^{-\ell}[ (t-L_-)(t-L_+)]^{-\frac{1}{2}} \dd t\,,
\end{align}}
having used the well-known discontinuity of the square-root.
Shifting integration variables for both integrals manifests their non-perturbative, exponentially suppressed nature.
\begin{align}
&\label{eq:appmedian}I_{\alpha,\ell}(N;\tau)= e^{ -4 L_- \sqrt{  N Y_{mn}(\tau) }}\int_{0}^{\infty \pm i \epsilon}   e^{ -4 t \sqrt{  N Y_{mn}(\tau) }} (t+L_-)^{-\ell}[ t(L_+-L_--t)]^{-\frac{1}{2}} \dd t\\
&\notag \mp i e^{ -4 L_+ \sqrt{  N Y_{mn}(\tau) }} \int_{0}^{\infty }   e^{ -4 t \sqrt{  N Y_{mn}(\tau) }} (t+L_+)^{-\ell}[ t (t+L_+-L_-)]^{-\frac{1}{2}} \dd t\,.
\end{align}
Lastly, we simply need to expand both integrands in the above equation for small-$t$ and perform the Borel integral term by term thus yielding the formal transseries expansion
\begin{align}
&\label{eq:median}I_{\alpha,\ell}(N;\tau)=\\
&\notag \sum_{k=0}^\infty a_k(\alpha,\ell)  e^{ -4 L_- \sqrt{  N Y_{mn}(\tau) }}\, [N Y_{mn}(\tau) ]^{-k-\frac{1}{2}}
 \mp i \sum_{k=0}^\infty \tilde{a}_k(\alpha,\ell) e^{ -4 L_+ \sqrt{  N Y_{mn}(\tau) }} \, [N Y_{mn}(\tau) ]^{-k-\frac{1}{2}}\,,
\end{align}
where the factorially divergent coefficients $a_k(\alpha,\ell) $ and $\tilde{a}_k(\alpha,\ell) $ can be easily derived from \eqref{eq:appmedian}.

We stress that the seemingly ambiguous expression \eqref{eq:median} has to be understood as a median resummation. The first asymptotic series of perturbative corrections on top of the non-perturbative scale $e^{ -4 L_- \sqrt{  N Y_{mn}(\tau) }}$ is non-Borel summable and one must perform a directional Borel integral $\mathcal{S}_{\pm \epsilon}$ to resum it. The choice of transseries parameter $\mp i$ multiplying the second contribution in \eqref{eq:median}, which is exponentially suppressed when compared to the first term, is precisely correlated with the choice of lateral resummation, $\mathcal{S}_{\pm \epsilon}$, for the first contribution.
When the formal transseries is resummed as such it yields the original well-defined integral \eqref{eq:medianresum}, which is a real-analytic function of $NY_{mn}(\tau)$ in the physical domain ${\rm Re}(NY_{mn}(\tau))>0$.
From \eqref{eq:median} we conclude that the Stokes automorphism \eqref{eq:StokesApp} for the Borel transform \eqref{eq:B0sum} of the leading correction to the large charge expansion in the limit where $N\gg1$ with $p=\alpha N^2$ and $\alpha >0$ must in fact contain the non-perturbative effects \eqref{eq:NPscales} presented in the main body of this work.

Following the discussion presented in \cite{Dorigoni:2024dhy}, we stress that the first type of non-perturbative corrections proportional to the ubiquitous non-perturbative scale $e^{ -4 t_{\rm I} \sqrt{  N Y_{mn}(\tau) }}=e^{ -4  \sqrt{  N Y_{mn}(\tau) }}$ contained in $\mathcal{C}_{\alpha,N\!P}^{{\rm I},(0)}$ as in \eqref{eq:C1NP} do contribute to the transseries expansion of the doubly scaled integrated correlator \eqref{eq:Cg} with a purely imaginary factor, i.e. the corresponding transseries parameter is purely imaginary.
However, this story becomes subtler for the novel non-perturbative sectors $\mathcal{C}_{\alpha,N\!P}^{{\pm},(0)}$ presented in \eqref{NP_terms_pNsq}  and proportional to the respective non-perturbative scales $e^{ -4 L_+ \sqrt{  N Y_{mn}(\tau) }}$ and $e^{ -4 L_- \sqrt{  N Y_{mn}(\tau) }}$.

To clarify this difference between non-perturbative effects it is easier to refer back to the spectral representation \eqref{eq:Cg} and consider the leading order  spectral overlap given in \eqref{eq:A0spec}.
As discussed in \cite{Dorigoni:2024dhy}, from a spectral decomposition point of view the non-perturbative corrections arise as saddle-point contributions to the spectral integral in the regime ${\rm Re}(s) \to \infty$.
We then need to compute the large-$s$ expansion of the term in \eqref{eq:A0spec} which is proportional to the hypergeometric function, since this is the part of the spectral overlap which gives rise to the non-perturbative corrections \eqref{Ialpha_definition} here studied.

From the integral representation of the hypergeometric function we derive the asymptotic expansion
\begin{equation}
 \, _2F_1(1-s,s;1;-\alpha ) \sim \frac{L_-^{1-2s}}{2 \sqrt{\pi }[\alpha (\alpha+1)]^{\frac{1}{4}} \sqrt{s}}(1+O(s^{-1}))+i \frac{ L_+^{1-2s}}{2 \sqrt{\pi }[\alpha (\alpha+1)]^{\frac{1}{4}} \sqrt{s}}(1+O(s^{-1}))\,,\label{eq:asy2F1}
\end{equation}
which is valid for ${\rm Re}(s) \to \infty$.
We note that since $0<L_- = \sqrt{\alpha+1} -\sqrt{\alpha}<1$ while instead $L_+ = L_-^{-1} = \sqrt{\alpha+1} +\sqrt{\alpha}>1$ for $\alpha>0$,  we have that the first term of \eqref{eq:asy2F1} is dominant for ${\rm Re}(s) \to + \infty$, while for ${\rm Re}(s) \to - \infty$ the second term becomes the dominant one.

Given the asymptotic expansion \eqref{eq:asy2F1} it is straightforward to check that the spectral integral \eqref{eq:Cg} has two saddle-points at large ${\rm Re}(s) \to +\infty$.
The first term in \eqref{eq:asy2F1} gives rise to a saddle point located at
\begin{equation}
s=s^{\star}_{-} = 2\, L_- \sqrt{N Y_{mn}(\tau)} \,,
\end{equation}
with a corresponding ``on-shell'' action given by
\begin{equation}
e^{ - 4 L_- \sqrt{N Y_{mn}(\tau)}}\,.
\end{equation}
However, we notice that in this limit the trigonometric factor $\tan(\pi s)$ present in \eqref{eq:A0spec} reduces precisely to the transseries parameter
\begin{equation}
\tan(\pi s^{\star}_{-}) \to \pm i \,, \qquad {\rm arg}(N) \gl 0\,.
\end{equation}

In a similar fashion, 
the second term in \eqref{eq:asy2F1} produces a different saddle point,
\begin{equation}
s=s^{\star}_{+} = 2\, L_+ \sqrt{N Y_{mn}(\tau)} \,,
\end{equation}
whose action is now given by
\begin{equation}
e^{ - 4 L_+ \sqrt{N Y_{mn}(\tau)}}\,.
\end{equation}
The trigonometric factor $\tan(\pi s)$ reduces again to $\pm i $ when $s\to s^{\star}_{+}$, but crucially the second term in \eqref{eq:asy2F1} is accompanied by a purely imaginary overall coefficient, thus making this second type of non-perturbative corrections a real contribution to the large-$N$ transseries expansion of \eqref{eq:Cg}.
This difference between the two saddles is precisely the spectral representation incarnation of the transseries \eqref{eq:median}.

These seemingly minor differences between non-perturbative sectors are in fact consequences of a rather beautiful semi-classical holographic picture as we discuss in Section \ref{sec:LLM}.

\section{Non-perturbative effects in the regime $p=\alpha N^\gamma$ with $\gamma>2$}
\label{sec:appNP2}

In this appendix we derive the non-perturbative effects presented in Section \ref{sec:gamma3} for the integrated correlator in the very heavy regime where $N\gg1$ while the charge of the maximal-trace operator is taken to be $p = \alpha N^\gamma$ with $\gamma>2$ and $\alpha >0$ fixed.

In Section \ref{sec:gamma3} we showed that when expanded in the heavy-charge limit, the integrated correlator can be divided into three contributions \eqref{eq:C000}, \eqref{eq:C0002} and \eqref{eq:C0003}.
While the analysis of \eqref{eq:C0002} and \eqref{eq:C0003} is  straightforward, that of \eqref{eq:C000} requires an interesting modification of the modular invariant resurgence methods described in Section \ref{Mod_inv_resurgence_section} which we now present.

For convenience we rewrite here the quantity defined in  \eqref{eq:C000} while dropping all of its subscripts as this will be the only function analysed in this appendix. Hence we consider
\begin{equation}\label{eq:C000app}
    \mathcal{C}(\Lambda;\tau) = \intRepsilon \frac{2^{1-2s}\Gamma(s-2)\Gamma(s-1)\Gamma(s)\tan(\pi s)^2}{\pi \Gamma(s-\frac{1}{2})^2} \Lambda^{-s}\eisen{s}\frac{\dd s}{2\pi i}\,.
\end{equation}
We then wish to extract the large-$\Lambda$ expansion of the above integral where we remind the reader that presently
\begin{equation}
\Lambda = 4\alpha N^{\gamma-1}\,.\label{eq:LambdaApp}
\end{equation}
Importantly, the integrand of \eqref{eq:C000app} has once again poles located at $s=k+\frac{3}{2}$ with $k\in \mathbb{N}$ but these are now double poles, hence the modular resurgence analysis presented in section \ref{Mod_inv_resurgence_section} cannot be applied in a straightforward manner.

However, to compute the asymptotic large-$\Lambda$ perturbative expansion of \eqref{eq:C000app} we can still proceed in the same way and collect residues from the double poles by closing the contour of integration to the right half-plane ${\rm Re}(s)>1$.
This yields the asymptotic expansion
\begin{equation}\label{LargeGammaNewPertSeries}
    \mathcal{C}(\Lambda;\tau)\sim \mathcal{C}_{P}(\Lambda;\tau) = \sum_{k=0}^\infty \tilde{b}_k \Lambda^{-k-\frac{3}{2}}\left[ \tilde{E}^*\Big(k+\tfrac{3}{2};\tau\Big)-\log(\Lambda)\eisen{k+\tfrac{3}{2}}\right]+\sum_{k=0}^\infty b_k \Lambda^{-k-\frac{3}{2}}\eisen{k+\tfrac{3}{2}}\,,
\end{equation}
where $b_k$ and $\tilde{b}_k$ are factorially growing coefficients given by
\begin{align}
{\tilde{b}}_k &\label{eq:btk} =- \frac{ \Gamma \left(k-\frac{1}{2}\right) \Gamma \left(k+\frac{1}{2}\right) \Gamma \left(k+\frac{3}{2}\right)}{\pi ^3 2^{2{k+1}}\Gamma (k+1)^2}\,,\\
b_k &\label{eq:bk} =-  \left.\frac{\partial}{\partial s}\left( \frac{ \Gamma \left(s-\frac{1}{2}\right) \Gamma \left(s+\frac{1}{2}\right) \Gamma \left(s+\frac{3}{2}\right)}{\pi ^3 2^{2{s+1}}\Gamma (s+1)^2}\right)\right\vert_{s=k} \,.
\end{align}
Due to the presence of double poles, the perturbative expansion of \eqref{eq:C000app} does contain a new family of modular invariant functions given by the derivative of Eisenstein series with respect to its index, i.e.
\begin{equation}\label{eq:Etilde}
\tilde{E}^*(s;\tau)\coloneq\partial_s \eisen{s}\,.
\end{equation}

To evade the difficulties associated with the resummation of the perturbative expansion \eqref{LargeGammaNewPertSeries} containing this new family of modular invariant functions, we introduce a regulator $\epsilon$ as to separate the double poles in \eqref{eq:C000app} and thus define the auxiliary function
\begin{equation}\label{eq:appaux}
    \mathcal{C}(\Lambda,\epsilon;\tau) = \intRepsilon \frac{2^{1-2s}\Gamma(s-2)\Gamma(s-1)\Gamma(s)\tan(\pi s)\tan(\pi (s+\epsilon))}{\pi \Gamma(s-\frac{1}{2})^2} \Lambda^{-s}\eisen{s}\frac{\dd s}{2\pi i}\,.
\end{equation}
The integrand of the above expression has two infinite families of single poles located at $s =k+ \frac{3}{2}$ as well as at $s=k+\frac{3}{2}-\epsilon$ with $k\in\mathbb{N}$. 
To compute the large-$\Lambda$ perturbative expansion of \eqref{eq:appaux} we proceed as before and push the contour of integration to the right half-plane while collecting residues thus arriving at
\begin{align}
\mathcal{C}(\Lambda,\epsilon;\tau)& \sim \mathcal{C}_P(\Lambda,\epsilon;\tau)=\mathcal{C}^{(1)}_{P}(\Lambda,\epsilon;\tau)+\mathcal{C}^{(2)}_{P}(\Lambda,\epsilon;\tau)\,,
\end{align}
where we separated the contributions from the two infinite families of residues in
\begin{align}
    \mathcal{C}^{(1)}_{P}(\Lambda,\epsilon;\tau) &\label{eq:C1app}= -\frac{\cot(\pi \epsilon)}{\pi^2}\sum_{k=0}^\infty\frac{\Gamma(k-\frac{1}{2})\Gamma(k+\frac{1}{2})\Gamma(k+\frac{3}{2})}{4^{k+1}\Gamma(k+1)^2}\Lambda^{-k-\frac{3}{2}}\eisen{k+\tfrac{3}{2}}\,,\\
    \mathcal{C}^{(2)}_{P}(\Lambda,\epsilon;\tau) &\label{eq:C2app}= \frac{\cot(\pi \epsilon)}{\pi^2}\sum_{k=0}^\infty\frac{\Gamma(k-\frac{1}{2}-\epsilon)\Gamma(k+\frac{1}{2}-\epsilon)\Gamma(k+\frac{3}{2}-\epsilon)}{4^{k+1-\epsilon}\Gamma(k+1-\epsilon)^2}\Lambda^{-k-\frac{3}{2}+\epsilon}\eisen{k+\tfrac{3}{2}-\epsilon}\,.
\end{align}

Using the integral representation \eqref{Eisen_largeNp_intRep} we compute the Borel transforms associated with both series \eqref{eq:C1app} and \eqref{eq:C2app},
\begin{align}
    \mathcal{B}[\mathcal{C}^{(1)}_P](t;\epsilon) &\label{eq:B1app}= 2t^2\cot(\pi \epsilon) \tFt{-\tfrac{1}{2}}{\tfrac{1}{2}}{\tfrac{3}{2}}{1}{2}{t^2}\,,\\
    \mathcal{B}[\mathcal{C}^{(2)}_P](t;\epsilon) &\label{eq:B2app}= \frac{2t^{2-2\epsilon}\cot(\pi\epsilon)}{\pi^{\frac{3}{2}}}\frac{\Gamma(-\frac{1}{2}-\epsilon)\Gamma(\frac{1}{2}-\epsilon)\Gamma(\frac{3}{2}-\epsilon)}{\Gamma(1-\epsilon)\Gamma(1-\epsilon)\Gamma(2-\epsilon)}{}_4F_3(1,{-}\tfrac{1}{2}{-}\epsilon,\tfrac{1}{2}{-}\epsilon,\tfrac{3}{2}{-}\epsilon;1{-}\epsilon,1{-}\epsilon,2{-}\epsilon\vert t^2)\,.
\end{align}
We note that the resummation of the original perturbative series \eqref{LargeGammaNewPertSeries} can now be defined through an appropriate limit. Both Borel transforms \eqref{eq:B1app}-\eqref{eq:B2app} are singular as $\epsilon\to 0$ but their sum has a finite limit as $\epsilon\to0$,
\begin{equation}
    \mathcal{B}[\mathcal{C}_{P}](t) = \lim_{\epsilon\to 0}\left(\mathcal{B}[\mathcal{C}^{(1)}_{P}](t;\epsilon)+\mathcal{B}[\mathcal{C}^{(2)}_{P}](t;\epsilon)\right) = f_1(t)+\log(t)f_2(t)\,.\label{eq:BorelApp}
\end{equation}
The two functions $f_1(t),f_2(t)$ can be expressed as power series in the original coefficients $\tilde{b}_k$ and $b_k$ \eqref{eq:btk}-\eqref{eq:bk},
\begin{equation}
    f_1(t) = \sum_{k=0}^\infty \Big[b_k \frac{\Gamma(k+\frac{3}{2})}{\Gamma(2k+3)}4^{2k+2}+\tilde{b}_k\chi\big(k+\tfrac{3}{2}\big)\Big]t^{2k+2}\,,\quad f_2(t) = \sum_{k=0}^\infty \tilde{b}_k\frac{2\Gamma(k+\frac{3}{2})}{\Gamma(2k+3)}(4t)^{2k+2}\,,
\end{equation}
where we defined $\chi(s) \coloneqq \partial_s\big(2^{4s-2} \Gamma(s)/\Gamma(2s)\big)$.
It is easy to show that these series have finite radius of convergence $|t|<1$.
Concretely we have,
\begin{align}
    f_1(t) &\label{eq:f1}= \frac{-2t^2}{\pi^{\frac{5}{2}}} \left[\left( \tfrac{\partial}{\partial \epsilon_1} +  \tfrac{\partial}{\partial \epsilon_2} \right)
    2^{2 \epsilon_1-2 \epsilon_2} \Gamma \left(\epsilon_2-\tfrac{1}{2}\right) \Gamma \left(\epsilon_2+\tfrac{1}{2}\right) \Gamma \left(\epsilon_2+\tfrac{3}{2}\right)\right.\\
    &\phantom{=}\notag \left.\times \, _4\tilde{F}_3\left(1,\epsilon_2-\tfrac{1}{2},\epsilon_2+\tfrac{1}{2},\epsilon_2+\tfrac{3}{2};\epsilon_2+1,\epsilon_2+1,\epsilon_1+2;t^2\right)
    \right]_{\epsilon_1=\epsilon_2=0}\,,\\ 
    f_2(t) &\label{eq:f2}= \frac{4t^2}{\pi}\tFt{-\tfrac{1}{2}}{\tfrac{1}{2}}{\tfrac{3}{2}}{1}{2}{t^2}\,,
\end{align}
where $\,_p\tilde{F}_q$ denotes the regularised hypergeometric function.

At this point we proceed with the resummation of the perturbative series as presented in Section \ref{Mod_inv_resurgence_section}.
Given the Borel transform \eqref{eq:BorelApp}, we define its directional modular invariant Borel resummation as
\begin{equation}
    \mathcal{S}_\theta[\mathcal{C}_{P}](\Lambda;\tau) = \int_0^{e^{i\theta}\infty}\mathcal{B}[\mathcal{C}_{P}](t)\mathcal{E}(\sqrt{\Lambda}t;\tau)\dd t\,.\label{eq:directionalApp}
\end{equation}
From the explicit expressions \eqref{eq:f1}-\eqref{eq:f2} it appears manifest that the integral above is well-defined for ${\rm Re}(\Lambda)>0$ provided that direction of integration $\theta \in (-\pi/2,\pi/2)$ with $\theta \neq 0$, since $\theta=0$ is a Stokes direction. 

Schematically the discontinuities of the functions $f_1(t)$ and $f_2(t)$ defined in \eqref{eq:f1}-\eqref{eq:f2} take the form
\begin{equation}\label{eq:discApp}
   {\rm Disc}_0 f_a(t) \coloneqq \lim_{\epsilon\to 0^+} [f_a(t+i \epsilon)-f_a(t-i \epsilon)] = -2\pi i g_a(1-t) = -2\pi i \sum_{k=0}^\infty \frac{d^{(a)}_k}{k!} (t-1)^k\,,
\end{equation}
where $a\in \{1,2\}$ and the above expression is valid for $t>1$.
The coefficients $d^{(a)}_k$ and hence the discontinuities of the functions $f_a(t)$ with $a\in\{1,2\}$ can be derived from the integral representation of the hypergometric functions appearing in \eqref{eq:f1}-\eqref{eq:f2}.
We will not present the precise details of the calculation here since we are only interested in understanding the scale of the non-perturbative corrections rather than the small fluctuations around them, especially given that it is presently not even known whether the non-perturbative scale we will shortly derive has any holographic interpretation in this very heavy large-charge regime.

From the directional Borel resummation \eqref{eq:directionalApp}, we use \eqref{eq:discApp} to derive the discontinuity in lateral resummations,
\begin{align}
    \big(\mathcal{S}_+-\mathcal{S}_-\big)\big[\mathcal{C}_P\big]\big(\Lambda;\tau\big) &= (-2\pi i)\int_1^\infty \big(g_1(1-t)+\log(t)g_2(1-t)\big)\mathcal{E}(\sqrt{\Lambda}t;\tau)\dd t\\
    &\notag  =:(-2\pi i )\mathcal{S}_0[\mathcal{C}_{N\!P}](\Lambda;\tau)\,.
\end{align}
We then proceed to shift the contour of integration $t\to t+1$, and evaluate the $t$-integral term by term using \eqref{eq:discApp} thus manifesting the presence of a novel non-perturbative scale,
\allowdisplaybreaks{
\begin{align}
    \mathcal{S}_0[\mathcal{C}_{N\!P}](\Lambda;\tau) &\notag = \int_0^\infty \big[g_1(-t)+\log(t+1)g_2(-t)\big]\mathcal{E}(\sqrt{\Lambda}(t+1);\tau)\dd t
    \\*
    &\label{eq:NPapp} = \sum_{k=0}^\infty d_k \Dfn{\Lambda}{\tfrac{k+1}{2}}+\sum_{k=1}^\infty \sum_{\ell=1}^k \frac{(-1)^{\ell+1}k!}{\ell(k-\ell)!}\tilde{d}_{k-\ell}\Dfn{\Lambda}{\tfrac{k+1}{2}}\,.
\end{align}}
Interestingly, although the starting asymptotic perturbative expansion \eqref{LargeGammaNewPertSeries} did contain a new class of modular-invariant functions $\tilde{E}(s;\tau)$ defined in \eqref{eq:Etilde}, the form of the non-perturbative corrections remains unchanged and only contains functions of the form $\Dfn{\Lambda}{\frac{k+1}{2}}$.

We stress again that the non-perturbative corrections obtained in \eqref{eq:NPapp} define a new non-perturbative scale
 for the integrated correlator in the very heavy regime where $N\gg1$ and $p = \alpha N^\gamma$ with $\gamma>2$ and $\alpha>0$ fixed.
Reinstating the $N$ dependence using \eqref{eq:LambdaApp}, we can write schematically
\begin{equation}
\mathcal{C}_{N\!P}(N,\alpha;\tau)  \sim \exp\left( - 8 N^{\frac{\gamma-1}{2}} \sqrt{\alpha Y_{mn}(\tau)} \right) \,,\label{eq:CNPApp}
\end{equation}
with $Y_{mn}(\tau) = \pi|n\tau+m|^2 /\tau_2$ and $(m,n)\in \mathbb{Z}^2 \setminus\{(0,0)\}$.
It would be worth analysing more in detail whether there is a holographic interpretation of the very heavy maximal-trace operators as to derive a possible semi-classical interpretation for these non-perturbative effects.

\section{Coherent-state vs maximal-trace with $p=\alpha N^2$ at large $N$ and genus one}
\label{comparison_Nsq_coherent_app}

In Section \ref{sec:LLM} we found that at the leading order in the large $N$ expansion the Borel transform for the coherent-state integrated correlator with parameter $\alpha$ and that of the maximal-trace integrated correlator in the double-scaling regime $p=\tilde{\alpha}N^2$ are connected by the very simple reparametrisation $\tilde{\alpha}= \frac{2\alpha^2}{1-2\alpha^2}$, as displayed in equation (\ref{eq:BorelId}). In this appendix, we show at the next order in $1/N$ such a simple relation ceases to hold.

We begin by looking at the genus-one integrated HHLL correlator $\mathcal{C}^{(1)}_{\HHLL}(N,\alpha;\tau)$ as defined in \eqref{eq:Cgenus}. At large-$N$ this quantity admits the perturbative asymptotic expansion
\begin{align}
    &\notag\mathcal{C}^{(1)}_{\HHLL}(N,\alpha;\tau) \sim \mathcal{C}^{(1)}_{\HHLL,P}(N,\alpha;\tau) = \\
    &\sum_{k=0}^\infty \frac{(2k+13)(k+1) \Gamma(k+\frac{1}{2})\Gamma(k+\frac{5}{2})}{24\pi^{\frac{3}{2}}\cdot 2^{2k+3}\Gamma(k+3)}\left[1-\tFo{k+\tfrac{3}{2}}{-k-\tfrac{1}{2}}{1}{\tfrac{2\alpha^2}{2\alpha^2-1}}\right]N^{-k-\frac{3}{2}}\eisen{k+\tfrac{3}{2}}\,.
\end{align}
Notice that this expansion is structurally very similar to the genus-0 case (\ref{eq:C0asy}), the main difference coming from the factorially divergent perturbative coefficients being multiplied by a slightly different rational function of $k$. From the definition \eqref{eq:BorelDef}, we see that its associated Borel transform simplifies to
\begin{equation}
\label{eq:B1HHLL}
    \mathcal{B}[\mathcal{C}^{(1)}_{\HHLL,P}](\alpha,t) = \sum_{k=0}^\infty \frac{(2k+13)\Gamma(k+\frac{1}{2})\Gamma(k+\frac{5}{2})}{24\pi(k+2)\Gamma(k+1)^2}\left[1-\tFo{k+\tfrac{3}{2}}{-k-\tfrac{1}{2}}{1}{\tfrac{2\alpha^2}{2\alpha^2-1}}\right]t^{2k+2}\,.
\end{equation}
If we compare the above expression with the similar Borel series  \eqref{eq:BorelId} for the leading contribution to the HHLL correlator we easily see that the two expressions are related by a simple integro-differential operator in the $t$ variable.
As a consequence we have that at the sub-leading order there are no fundamental modifications to the singularity structure of the Borel transform, and hence to the exponentially suppressed non-perturbative corrections to the sub-leading contribution.

The expression for the sub-leading contribution to the regime $p={\tilde{\alpha}} N^2$ is far more involved.
If we specialise the general expression \eqref{eq:Cg} to $g=1$ we find the spectral representation
\begin{equation}
    \mathcal{C}_{{\tilde{\alpha}}}^{(1)}(N;\tau) = 2\intRepsilon \mathcal{A}^{(1)}(s;{\tilde{\alpha}}) N^{-s}\eisen{s}\frac{\dd s}{2\pi i}\,.\label{eq:C1subapp}
\end{equation}
The spectral overlap $\mathcal{A}^{(1)}(s;{\tilde{\alpha}})$ has been presented in \eqref{eq:A1} in terms of the building blocks \eqref{eq:M0}-\eqref{eq:M1} and \eqref{eq:G0}-\eqref{eq:G1} and it can be simplified to\allowdisplaybreaks{
\begin{align}
    \notag\mathcal{A}^{(1)}(s;{\tilde{\alpha}}) = \frac{2^{-3-2s}(2s-1)\Gamma(s-2)\Gamma(s+1)}{3\sqrt{\pi}\Gamma(s-\frac{1}{2})}\left\lbrace 24(2{\tilde{\alpha}}+1)\tFo{2-s}{s+1}{2}{-{\tilde{\alpha}}}\phantom{\frac{(s)}{(s)}}\right.\\*
    \left.-24({\tilde{\alpha}}+1)\tFo{2-s}{s+1}{1}{-{\tilde{\alpha}}}+\frac{(2-s)(5+s)}{1+2s}\big[\tFo{s}{1-s}{1}{-{\tilde{\alpha}}}-1\big]\right\rbrace\tan(\pi s)\,.
\end{align}}

At this point we proceed as above. We move the contour of integration in \eqref{eq:C1subapp} to the right half-plane and pick up the residues from the poles located at $s=k+\frac{3}{2}$ with $k\in \mathbb{N}$. In this way we obtain the factorially divergent power series expansion $\mathcal{C}^{(1)}_{{\tilde{\alpha}},P}(N;\tau)$ to the sub-leading order maximal-trace correlator in the double scaling regime.
Rather than writing this expression, we present here directly its associated Borel transform computed using  \eqref{eq:BorelDef} which we simplified to
\begin{align}
    &\mathcal{B}[\mathcal{C}_{{\tilde{\alpha}},P}^{(1)}](t) \notag= \mathcal{B}\big[\mathcal{C}_{\HHLL}^{(1)}\big]({\alpha},t)\Big\vert_{{\alpha} = \sqrt{\frac{{\tilde{\alpha}}}{2({\tilde{\alpha}}+1)}}}+\\
    &\label{eq:B1Heavy}\sum_{k=0}^\infty \frac{4\Gamma(k-\frac{1}{2})\Gamma(k+\frac{5}{2})}{\pi\Gamma(k+1)^2} \left[(2{\tilde{\alpha}}+1)\tFo{k+\tfrac{5}{2}}{\tfrac{1}{2}-k}{2}{-{\tilde{\alpha}}}-({\tilde{\alpha}}+1)\tFo{k+\tfrac{5}{2}}{\tfrac{1}{2}-k}{1}{-{\tilde{\alpha}}}\right]t^{2k+2}\,.
\end{align}

It is striking to note that the first term in the above equation is precisely the Borel transform for the sub-leading HHLL correlator presented in \eqref{eq:B1HHLL} where the parameters between the two correlators are mapped using exactly the same (inverse) map found in \eqref{eq:BorelId} for the leading order contributions to both correlators.
However, we crucially see that the Borel transform \eqref{eq:B1Heavy} for the sub-leading contribution to the maximal-trace correlator in the double-scaling regime does contain additional terms which are not related to the corresponding contribution \eqref{eq:B1HHLL} to the HHLL correlator via a map in parameters ${\tilde{\alpha}} \to {\alpha} = f({\tilde{\alpha}})$.

\section{On the ${\rm AdS}$ bubble geometry}\label{app:AdSbubble}

In the main text, specifically in Section \ref{sec:coherent} 
and \ref{sec:LLM}, we referred to the ${\rm AdS}$ bubble geometry 
and the more general LLM geometries. As a guide to the reader, and in 
order to provide more context for our discussion, 
we collect  in this appendix a number of results on the gravity 
description of the ${\rm AdS}$ bubble.
Our discussions follows the original paper \cite{Liu:2007xj}. For further details we refer the reader to the related works \cite{Giusto:2024trt,Ganchev:2025dzn}
and \cite{Aprile:2011uq}.

As usual, the starting point for the dual holographic theory is ten-dimensional type 
IIB supergravity. In this context, we are interested 
in a truncation to five-dimensional ${\cal N} = 8$ gauged supergravity 
that retains the  three $U(1)$ gauge fields in the maximal torus of the 
$SO(6)$ gauge group, 
\begin{equation}\label{matrixAfields}
A= \left(\begin{array}{ccc} A_{3}  &  & \\ & A_{2}  &  \\ & & A_{1} \end{array}\right)\otimes i\sigma_2\,.
\end{equation}
The matter fields are instead arranged into the matrices, 
\begin{equation}\label{matrixT}
T=\left(\begin{array}{ccc} X_3 {\cal M}_3 &  & \\ & X_2 {\cal M}_2 &  \\ & & X_1 {\cal M}_1\end{array}\right),\qquad
{\cal M}_i= \left( \begin{array}{cc} \cosh\eta_i +\sinh\eta_i\cos\theta_i & \sinh\eta_i\sin\theta_i \\ \sinh\eta_i \sin\theta_i & \cosh\eta_i-\sinh\eta_i\cos\theta_i \end{array}\right)\,.
\end{equation}
The matrix $T$ has unit determinant, thus implying the condition $X_1X_2X_3=1$ since ${\rm det} ({\cal M}_i) = 1$ for $i=1,2,3$. Subject to this constraint, we can parametrise the three fields $X_{i}$ in terms of two neutral scalars, 
$\varphi_{a}$ with $a=1,2$, 
\begin{equation}
X_1=e^{-\frac{2}{\sqrt{6}}\varphi_1} ,\qquad X_2=e^{\frac{1}{\sqrt{6}}\varphi_1+\frac{1}{\sqrt{2}}\varphi_2},\qquad X_3=e^{\frac{1}{\sqrt{6}}\varphi_1-\frac{1}{\sqrt{2}}\varphi_2}\, .
\end{equation}
The three charged scalar fields {contained in the three matrices $\mathcal{M}_j$} are expressed through a Stueckelberg parametrisation $(\eta_j,\theta_j)$,~i.e.~$\phi_j = \eta_j e^{i \theta_j}$ with $j=1,2,3$. 

To summarise all the fields introduced so far we have: the metric, 
the gauge fields and the scalar fields. Importantly, the fields 
here considered all belong to the supergraviton multiplet.
We also note that {the reduced set of} fields $\{g_{\mu\nu},A_{i=1,2,3},
\varphi_{a=1,2}\}$ are in fact those corresponding to the so-called `STU' model which is an ${\cal N}=2$ supergravity coupled 
to two vector multiplets. The charged scalar fields {$(\eta_i,\theta_i)$}, 
on the other hand, lie outside of the  ${\cal N}=2$ supergravity formalism.

The five-dimensional action can be written as follows, 
\begin{equation}\label{lagrangian5}
{\cal L}_5= \sqrt{g}\, \Bigg[ R - \tfrac{1}{2} \sum_{a=1}^2 (\partial\phi_a)^2 
- \tfrac{1}{4}\sum_{i=1}^3 X_i^{-2} F^{i,\mu\nu}F^i_{\mu\nu}
-\tfrac{1}{2}\sum_{i=1}^3 [ (\partial\eta_i)^2 + \sinh^2\!\eta_i \,(\partial\theta_i+ q A_i)^2] - V\Bigg]\,,
\end{equation}
with $q=2$ and where the potential term $V$ is given by
\begin{equation}\label{superpotential}
V\coloneqq 2\Bigg[ 
\sum_{i=1}^3 \left(\frac{\partial W}{\partial \eta_i }\right)^2 + \left(\frac{\partial W}{\partial \varphi_1} \right)^2+ \left(\frac{\partial W}{\partial \varphi_2 }\right)^2\Bigg]-\frac{4}{3}W^2,\qquad
\qquad {\rm with}\quad W\coloneqq \frac{1}{2}\sum_{i=1}^{3}  q X_i\cosh(\eta_i)\,.
\end{equation}
Explicitly, the potential $V$ takes the form
\begin{equation}
V= \frac{2}{L^2} \Bigg[ \sum_{i=1}^{3} X_i^2 \sinh^2\! \eta_i - \sum_{i\neq j} X_i X_j \cosh\eta_i\,\cosh\eta_j\Bigg]\,.
\end{equation}
We shall use the superpotential {$W$} later on when we compute the holographic vacuum expectation values (\textit{vevs} in what follows) corresponding to the ${\rm Ads}$ bubble solution.

The Lagrangian \eqref{lagrangian5} has an ${\rm AdS}_5$ solution by setting to zero the scalar and gauge fields vevs,
\begin{center}
${\rm Ads}$ vacuum :  $\qquad\qquad \varphi_{a=1,2}=\eta_{j=1,2,3}=\theta_{j=1,2,3}=A_{j=1,2,3}=0$. 
\end{center}
In this case, the potential simplifies to $V=-12/L^2$ thus simply 
producing the negative cosmological constant of AdS.
By computing the quadratic fluctuations to the potential $V$ around 
the ${\rm AdS}$ vacuum, we see that the scalar fields {$\varphi_{a=1,2}$ 
and $\eta_{j=1,2,3}$} have a diagonal mass matrix with five identical 
eigenvalues given by  $m^2L^2=-4$. At the boundary, the Stueckelberg 
fields $(\eta_i,\theta_i)$ can be put 
together into a standard charged complex scalar.
Then, through the well known formula 
$m^2L^2=\Delta(\Delta-4)$ we find that these scalar 
fields correspond to the real and the charged components of the 
half-BPS chiral primary operators in ${\cal N}=4$ SYM 
with scaling dimension  $\Delta=2$.  Since on the supergravity side we are only 
retaining fields in the graviton multiplet, it follows that in the dual CFT the
corresponding operators must belong to the stress-tensor multiplet and thus 
are given by harmonics of the $20'$ representation of the $R$-symmetry group,~i.e. 
\begin{equation}
{\cal O}_2\sim C^a_{IJ}\, {\rm Tr}\,\Phi^I\Phi^J \,,
\end{equation}
where $\Phi_I$, with $I=1,\ldots, 6$, are the six scalar 
fields,of the $\mathcal{N}=4$ multiplet and with $C^{a=1,\ldots 20}_{IJ}$ 
a symmetric and traceless matrix. We shall come back to this point shortly.
The bulk gauge fields correspond to components of 
the $R$-symmetry current belonging to the same stress-tensor multiplet 
for which ${\cal O}_2$ is the bottom $su(4)$ component. 

Rather than imposing the usual ${\rm AdS}_5$ supergravity background, we are interested in studying different supersymmetric solutions to the equations of motion derived from \eqref{lagrangian5}. In 
particular, we are interested in solutions which are charged under the $U(1)$ gauge fields and which satisfy the BPS 
condition\footnote{We remind the reader that the spectrum
of dimensions of operators coincides with the spectrum of energies with respect to time in global ${\rm AdS}$.} \emph{energy}=\emph{charge}.
These BPS solutions have been studied in \cite{Cvetic:2000nc,Chong:2004ce}. 

The ${\rm Ads}$ bubble solution corresponds to the fields configurations
\begin{equation}\label{5dmetricAdSbubble0}
A_{1}=\frac{dt}{{\faH}(r)},\qquad \varphi_1=\sqrt{\frac{2}{3}}\log {\faH}(r) ,\qquad \eta_1= {\rm arccosh}\left[ \frac{1}{2r}\frac{d}{dr}\big[ r^2 {\faH(r)}\big]\right],\qquad \theta_1=0\,,
\end{equation}
while all the remaining fields have been set to zero, i.e. $A_{2}=A_{3}=0,\ \varphi_2=0,\ \eta_{j=2,3}=\theta_{j=2,3}=0$. The five-dimensional metric is given by 
\begin{equation}\label{5dmetricAdSbubble}
ds^2_5= g_{\mu\nu} dx^{\mu} dx^{\nu}= - {\faH}(r)^{-\frac{2}{3}} \left(1+{r^2}{} {\faH}(r)\right) dt^2 + {\faH}(r)^{\frac{1}{3}} \left( \frac{dr^2}{1+{r^2}{}{\faH}(r)} + r^2 d\Omega_3^2 \right)\,,
\end{equation}
where the auxiliary function $\faH$ takes the form
\begin{equation}
{\faH}(r)=-\frac{1}{r^2} +\sqrt{ \left(1+\frac{1}{r^2}\right)^{\!2} +\frac{2Q}{r^2}}\,, \label{eq:faH}
\end{equation}
with $Q$ a real parameter. We will later interpret $Q$ through the AdS/CFT dictionary.

Since the boundary of our spacetime is the region where $r\to \infty$, 
we can easily expand in Taylor the above expression
\begin{equation}
{\faH}(r)=1+\frac{Q}{r^2} +O(r^{-4}) \,, \qquad\qquad  r\to \infty\,,
\end{equation}
thus implying that the metric \eqref{5dmetricAdSbubble} asymptotes ${\rm AdS}$ 
in global coordinates as $r \to \infty$. The centre of the space is located 
at $r=0$, where the function $\faH(r)$ admits the expansion,
\begin{equation}
{\faH}(r)=1+Q-\frac{1}{2}[ Q(2+Q)] r^2 +O(r^4) \,, \qquad \qquad  r\to 0\,.
\end{equation}
From \eqref{5dmetricAdSbubble} we then note that the metric is smooth at $r=0$. 

From a bottom-up perspective, the function ${\faH}(r)$ is defined 
as the solution to a second order ODE obtained from Einstein's equation.
It is important to stress that from this point of view, the function 
\eqref{eq:faH} is actually a particular element of a larger family 
of solutions to this ODE. The general solution, here denoted by 
${\faH}_C(r)$, is labelled by an additional constant of integration 
$C$ and takes the form 
\begin{equation}\label{faHwithC}
{\faH}_C(r)=-\frac{1}{r^2} +\sqrt{ \left(1+\frac{1}{r^2}\right)^{\!2} +\frac{2Q_\sstar}{r^2}+\frac{C^2-1}{r^4}}\,.
\end{equation}
The ${\rm AdS}$ bubble corresponds to the particular case $C=1$ and $ Q_\sstar=Q$ and 
it is the only solution which gives rise to a smooth geometry at the centre of space $r=0$. 

We note that if instead of imposing the fields configurations \eqref{5dmetricAdSbubble0} 
{for the field $\eta_1$} we were to require that $\eta_1=0$, we would have 
found a different value for the integration constant $C\neq 1$ in \eqref{faHwithC}, 
namely $C=Q_\sstar+1$ so that 
\begin{equation}
\eta_1=0
\qquad\rightarrow\qquad
{\faH}_{\sstar}(r)=1+\frac{Q_{\sstar}}{r^2}\,.
\end{equation}
This is another known solution to the equations of motion derived from 
\eqref{lagrangian5} sometimes called \textit{superstar}. 
However, we note that  in this case the metric \eqref{5dmetricAdSbubble} 
corresponding to the superstar has a singularity at the centre of space $r=0$.

Since we are also interested in the dual CFT side, we highlight 
the holographic dictionary for each of the fields 
$\{g_{\mu\nu}, A_{1}, \varphi_1,\eta_1\}$ turned on 
in \eqref{5dmetricAdSbubble0}-\eqref{5dmetricAdSbubble}. 
In particular we have the mapping:
\vspace{-0.2cm}
\begin{itemize}
\item[$\bullet$]
the metric tensor $g_{\mu\nu}$ is dual to the stress tensor $T_{\mu\nu}$;
\item[$\bullet$]
the gauge field $A_{1}$ is dual to the time component of the $U(1)$ factor for the $R$-symmetry current $J_\mu$;
\item[$\bullet$]
the two scalar fields $\varphi_1$ and $\eta_1$ are dual to the operators 
\begin{align}
\varphi_1\sim&\label{eq:phiPhi}\   {\rm tr}( 2 X\!\bar{X} - Y \bar{Y}- Z\bar{Z})\rule{1cm}{0pt} {\rm \,diagonal}\\
\eta_1 e^{i\theta_1}\sim& \label{eq:etPhi} \  {\rm tr}( X^2 ) 
\rule{3.35cm}{0pt} {\rm non-diagonal}
\end{align}
where $X\coloneqq \Phi_1+i\Phi_2$, {$Y\coloneqq \Phi_3+i\Phi_4$ and $Z\coloneqq \Phi_5+i\Phi_6$ }. 
\end{itemize}

The notation `diagonal' and `non-diagonal' makes reference 
to the two three-point function topologies 
that we introduced in \eqref{topologies}. 
With this nomenclature, we have in mind the distinct configurations for the two distinct three-point functions
\begin{equation}\label{topology2}
\begin{array}{ccc}
\begin{tikzpicture}[scale=1.3]  
\def\shift{.35}

\def\latoxuno{-.35}
\def\latoxdue{-.38+1.5}
\def\latoyuno{.7}
\def\latoydue{-.25}

\foreach \x in {0,.07}
\draw (\latoxuno-.02+\x,\latoyuno) -- (\latoxuno-.02+\x,\latoydue);

\foreach \x in {.14,.21,.28}
\draw[fill=black] (\latoxuno+.03+\x,\latoyuno/2+\latoydue/2) circle (.3pt);

\foreach \x in {.42,.49,.56}
\draw (\latoxuno+.02+\x,\latoyuno) --(\latoxuno+.02+\x,\latoydue);

\draw (\latoxuno-.02+.68,\latoyuno) -- (\latoxuno-.02+1.2, \latoyuno/2+\latoydue/2);

\draw (\latoxuno-.02+.68,\latoydue) -- (\latoxuno-.02+1.2, \latoyuno/2+\latoydue/2);

\draw  (\latoxuno+.25,\latoydue-.2) node[scale=.6] {$[\bar{X}... \bar{X} \bar{X}]$};
\draw  (\latoxuno+.25,\latoyuno+.2) node[scale=.6] {$[X... X X]$};

\draw (\latoxuno-.02+1.5, \latoyuno/2+\latoydue/2) node[scale=.6] {$X \bar{X}$};
\end{tikzpicture} 
 & \rule{2cm}{0pt} &
\begin{tikzpicture}[scale=1.3]  
\def\shift{.35}

\def\latoxuno{-.35}
\def\latoxdue{-.38+1.5}
\def\latoyuno{.7}
\def\latoydue{-.25}

\foreach \x in {0,.07}
\draw (\latoxuno-.02+\x,\latoyuno) -- (\latoxuno-.02+\x,\latoydue);

\foreach \x in {.14,.21,.28}
\draw[fill=black] (\latoxuno+.03+\x,\latoyuno/2+\latoydue/2) circle (.3pt);

\foreach \x in {.42,.49}
\draw (\latoxuno+.02+\x,\latoyuno) --(\latoxuno+.02+\x,\latoydue);

\draw (\latoxuno-.02+.68-.01,\latoyuno+.01) -- (\latoxuno-.02+1.2, \latoyuno/2+\latoydue/2);

\draw (\latoxuno-.02+.68,\latoyuno+.08) -- (\latoxuno-.02+1.2, \latoyuno/2+\latoydue/2+.08);

\draw  (\latoxuno+.25,\latoydue-.2) node[scale=.6] {$[\bar{X}... \bar{X} \bar{X}]$};
\draw  (\latoxuno+.25,\latoyuno+.2) node[scale=.6] {$[X... X X]$};

\draw (\latoxuno-.02+1.35, \latoyuno/2+\latoydue/2) node[scale=.6] {$X^2$};
\end{tikzpicture} 
\end{array}
\end{equation}
In an asymptotically AdS background, the normalisable mode of a bulk field away from the boundary
computes the expectation value of the corresponding dual operator 
in the heavy background. This is the case for the AdS bubble and the fields listed above.  
As we will shortly review, these expectation values can then be translated to three-point functions. 
In this language, the reference to `diagonal' and `non-diagonal' has the same meaning as in \eqref{topologies}.

For later sections, we find it convenient to rephrase part of the supergravity field content 
in the language of ${\cal N}=2$ gauged supergravity. In particular, the fields $\{g_{\mu\nu},
A_{i=1,2,3},\varphi_{a=1,2}\}$ correspond to the ${\cal N}=2$ graviton multiplet coupled 
to two vector multiplets, thus in total we have  one gravi-photon, two gauge fields and two 
neutral scalar fields. In our setup \eqref{5dmetricAdSbubble0}, we have turned on only a single 
gauge field and since the fields $A_{1}$ and $\varphi_1$ belong to the same multiplet we expect 
them to carry the same information regarding the holographic vevs they will source. 
In contrast, we expect the charged scalar fields to give rise to independent CFT data. 

\subsection{The two-point function frame and the vevs}

The metric given in \eqref{5dmetricAdSbubble} 
is asymptotic to AdS in global coordinates when $r\to \infty$. 
Correspondingly, in this frame the dual CFT is defined on an $\mathbb{R}\times {\rm S}^3$ manifold. In the context of 
holographic correlators, it is useful to perform a change of coordinates 
such that the geometry is asymptotic to ${\rm Ads}$ in the Poincare patch, so that the dual CFT lives on $\mathbb{E}^4$, 
the euclidean boundary of the Poincar\'e patch. This change of coordinates 
is achieved by a map called 
Global-to-Poincar\'e (GtP) given in \cite{Abajian:2023jye}, which reads 
\begin{equation}\label{GtPmap}
t=\frac{{i}}{2}\log(z^2+R^2)\,, \qquad \qquad r=\frac{R}{z}\,.
\end{equation}
{Here $z>0$ denotes the usual ${\rm AdS}$ radial coordinate while $R\geq0 $ corresponds to the radial coordinate on the
Poincar\'e boundary.}

Under this map, the geodesic at the centre of global ${\rm Ads}$, defined by 
$r=0$ and all times $ t \in \mathbb{R}$, is mapped to the geodesic that 
shoots from the origin of $\mathbb{E}^4$ towards the point 
at infinity passing through the bulk, thus having $R=0$ for all $z>0$. 
It is important to notice that the metric has two marked points 
at the Poincar\'e boundary. In our case $0$ and $\infty$. Crucially, 
we interpret these two marked points as the insertion 
of the heavy operators which source the bulk geometry. In this 
frame the ${\rm AdS}$ bubble can be understood as a two-point function geometry while the 
holographic vev of a light field corresponds to the three-point function between 
the two heavy operators and the corresponding 
light operator in the dual CFT.

In this section we generalise the procedure of \cite{Abajian:2023jye}  
as to include a gauge field. We shall apply the GtP map \eqref{GtPmap} to the 
${\rm AdS}$ bubble solution given in \eqref{5dmetricAdSbubble0}-\eqref{5dmetricAdSbubble},
and work with coordinates $(z,R)$. 
In \cite{Abajian:2023jye} this frame has been referred to as  `the cone' since at fixed-$r$ 
the map $r=\frac{R}{z}$ provides
a foliation of the geometry given by concentric cones in the coordinates 
$(z,R)$. 

The action of the GtP map \eqref{GtPmap}  on the scalar fields is trivial. 
Less so is the effect of the GtP map on the metric and the gauge field.
For the metric defined in \eqref{5dmetricAdSbubble} we find
\begin{equation}
ds_5^2= \frac{1}{z^2}\left[ \frac{dz^2}{h} + h\left(dR+\frac{R}{z} v_z dz\right)^2+ {{\faH}}^{\frac{1}{3}} R^2 d\Omega_3^2\right]\,,
\end{equation}
with $H= H(\tfrac{R}{z})$ and where we introduced the auxiliary quantities
\begin{equation}
h = h(z,R)\coloneqq \left( \frac{1}{g_{tt}}+ \frac{r^2 \faH^{\frac{1}{3}}g_{tt} }{(1+r^2)^2} \right)\Bigg|_{r=\frac{R}{z}}\,,\qquad \qquad
v_z=v_z(z,R) = \frac{1}{h}\left( -\frac{1}{g_{tt}}+ \frac{ \faH^{\frac{1}{3}}g_{tt} }{(1+r^2)^2}\right)\Bigg|_{r=\frac{R}{z}}\,.
\end{equation}
For the gauge field given in \eqref{5dmetricAdSbubble0} we instead have
\begin{equation}\label{gauge_field1}
A_1(z,R)=-\frac{i}{\faH(\frac{R}{z})}\frac{ R dR +  z dz}{(R^2+z^2)} = -\frac{idR}{\faH(\frac{R}{z})} + d\left( \int_1^{\frac{R}{z}}\frac{i}{\faH(r)(r^2+1)} \frac{\dd r}{r}\right)\,.
\end{equation}

At this point, it is useful to comment on the effect caused by gauge transformations as these will be needed when we study the holographic vevs. 
In particular, we note that it should be possible to remove the total derivative appearing in the 
above \eqref{gauge_field1} by performing a suitable gauge transformation. 
However, in the ${\rm AdS}$ bubble solution \eqref{5dmetricAdSbubble0} 
we have turned on the scalar field $\eta_1$ which is charged under the gauge field $A_1$. 
Under a gauge transformation we want the equations of motion to remain unchanged hence, 
to compensate for a gauge transformation, we need
to modified the solution in \eqref{5dmetricAdSbubble0} by turning on the Stueckelberg field $\theta_1$, which appears 
in the Lagrangian \eqref{lagrangian5} via the coupling $(\partial\theta_1+ q A_1)^2$. 

Before being able to relate the background supergravity solution to the dual CFT 
side we have to discuss one last issue. Holographic vevs are computed by going to the 
so-called FG gauge, as to ensure a good notion of ${\rm AdS}$ boundary \cite{Skenderis:2002wp}.
The FG gauge is defined by requiring that the metric takes the following form,
\begin{equation}\label{eq:FGmet}
ds_5^2= \frac{d\mathbb{z}^2}{\mathbb{z}^2} + \mathbb{h}_{ij}(\mathbb{z},\mathbb{r}) d\mathbb{x}^i d\mathbb{x}^j \,,
\end{equation}
where the metric $\mathbb{h}_{ij}$ denotes the boundary metric. 
The first few orders of the FG expansion are 
\begin{align}
\label{FGexpansion1}
z(\mathbb{z},\mathbb{r})&= \mathbb{z} \left[ 1+ \frac{Q}{6} \frac{ \mathbb{z}^2 }{ \mathbb{r}^2 } -\frac{ Q(12+Q) }{36}\frac{\mathbb{z}^4}{\mathbb{r}^4} +\frac{35 Q^2(6+Q)+24 Q(45-2Q)}{1944} \frac{\mathbb{z}^6}{\mathbb{r}^6} + O\left( \frac{\mathbb{z}^8}{\mathbb{r}^8}\right) \right] \,, \\[.2cm]
\label{FGexpansion2}
R(\mathbb{z},\mathbb{r})&= \mathbb{r}\left[1 +  0 \cdot \frac{ \mathbb{z}^2 }{ \mathbb{r}^2 } - \frac{Q}{6} \frac{\mathbb{z}^4}{\mathbb{r}^4} +\frac{Q(12+Q)}{27}\frac{\mathbb{z}^6}{\mathbb{r}^6} + O\left( \frac{\mathbb{z}^8}{\mathbb{r}^8} \right)  \right]\,.
\end{align}
The differentials, $dz,dR$ are related to $d\mathbb{z},d\mathbb{r}$ 
by a non diagonal matrix. We will not write $\mathbb{h}_{ij}$ explicitly, 
but the first few orders in the $\mathbb{z}$ expansion can be  obtained in a straightforward manner from the formulae given above. 

Equipped with the FG expansion, we shall now use the holographic dictionary and compute the 
boundary expansion of the various fields of the ${\rm AdS}$ bubble 
thanks to which we can read off the vevs of the dual CFT operators.

\vspace{0.2cm}
{\bf \underline{The holographic vev for the stress tensor}}

Let us introduce the extrinsic curvature $\mathbb{K}_{ij}(\mathbb{r})$ of the 
boundary metric $\mathbb{h}_{ij}(\mathbb{r})$, 
\begin{equation}
\mathbb{K}_{ij}\coloneqq \frac{1}{2}\, \mathbb{z}\partial_{\mathbb{z}} \mathbb{h}_{ij}\,,\qquad\qquad {\rm and}\qquad \mathbb{K}\coloneqq \mathbb{h}^{ij} \mathbb{K}_{ij}\,.
\end{equation}
The extrinsic curvature is related to the boundary stress tensor
but unfortunately it diverges as $\mathbb{z}\rightarrow 0$ and requires a renormalisation scheme.
In the particular case where the boundary is flat, we have that the vev of the holographic stress energy tensor is simply given by
\cite{Batrachenko:2004fd},
\begin{equation}
\langle\!\langle T_{ij}(\mathbb{r})\rangle\!\rangle_{\rm AdS} = 
\frac{1}{8\pi G_N}\lim_{\epsilon\rightarrow 0}\  \frac{1}{\epsilon^{2}}
\Bigg[ \mathbb{K}_{ij} -\mathbb{K}\, \mathbb{h}_{ij} -  \mathbb{h}_{ij} W \Bigg]_{\mathbb{z} = \epsilon}\,,
\end{equation}
{where $G_N$ denotes the five-dimensional Newton constant.}
We note that the last term inside the limit is precisely the contribution originating from the 
counter-terms, and crucially it depends on the superpotential $W$ defined
in \eqref{superpotential}. The final result reads
\begin{equation}\label{holovevT}
\langle\!\langle\, T(\mathbb{r})\,\rangle\!\rangle_{\rm AdS}= \frac{Q}{8\pi G_N}\frac{1}{\mathbb{r}^4}\ 
\left(-{d\mathbb{r}^2} + \frac{\mathbb{r}^2}{3}d\Omega_3^2
\right) \,.
\end{equation}
Note that $\langle\!\langle\, T\,\rangle\!\rangle_{\rm AdS}$ is conserved and traceless (with respect to the boundary metric). 

\vspace{0.2cm}
{\bf \underline{The holographic vev for the $R$-symmetry current}}

The vev of the $U(1)$ part of the $R$-symmetry current is related to the boundary expansion of the gauge field $A_1$, through the relation 
\begin{equation}\label{eq:Jvev}
\langle\!\langle J(\mathbb{r})\,\rangle\!\rangle_{\rm AdS} = \frac{1}{8\pi G_N}\lim_{\epsilon\rightarrow 0}\  \frac{1}{\epsilon^{2}}\Big[ A_1\, \Big]_{\mathbb{z} = \epsilon}\,.
\end{equation}
The gauge field $A_1$ was presented in \eqref{gauge_field1} and reads\footnote{We remind the reader that total derivative appearing in \eqref{gauge_field1} has been removed by performing a gauge transformation.}
\begin{equation}
A_1(z,R)= -\frac{i\,dR}{\faH(\frac{R}{z})}\,.
\end{equation}
When plugging the FG expansion \eqref{FGexpansion1}-\eqref{FGexpansion2} in the above equation, we find that at leading order the gauge field reduces to a pure gauge. This term can be removed by considering a further gauge transformation before going to the FG frame,
\begin{equation}
A_1(z,R)\rightarrow A_1(z,R)= -\frac{i\,dR}{\faH(\frac{R}{z})}+ d\left( i \log(R) \right)= \frac{i(-1+\faH(\frac{R}{z}))}{\faH(\frac{R}{z})}\frac{dR}{R}\,.
\end{equation}

We can then substitute the FG expansion \eqref{FGexpansion1}-\eqref{FGexpansion2} in the above expression and compute the limit
\eqref{eq:Jvev}, to derive the $U(1)$ $R$-symmetry current vev,
\begin{equation}\label{holovevJ}
\langle\!\langle J(\mathbb{r})\,\rangle\!\rangle_{\rm AdS}=
\frac{Q}{8\pi G_N}\frac{i\,d\mathbb{r}}{\mathbb{r}^3} \,.
\end{equation}
We note that $\langle\!\langle J\,\rangle\!\rangle_{\rm AdS}$ is conserved as expected.

We emphasise that the FG expansion \eqref{FGexpansion1}-\eqref{FGexpansion2} has been fixed 
solely by requiring that the metric takes the form \eqref{eq:FGmet}.
The fact that the same quantity $Q/(8\pi G_N)$ controls both the vev of the stress tensor  
\eqref{holovevT} and the $R$-symmetry current \eqref{holovevJ} is a consequence of the 
aforementioned BPS condition \emph{energy}=\emph{charge}.
We will elaborate more on this point in the next section. 

\vspace{0.2cm}
{\bf \underline{The holographic vev for the scalar fields}}

The last fields left to discuss are the scalars.
We obtain the boundary expansion of the scalar fields by simply plugging in the FG expansion \eqref{FGexpansion1}-\eqref{FGexpansion2} in the solution \eqref{5dmetricAdSbubble0}. As a result we find the boundary vevs,
\begin{equation}
\label{3ptcouplingHHscalars}
\langle\!\langle \varphi_1 (\mathbb{r})\rangle\!\rangle_{\rm AdS} = 
\frac{1}{8\pi G_N}\sqrt{\frac{2}{3}} \frac{Q}{\mathbb{r}^2}
\,, \qquad\qquad 
\langle\!\langle \eta_1 (\mathbb{r}) \rangle\!\rangle_{\rm AdS} =  \frac{1}{8\pi G_N}\frac{\sqrt{Q(2+Q)}}{\mathbb{r}^2}\,.
\end{equation}

We now move to discuss how to interpret the vevs here presented from the dual CFT side. 

\subsection{Matching with CFT data}

As discussed at the beginning of the previous section, the ${\rm AdS}$ bubble geometry 
should really be understood as a two-point function geometry generated by the insertion 
of two heavy operators in the dual CFT. Consequently, the holographic vevs of the light fields
we have just computed from the geometry correspond on the CFT side to three-point functions between 
the light operators and the two heavy operators sourcing ${\rm Ads}$ bubble. 
In this section we wish to match the holographic vevs presented in the previous section 
with three-point functions in the dual conformal theory. To this end we shall borrow 
some well-known CFT results from \cite{Osborn:1993cr,Freedman:1998tz}. 

In a CFT three-point functions are very constrained quantities, in particular their spacetime dependence 
is fully determined by conformal invariance. We begin by considering three-point functions 
involving either the $R$-symmetry current or the stress energy tensor. Let $d$ be the number of (boundary) 
spacetime dimensions, and denote by $\mathbb{O}_{\Delta}$ a generic eigenstate of the 
dilation operator. The normalised three-point functions read, 
\begin{align}
\frac{
\langle \mathbb{O}_{\Delta}(x_1)\overline{\mathbb{O}}_{\Delta}(x_2) J^{}_i(x_3)\rangle}{ \langle \mathbb{O}_{\Delta}(x_1)\overline{\mathbb{O}}_{\Delta}(x_2)\rangle } &= i{g_{{H\!H}\!J}}\frac{ (x_{12}^2 )^{\frac{d-1}{2} }}{ (x_{13}^2 x_{23}^2 )^{\frac{d-1}{2}}} (X_{12})_i \,, \label{eq:3ptJ}\\
\frac{
\langle \mathbb{O}_{\Delta}(x_1)\overline{\mathbb{O}}_{\Delta}(x_2) 
T_{ij}(x_3)\rangle}{ \langle \mathbb{O}_{\Delta}(x_1)\overline{\mathbb{O}}_{\Delta}(x_2)\rangle } & = g_{{H\!H}T}
\frac{ (x_{12}^2)^{\frac{d}{2}} }{ (x_{13}^2x_{23}^2)^{\frac{d}{2}} } \left[ -(X_{12})_i (X_{12})_j+ \frac{1}{d} \delta_{ij} (X_{12})^2 \right] \label{eq:3ptT}\,,
\end{align}
where we introduced the quantity,
\begin{equation}
(X_{12})_i\coloneqq \frac{ (x^2_{13}x^2_{23})^{\frac{1}{2}} }{ (x^2_{12})^{\frac{1}{2} }}\left[ -\frac{(x_{13})_i}{ x^2_{13} } + \frac{(x_{23})_i}{ x^2_{23} }  \right] \,.
\end{equation}
Similarly, the three-point function between two generic  $\mathbb{O}_{\Delta}$ 
and a scalar field $O_S$ with dimension $\Delta_S$ is simply,
\begin{equation}
\frac{
\langle \mathbb{O}_{\Delta}(x_1)\overline{\mathbb{O}}_{\Delta}(x_2) O_S(x_3) \rangle}{ \langle \mathbb{O}_{\Delta}(x_1)\overline{\mathbb{O}}_{\Delta}(x_2)\rangle } = g_{{H\!H}O}  \frac{ (x^2_{12})^{\frac{\Delta_S}{2}} }{  (x^2_{13}x^2_{23} )^{\frac{\Delta_S}{2}}} \label{eq:3ptS}\,.
\end{equation}

We now consider the above three-point functions \eqref{eq:3ptJ}-\eqref{eq:3ptT}-\eqref{eq:3ptS} 
in the same two-point function frame as  the ${\rm Ads}$ bubble.
That is, we interpret the insertions $ \mathbb{O}_{\Delta}(x_1) $ and $\overline{\mathbb{O}}_{\Delta}(x_2)$ 
as the two heavy operators sourcing the geometry, while the third operator will acquire a `vev' in this background.
Thus, we define the CFT `vevs' via the limit\footnote{We divide by the two point function of the
heavy operators so that the vev does not depend from their normalisation. This is equivalent 
to considering the insertions of the two heavy operators at $0$ and $\infty$ in radial quantisation. }
\begin{equation}\label{limitHapp_holocorr}
\langle\!\langle {\cal O} (\mathbb{x}) \rangle\!\rangle_{\rm CFT}\coloneqq
\lim_{\substack{  x\rightarrow \infty} } 
\frac{ \langle \mathbb{O}_{{ H}}(0)\overline{\mathbb{O}}_{{ H}}(x) {\cal O}(\mathbb{x}) \rangle}{ \langle \mathbb{O}_{{ H}}(0)\overline{\mathbb{O}}_{{ H}}(x) \rangle }\,.
\end{equation}

From~\eqref{eq:3ptJ}-\eqref{eq:3ptT}  we then find the vevs for the~$R$-current and the stress tensor
\begin{align} \label{CFTvevsforms}
\langle\!\langle J_i(\mathbb{x}) \rangle\!\rangle_{\rm CFT} = 
\frac{i{g_{{H\!H}\!J}}}{|\mathbb{x}|^{d-1}}\cdot \frac{\mathbb{x}_i}{|\mathbb{x}|}\,,\qquad\qquad
\langle\!\langle T_{ij}(\mathbb{x}) \rangle\!\rangle_{\rm CFT}= 
\frac{g_{{H\!H}T}}{ |\mathbb{x}|^d  } \left( -\frac{\mathbb{x}_i\mathbb{x}_j}{|\mathbb{x}|^2} + \frac{1}{d} \delta_{ij} \right)\,.
\end{align}
 The three-point couplings $g_{{H\!H}T}$ and $g_{{H\!H}\!J}$ are fixed in terms of the dimension, ${\Delta_{ H}}$, and the charge, $p$, of the heavy operators considered, namely
\begin{equation}\label{resultspetkou}
g_{{H\!H}\!J}=\frac{p}{S_d}\,,\qquad \qquad g_{{H\!H}T}=\frac{d}{d-1}\frac{{\Delta_{ H}} }{S_d}\,,\qquad \qquad {\rm with}\quad S_d\coloneqq \frac{
2\pi^{\frac{d}{2}}}{\Gamma[\frac{d}{2}]}\,.
\end{equation}

Thanks to the above results, we will now reinforce our previous claims where we 
related the back-reacted geometry with the maximal-trace operators on the CFT side.
As already commented, we know that the consistent truncation only contains fields 
from the graviton multiplet and therefore on the CFT side this heavy background can only 
correspond to the insertion of a superposition of multi-graviton operators, 
i.e.~maximal-trace operators  $\Omax{p}$ with $p=0\,{\rm mod}\,2$. 
As emphasised above, the results \eqref{resultspetkou} 
are valid when the heavy operator $\mathbb{O}_{{ H}}$ inserted is an eigenstate of the dilation operator, 
therefore it would seem that the present discussion is not immediately applicable to the case where $\mathbb{O}_{{ H}}$ corresponds to a coherent state operator.

However, this issue can be bypassed thanks to a couple of useful observations. Let us focus our attention first towards 
the scalars vevs. We know that there are only two possible topologies which can contribute to the three-point function of 
two maximal-trace operators and ${\cal O}_2$, see \eqref{topologies}. In fact these two different topologies are precisely realised when we consider 
the expectation values\footnote{With a slight abuse of notation we denote here the CFT operators by their corresponding scalar fields on the dual holographic side as given in \eqref{eq:phiPhi}-\eqref{eq:etPhi}.} $\langle\!\langle \varphi_1 \rangle\!\rangle_{\rm CFT}\propto g_{H\!H\varphi_1}$ and $\langle\!\langle \eta_1 \rangle\!\rangle_{\rm CFT}\propto g_{H\!H\eta_1}$ as discussed in \eqref{topology2}.
In particular, given that the three-point function $g_{H\!H\eta_1}\neq 0$ originates from a non-diagonal topology and since this three-point function actually reduces to a two-point function, 
we can deduce in this way which components of the maximal trace operator $\Omax{p}$ the coherent state is made of. 

Furthermore, from the point of view of three-point couplings we have that $g_{H\!H\!J}$, $g_{H\!HT}$ and $g_{H\!H\varphi_1}$ 
are \emph{diagonal} with respect to the heavy operators, 
see again~\eqref{topology2}, hence only the diagonal topology contributes to these CFT vevs. 
Therefore we conclude that the formulae in \eqref{resultspetkou} are valid even in the case 
where the heavy operator $\mathbb{O}_{{ H}}$ is the coherent-state operator.

We can now match against the supergravity expectations for the vevs of the $R$-symmetry current
and stress tensor. First of all, by going to radial coordinates on the Euclidean boundary 
we obtain a perfect match between the spacetime dependence of the CFT vevs presented 
in \eqref{CFTvevsforms} and their corresponding holographic versions given in \eqref{holovevJ}-\eqref{holovevT}. 
Furthermore we deduce that the CFT data and the gravity parameters must be related by   
\begin{equation}\label{eq:BPScond}
\frac{p}{S_4}=\frac{{\Delta_{ H}}}{S_4}=\frac{Q}{8\pi G_N} \qquad\Rightarrow\qquad {\Delta_{ H}}=p= N^2 \frac{Q}{2}\,,
\end{equation}
where we used the AdS/CFT dictionary to rewrite the five-dimensional Newton's constant in terms of CFT parameters,~i.e. $8\pi G_N
={4\pi^2 L^5}/{N^2}$.
We shall set the geometry scale $L=1$ from here onward. 

In Section \ref{sec:N2} we considered the maximal-trace operator $\mathbb{O}_{ H} 
\sim [\mathcal{O}_2]^{p/2}  $ with $\Delta = p= {\tilde \alpha} N^2$.  We then deduce from 
\eqref{eq:BPScond} that the supergravity parameter $Q$ is related to $\tilde{\alpha}$ via
\begin{equation}\label{relationtildealphaQ}
Q = 2 \tilde \alpha\,,
\end{equation}
or alternatively, using the map in parameters $\tilde{\alpha}=2\alpha^2/(1-2\alpha^2)$, 
\begin{equation}\label{relationalphaQ}
Q = \frac{4\alpha^2}{(1-2\alpha^2)}\,.
\end{equation}

Lastly, we consider the CFT vev \eqref{limitHapp_holocorr} of a scalar operator $O_S$ with 
dimension $\Delta_S$, which thanks to the three point function \eqref{eq:3ptS} it must take the form
\begin{equation}
\langle\!\langle O(\mathbb{x}) \rangle\!\rangle_{\rm CFT} =   \frac{ g_{{H\!H}O} }{  \ |\mathbb{x}|^{\Delta_S}}\,.\label{eq:vevSc}
\end{equation}
Holographically, we know that the vevs of the scalars
$\varphi_1$ and $\eta_1$ must be dual to scalar CFT operators of dimension $\Delta_S=2$. 
Comparing the holographic vevs \eqref{3ptcouplingHHscalars} against the CFT result \eqref{eq:vevSc} 
we immediately see that the spacetime dependence is perfectly matched on both sides.
Furthermore, it is immediate to read off the couplings 
\begin{equation}
g_{{H\!H}\varphi_1}=\sqrt{\frac{2}{3}} \frac{Q}{8\pi G_N}\,,\qquad\qquad  g_{{H\!H}\eta_1}= \frac{\sqrt{Q(2+Q)}}{8\pi G_N}\,.
\end{equation}
As mentioned, $\varphi_1$ sits in the same $\mathcal{N}=2$ graviton multiplet as the 
gauge field $A_1$ therefore the two vevs are equivalent and we find no new information compared to $g_{H\!H\!J}$. 
On the other hand, we can rewrite the three-point coupling $g_{{H\!H}\eta_1}$ making use of  
\eqref{relationtildealphaQ} to find the new constraint\footnote{Up to a normalisation factor, 
this is the same result derived in \cite[section 3.4]{Anempodistov:2025maj}.}
\begin{equation}
g_{{H\!H}\eta_1}=\frac{2}{8\pi G_N}\sqrt{ \frac{Q}{2}\left(1+\frac{Q}{2}\right)} = \frac{N^2}{2\pi^2}\sqrt{\tilde \alpha(1+\tilde \alpha)}\,.
\end{equation}
Using the map in parameters $\tilde{\alpha}=2\alpha^2/(1-2\alpha^2)$ this equation can be rewritten as
\begin{equation}\label{eq:ghhet}
g_{{H\!H}\eta_1}= \frac{\sqrt{2}N^2}{\pi^2}\left(\frac{\alpha}{1-2\alpha^2}\right)\,.
\end{equation}
Remembering from \eqref{eq:etPhi}  that the field $\eta_1$ is holographically dual to $\mathcal{O}_2$ 
we can then compare the above three-point functions against the corresponding non-diagonal contribution 
in the three-point function of $\mathcal{O}_2$ and two coherent-state operators.
Up to an overall numerical normalisation (i.e. independent from the parameter $\alpha$) for $\eta_1$, 
we find a perfect match between \eqref{eq:ghhet} and \eqref{terza} as well as the results of 
\cite{Giusto:2024trt,Aprile:2025hlt}.

\subsection{Ten-dimensional uplift and giant magnons}\label{10dmetricPope}

As anticipated, the five-dimensional geometry here considered can be realised as a consistent truncation of type IIB supergravity.
In \cite{Cvetic:2000nc}, the authors showed that it is possible to uplift the five-dimensional ${\rm Ads}$ bubble solution \eqref{5dmetricAdSbubble0}-\eqref{5dmetricAdSbubble} to the metric 
and the five-form field of type IIB ten-dimensional supergravity. 
Focussing on the $10$d metric part we have \cite[eq.(1)]{Cvetic:2000nc}
\begin{equation}\label{metricPope}
ds^2_{10}= G_{\mu\nu} dX^\mu dX^\nu=  \sqrt{ \Delta }\, ds_{5}^2 +  \frac{1}{\sqrt{\Delta}} (T_{ij})^{-1} D\mu^i D\mu^j\,.
\end{equation}
The coordinates $\mu^i$ with $i=0,...,5$ are subject to the constraint $\mu^i\mu^i=1$ and thus corresponds to the coordinates on the ${\rm S}^5$ factor.
Their covariant derivative is defined by
\begin{equation}
D\mu^i\coloneqq d\mu^i + A^{ij} \mu^j\,, \qquad \qquad \Delta\coloneqq T_{ij}\mu^i\mu^j\,.
\end{equation}
Here, the $6\times 6$ matrices $A^{ij}$ and $T_{ij}$ are those given in \eqref{matrixAfields} and \eqref{matrixT}, respectively.

We focus now on the~$U(1)$ truncation~\eqref{5dmetricAdSbubble0}.
In this case the~$6\times 6$ matrix~$T_{ij}$ reduces to
\begin{equation}
T= {\faH}^{\frac{1}{3}}  \left(\begin{array}{cc} 
\mathbb{1}_4& 0 \\
0 & \frac{1}{{\faH}} \left[\begin{array}{cc} e^{\eta_1} & 0 \\  0 & e^{-\eta_1} \end{array}\right]
\end{array}\right)\,.
\end{equation}
It is also convenient to parametrise the ${\rm S}^5$ by
\begin{equation}\label{coordinatesmucon_i}
\mu^0 + i \mu^3 = \cos\theta \cos\chi e^{i \psi_1},\qquad
\mu^1 + i \mu^2 = \cos\theta \sin\chi e^{i \psi_2},\qquad 
\mu^4+i \mu^5 = \sin\theta e^{i\phi}\,.
\end{equation}
We can identify an ${\rm S}^3\subset {\rm S}^5$ whose metric is given by $d{\tilde{\Omega}}_3^2= d\chi^2 + \cos^2\!\chi \,d\psi_1^2  + \sin^2\!\chi\,d\psi_2^2$. 
Using the coordinates defined above, we rewrite the metric on the internal manifold as
\begin{equation}
\frac{1}{\sqrt{\Delta}} (T_{ij})^{-1} D\mu^i D\mu^j =\frac{  
 \cos^2\!\theta\, d{\tilde{\Omega}}_3^2 + \sin^2\!\theta\, d\theta^2 + {\faH} \, (d\theta,D\phi).\mathbb{T}.(d\theta,D\phi) }{ \sqrt{ {\faH} \cos^2\!\theta +\mathbb{E}_{\eta_{1}}(\phi) \sin^2\!\theta  }}\,.
\end{equation}
In the above formula we have $D\phi = d\phi + A_1$ where $A_1$ is given in \eqref{5dmetricAdSbubble0}, and we introduced
\begin{align}
\mathbb{T}&=\left(
\begin{array}{cc} 
\cos^2\!\theta\, \mathbb{E}_{\eta_{1}}(\phi+\frac{\pi}{2}) & \frac{1}{2}\sin(2\theta)\sin(2\phi)\sinh(\eta_{1}) \,, \\[.2cm]
\frac{1}{2}\sin(2\theta)\sin(2\phi)\sinh(\eta_{1}) & \sin^2\!\theta\, \mathbb{E}_{\eta_{1}}(\phi)
\end{array} \right)\,, \\[.4cm]
\mathbb{E}_{\eta_{1}}(\phi)&=e^{{\eta_{1}}} \cos^2\!\phi + e^{-{\eta_{1}}}\sin^2\!\phi\,.
\end{align}
As it will shortly be important, we note that as function of the variable $\phi$ the quantity $\mathbb{E}_{{\eta_{1}}}(\phi)$ parametrises an ellipse whose shape depends on the value of the field $\eta_1$. 

Armed with all of the above definitions, we can finally derive the form of the Nambu-Goto action for 
the ${\rm AdS}$ bubble giant magnons presented without derivation in equation \eqref{rew_MH_magnons}.
We start by re-writing the Nambu-Goto action 
\begin{equation}\label{eq:NGapp}
S=\frac{ \sqrt{\lambda} }{ 2\pi }\int\!\!\dd t \,\dd\sigma\, \sqrt{ - {\rm det}( g_{ab}) }\,,
\qquad\qquad {\rm with} \quad g_{ab}=  G_{\mu\nu}\partial_{a}X^{\mu}\partial_{b}X^{\nu}\,.
\end{equation}
The induced metric is computed from \eqref{metricPope}
by specialising the coordinates $X^{\mu}$ to the embedding of the giant magnon. 

The worldsheet time is identified with the global time $t$ and the giant magnon sits at $r=0$ 
and $\chi=\psi_1=\psi_2=0$. Its motion is then parametrised by the worldsheet coordinate $\sigma$, 
which is identified with $\sigma= \phi$. Hence the magnon draws a non trivial profile in the remaining 
angle variable $\theta= \theta(\sigma)$. The relevant part of the ten-dimensional metric \eqref{metricPope} is thus
\begin{equation}\label{eq:gmApp}
    ds^2\Big\vert_{\rm giant-magnon}\!\!=\  - \sqrt{{\faH} \cos^2\!\theta + \mathbb{E}_{{\eta_{1}}}(\phi) \sin^2\!\theta }\,\frac{f dt^2}{\faH}   + 
   \frac{ \sin^2\!\theta\, d\theta^2 + 
    {\faH} \, (d\theta,D\phi).\mathbb{T}.(d\theta,D\phi)}{\sqrt{{\faH} \cos^2\!\theta +   \mathbb{E}_{{\eta_{1}}}(\phi)\sin^2\!\theta}} \,.
\end{equation}
Although the magnon is localised at $r=0$, in the above expression we kept the auxiliary function 
$f(r)\coloneqq 1+r^2 \faH(r)$ since we want to compute the action in the more general case for which the 
function $\faH(r)$ take the form given in \eqref{faHwithC} and in particular it may be singular as $r\to 0$. 
We will specify $r=0$ at the end of our calculation.

At this point we simply need substituting \eqref{eq:gmApp} in the Nambu-Goto action  \eqref{eq:NGapp} 
to arrive at the giant magnon action
\begin{equation}\label{ActiongeneralGM}
S=\frac{\sqrt{\lambda}}{2\pi} \int\!\! \dd t\, \dd \sigma \,\sqrt{f}\sqrt{ 
\mathbb{E}_{{\eta_{1}}}(\sigma) s^2(\sigma) 
-\mathbb{E}'_{{\eta_{1}}}(\sigma) s(\sigma)s'(\sigma)
+\mathbb{E}_{{\eta_{1}}}(\sigma+\tfrac{\pi}{2}) s'(\sigma)^2+ \frac{(f-1)}{f \faH} s'(\sigma)^2}\,,
\end{equation}
where we introduced the shorthand notation $s(\sigma)\coloneqq \sin\theta(\sigma)$ 
{and $'$ denotes a derivative with respect to the worldsheet coordinate $\sigma$.}

For the ${\rm AdS}$ bubble the function $\faH(r)$ takes the form \eqref{eq:faH} 
and it is in fact regular as $r\to 0$. This implies that in this case $f(r=0)=1$ and therefore 
the last term inside of the square root in \eqref{ActiongeneralGM} vanishes.
We then substitute the field  $\eta_1$ from the solution \eqref{5dmetricAdSbubble0}, 
and use the map in parameters \eqref{relationalphaQ} to compute 
explicitly the value of $\mathbb{E}_{{\eta_{1}}}(\sigma)\vert_{r=0}$,
\begin{equation}
\mathbb{E}_{{\eta_{1}}}(\sigma)\Big\vert_{r=0}=\frac{1+2\alpha^2 + 2\sqrt{2}\alpha \cos(2\sigma)}{1-2\alpha^2}\,.
\end{equation}
Plugging the above expression in the action \eqref{ActiongeneralGM} 
and after some trivial algebra we finally arrive at
\begin{equation}\label{actionLLMGM}
    S=\frac{ \sqrt{\lambda} }{ 2\pi }\int\!\!\dd t \, \dd\sigma\, \sqrt{ 
L^{*2}_{-} \left(\frac{d\mu^4}{d\sigma}\right)^2 +L^{*2}_{+} \left(\frac{d\mu^5}{d\sigma}\right)^2 }\,, \qquad\qquad {\rm with} \quad L^*_{\pm}=\sqrt{\frac{1\pm \sqrt{2}\alpha}{1\mp \sqrt{2}\alpha}}\,,
\end{equation}
where $\mu^4=\sin\theta\cos\sigma$ and $\mu^5=\sin\theta\sin\sigma$, as follows from \eqref{coordinatesmucon_i}. This is precisely the result quoted in equation \eqref{rew_MH_magnons} of the main text.

We conclude this appendix by discussing the giant magnon action \eqref{ActiongeneralGM} for the 
more general case where the function $H(r)$ is given by \eqref{faHwithC}. Following the same steps 
as above, and taking the limit $r\rightarrow 0$ at the end, we can show that even in this more general 
case the NG action takes exactly the form given in \eqref{actionLLMGM}, where the 
only modification is the replacement in the parameters
\begin{equation}
L^*_{\pm} \rightarrow \tilde{L}^*_{\pm} =  \sqrt{(1+Q) \pm \sqrt{(1+Q)^2-C^2}}
\end{equation}
We note that the parameters now satisfy the condition $\tilde{L}^*_{+}\tilde{L}^*_{-}   =C$. 
Therefore, by imposing $C=1$ we retrieve precisely the LLM condition ${L}^*_{+}{L}^*_{-}   =1$ 
which we know to correspond to the ${\rm AdS}$ bubble.

\bibliography{refs} 
\end{document}